\documentclass[10pt,journal]{IEEEtran}
\usepackage{array}
\usepackage{graphicx}
\usepackage{epsfig}
\usepackage{cite}
\usepackage{epstopdf}
\usepackage{amsfonts,amsmath,amsthm,amssymb,balance}
\usepackage{bbding}
\usepackage{multirow,balance}
\usepackage{bm}
\usepackage{makecell}
\usepackage[usenames,dvipsnames]{color}
\usepackage{colortbl}
\usepackage{float}
\usepackage{subfigure}
\usepackage{tabularx}
\usepackage{enumerate}
\usepackage[marginal]{footmisc}
\usepackage{url}
\usepackage[colorlinks,linkcolor=blue,citecolor=black]{hyperref}
\usepackage[hyphenbreaks]{breakurl}
\usepackage{pifont}

\makeatletter
\newif\if@restonecol
\makeatother

\usepackage[linesnumbered,ruled]{algorithm2e}
\usepackage{algpseudocode}

\graphicspath{{figures/}}
\def\BibTeX{{\rm B\kern-.05em{\sc i\kern-.025em b}\kern-.08em
		T\kern-.1667em\lower.7ex\hbox{E}\kern-.125emX}}
\usepackage{balance}

\begin{document}
	\title{A Survey on Resource Management in Joint Communication and Computing-Embedded SAGIN}
	
	\author{Qian~Chen,~\IEEEmembership{Member,~IEEE,}
		Zheng Guo,~\IEEEmembership{Graduate Student Member,~IEEE,}
		Weixiao~Meng,~\IEEEmembership{Senior~Member,~IEEE,} \\
		Shuai~Han, ~\IEEEmembership{Senior~Member,~IEEE,}
		Cheng~Li, ~\IEEEmembership{Senior~Member,~IEEE,}
		Tony Q. S. Quek,~\IEEEmembership{Fellow,~IEEE}

		\thanks{
			
			
			Q. Chen, Z. Guo, W. Meng and S. Han are with the School of Electronics and Information Engineering, Harbin Institute of Technology, Harbin 150001, China (email: joycecq@163.com; zguohit@163.com; wxmeng@hit.edu.cn; hanshuai@hit.edu.cn).
			
		C. Li is with the School of Engineering Science, Simon Fraser University, Burnaby, BC V5A 1S6, Canada, and also with Electrical and Computer Engineering, Memorial University, St. John's, NL A1B 3X5, Canada (e-mail: li\_cheng@sfu.ca).

   T. Q. S. Quek is with the Singapore University of Technology and Design, Singapore 487372, and also with the Yonsei Frontier Lab, Yonsei University, South Korea (e-mail: tonyquek@sutd.edu.sg).
   
   (\textit{Corresponding author: Weixiao Meng.})
			
			}
	}
	
	\markboth{} {Shell \MakeLowercase{\textit{et al.}}: Bare Demo of IEEEtran.cls for IEEE Journals}

	\maketitle
	
    \begin{abstract}
The advent of the 6G era aims for ubiquitous connectivity, with the integration of non-terrestrial networks (NTN) offering extensive coverage and enhanced capacity. As manufacturing advances and user demands evolve, space-air-ground integrated networks (SAGIN) with computational capabilities emerge as a viable solution for services requiring low latency and high computational power.
Resource management within joint communication and computing-embedded SAGIN (JCC-SAGIN) presents greater complexity than traditional terrestrial networks. This complexity arises from the spatiotemporal dynamics of network topology and service demand, the interdependency of large-scale resource variables, and intricate tradeoffs among various performance metrics. Thus, a thorough examination of resource management strategies in JCC-SAGIN is crucial, emphasizing the role of non-terrestrial platforms with processing capabilities in 6G.
This paper begins by reviewing the architecture, enabling technologies, and applications in JCC-SAGIN. Then, we offer a detailed overview of resource management modeling and optimization methods, encompassing both traditional optimization approaches and learning-based intelligent decision-making frameworks. Finally, we outline the prospective research directions in JCC-SAGIN. \\
    	
 \noindent
	\end{abstract}

	\IEEEpeerreviewmaketitle

\section{Introduction}\label{sec:introduction}
\subsection{Background and Motivation}
The advent of 5G has signified a pivotal shift in communication applications, transitioning from mobile Internet to the industrial Internet of Things (IoT), thereby facilitating the integration with various vertical industries. This successful transition to 5G has established a robust theoretical groundwork for the subsequent exploration and development in 6G technologies. Distinct from 5G's emphasis on communication among individuals, vehicles, and IoT, 6G aims to broaden connectivity from a terrestrial two-dimensional scope to a global three-dimensional one, thereby substantially minimizing the ``digital divide" among individuals. Research conducted by Machina Research anticipates that global IoT connections are poised to reach 27 billion by 2025, generating in excess of 2 Zettabytes of data. This surge in device scale and data volume necessitates enhanced support for services that are both latency-sensitive and computationally demanding, such as virtual reality (VR) and real-time video analysis. Additionally, the diversity in communication needs is becoming increasingly prevalent across a broader geographical expanse.

Terrestrial cellular networks, boasting high speeds, low latency, and extensive connectivity, can fulfill the communication requirements in typical urban and suburban locales. In the present information era, the communication scope has expanded to remote regions, including oceans, volcanoes, deserts, and forests. Nonetheless, the establishment of terrestrial networks (TN) in these areas is impeded by commercial and operational considerations. Presently, TN can cover a mere 7\% of the Earth's surface, leaving an estimated 3.4 billion individuals without network access. This falls significantly short of the objective of seamless global coverage. Hence, 6G envisions the incorporation of non-terrestrial networks (NTN), utilizing satellites, unmanned aerial vehicles (UAVs), and high altitude platforms (HAPs) to supplement terrestrial coverage. The construction of multidimensional space-air-ground integrated networks (SAGIN) is crucial to achieve comprehensive area coverage. For the sake of brevity, this paper refers to non-terrestrial platforms as space-air platforms (SAPs).

\begin{figure*}[!h]
	\centering
	\includegraphics[width = 0.9\textwidth]{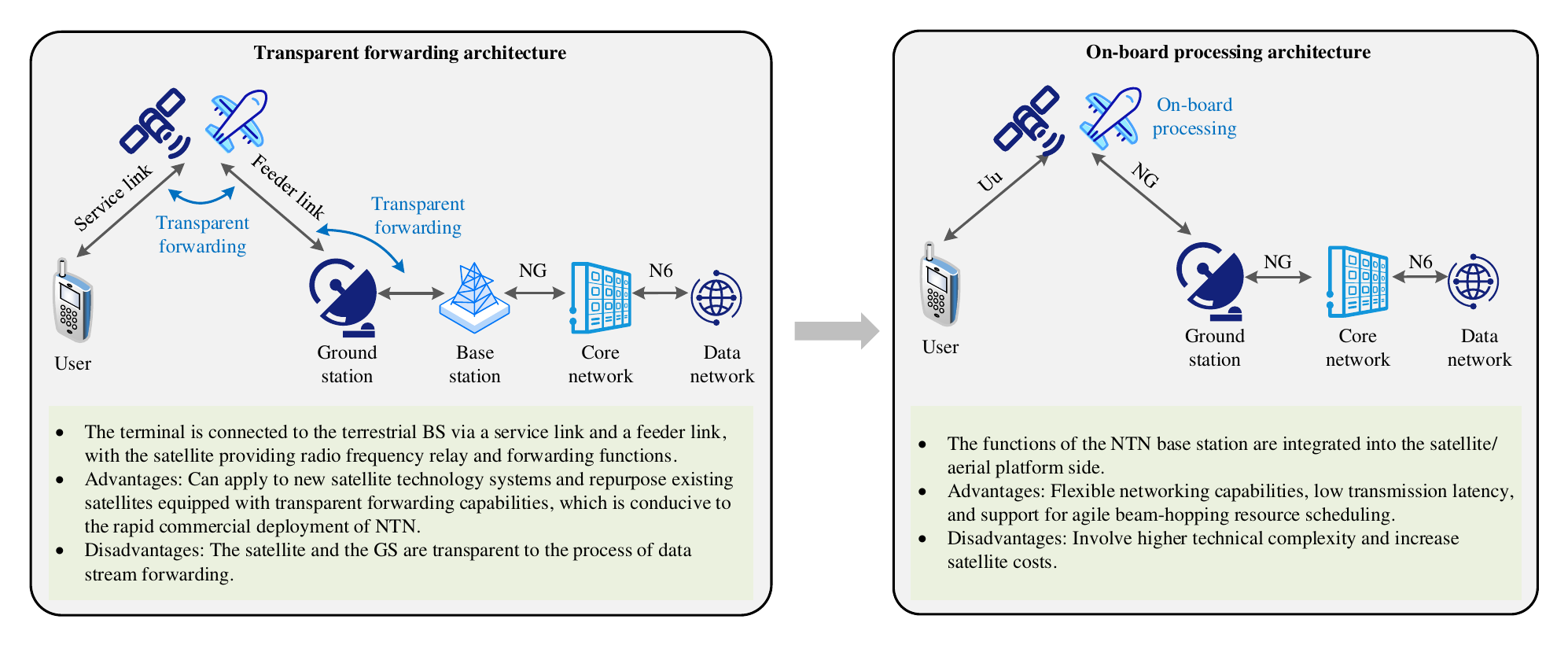}
	\caption{Two types of frameworks of NTN: Transparent forwarding and onboard processing. \label{fig:TransparentVsProcessing}}
\end{figure*}

Fig. \ref{fig:TransparentVsProcessing} delineates two predominant frameworks within NTN: transparent forwarding and onboard processing. The former, a traditional mode for SAPs, involves the mere amplification, conversion, and re-amplification of received signals by the SAPs' transparent transponders. This method is optimally suited for applications demanding high-speed data transmission, such as video streaming and bulk file transfers. Its primary advantage lies in the requirement for adjustments only to terrestrial equipment when altering communication links, obviating the need for satellite equipment upgrades, hence its widespread adoption. Recent advancements in manufacturing technology have enabled a shift to onboard processing. This mode allows for the direct processing of data on satellites, encompassing tasks like data compression, encryption, decryption, and routing. It is particularly useful for applications that require data processing capabilities, such as voice communications and internet access. In addition to the foundational elements of transparent forwarding, the processing transponder in this mode is supplemented with a demodulator, modulator, and signal processing unit, although this can introduce additional latency.

\begin{figure*}[h]
	\centering
	\includegraphics[width = 0.7\textwidth]{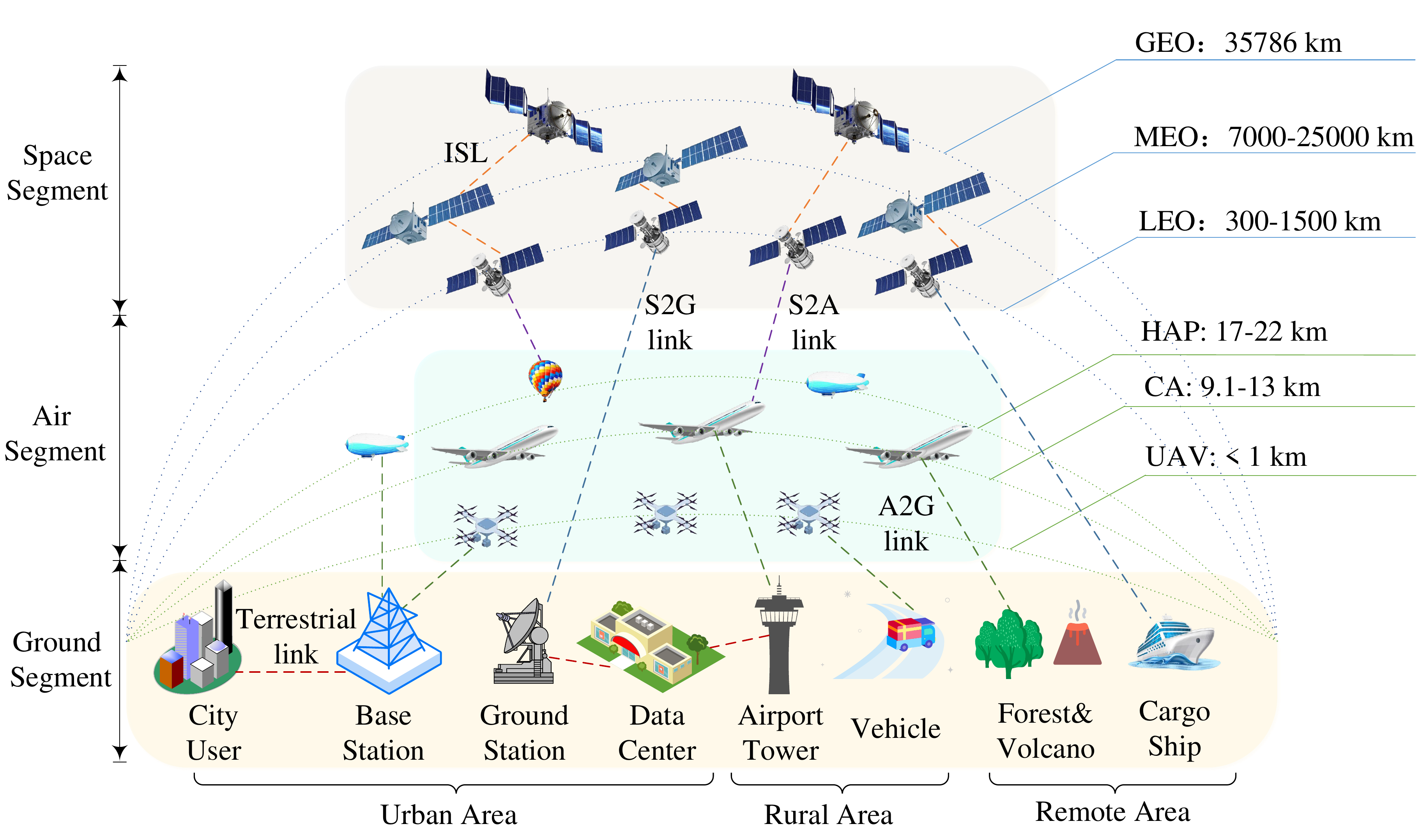}
	\caption{The architecture of JCC-SAGIN. \label{fig:network_architecture_CCSAGIN}}
\end{figure*}

With the implementation of onboard processing in SAPs, the possibility of integrating communication and computing within NTN becomes both feasible and promising. The architecture of joint communication and computing-embedded SAGIN (JCC-SAGIN) is illustrated in Fig. \ref{fig:network_architecture_CCSAGIN}. This approach is particularly advantageous for latency-sensitive or computation-intensive services, especially in remote areas. Such services, traditionally limited by the computational capacities of terrestrial IoT devices or reliant on remote cloud computing at ground stations (GSs), can now be offloaded to visible SAPs for edge computing. JCC-SAGIN, representing a significant advancement in the field, necessitates a comprehensive review and summarization of its research issues and challenges to fully realize its potential.

\subsection{Overview of Related Surveys and Our Contributions}

\subsubsection{Surveys on NTN}\label{subsec:survey_NTN}
Table \ref{tab:existing_survey_SAGIN} is a summary of the recent three years' surveys concerning NTN. A chronological analysis reveals an evolution in network architecture research, progressing from singular-layer NTN configurations such as satellite, HAP, or UAV networks, to more complex multi-layer NTN frameworks. Additionally, the focus of research has shifted from purely communication-centric issues to encompassing both communication and computing challenges.

\begin{table*}[!h]
	\centering
	\caption{Existing Surveys Related to non-terrestrial networks}\label{tab:existing_survey_SAGIN}
	\begin{tabular}{|m{0.6cm}<{\centering}|m{0.6cm}<{\centering}|m{1.1cm}<{\centering}|m{1.6cm}<{\centering}|m{11cm}<{\centering}|} \hline
		Ref. & Year & Network & SAP as Edge Computing Server & Contribution \\ \hline \hline
		\cite{8368236} & 2018 & SAGIN & \ding{55}&  The first extensive overview of the integration of space, air, and ground networks, delving into the research challenges, technologies, and applications in this emerging field.  \\ \hline
		\cite{9079470} & 2020 & Satellite & \ding{55}& A literature review encompassing a range of CubeSat missions and recent developments in constellation-and-coverage issues, channel modeling, physical techniques, and networking. \\ \hline
		\cite{8850067} & 2020&  Satellite & \ding{55}& A thorough survey that focuses on space information networks, particularly highlighting PLS.  \\ \hline
		\cite{8943319} & 2020 & UAV& \ding{51} &  An investigation into the effects at both cellular and systemic levels on UAV networks, presented from a cyber-physical system perspective. \\ \hline
		\cite{9442378} & 2021 & Satellite  & \ding{51} & Compile current solutions regarding the deployment of IoT devices in remote locations within satellite-based IoT systems. \\ \hline
		\cite{9210567} & 2021 & Satellite & \ding{51} & A detailed review capturing key innovation drivers, promising applications, and a comprehensive literature survey covering system aspects, air interfaces, medium access, networking, testbeds, and prototyping in satellite systems. \\ \hline
		\cite{9488323} & 2021 & UAV &  \ding{51} & An exhaustive survey on security-intensive drone applications, including the challenges and architectural solutions in UAV communication and computing-enabled networks. \\ \hline
		\cite{9508366} & 2021 & UAV &  \ding{51} & A complete review that enables connectivity applications of aerial vehicles, integrating advanced communication technologies. \\ \hline
		\cite{9380673} & 2021 & HAP & \ding{51} & A discussion on the technologies related to HAP energy and payload systems, radio resource management, handoff strategies, and physical layer techniques within HAP networks. \\ \hline
		\cite{9358097} & 2021 & HAP & \ding{51} & An overview of recent developments in ARAN, emphasizing the ongoing research directions toward 6G aerial radio access networks (ARAN). \\ \hline
		\cite{9768113} & 2022 & UAV  &  \ding{55}& A survey that examines UAV communications from an industrial viewpoint, analyzing the potential and limitations of 5G NR features in aerial devices and identifying promising 6G enablers for UAV communication. \\ \hline
		\cite{9598918} & 2022 & UAV & \ding{55}& A detailed analysis focusing on mmWave beamforming-enabled UAV communications and networking. \\ \hline
		\cite{9628162} & 2022 & SAGSIN & \ding{55} & The first survey to critically review state-of-the-art security in SAGSIN, including discussions on cross-layer attacks and security countermeasures. \\ \hline
		\cite{10045716} & 2023 & NTN & \ding{55} & A comprehensive survey reviewing both single-tier and multi-tier scenarios in aerospace integrated networks. \\ \hline
		\textbf{Ours} & 2024 & SAGIN & \ding{51} & An all-encompassing overview that discusses the network and computing architecture, resource management strategies, security issues, and future research directions in JCC within SAGIN. \\ \hline
	\end{tabular}
\end{table*}

Regarding satellite network surveys, Saeed \textit{et al.} \cite{9079470} explored various dimensions including constellation-and-coverage concerns, channel modeling, physical techniques, and networking. Li \textit{et al.} offered insights into physical layer security (PLS) techniques within space information networks \cite{8850067}. Subsequent studies began to incorporate scenarios where satellites are equipped with processing capabilities. Centenaro \textit{et al.} \cite{9442378} examined practical challenges in satellite-based IoT systems, while Kodheli \textit{et al.} \cite{9210567} reviewed upper-level system design considerations in the booming space era.

In the domain of aerial networks, the focus has been primarily on UAV and HAP networks. Wang \textit{et al.} \cite{8943319} investigated the interplay between communication, computation, and control in UAV networks, presenting a comprehensive survey from a cyber-physical systems perspective. Various aspects of UAV networks were explored, including security solutions in communication and computing-enable networks \cite{9488323}, connective applications \cite{9508366}, industrial applications \cite{9768113}, and the utilization of mmWave beamforming \cite{9598918}. Additionally, pivotal technologies in HAP networks concerning resource management, handoff strategies, and PLS technologies were discussed \cite{9380673} \cite{9358097}. Shirin Abkenar \textit{et al.} provided an exhaustive review of mobile edge computing (MEC) nodes, covering vehicular, spatial, aerial, and maritime nodes \cite{9910565}.

Following Liu \textit{et al.}'s foundational survey on SAGIN in 2018 \cite{8368236}, research expanded to include the integration challenges of heterogeneous networks. The first survey on security issues within space-air-ground-sea integrated networks (SAGSIN) highlighted cross-layer attacks and security countermeasures \cite{9628162}. Zhou \textit{et al.} contributed a thorough literature overview of both single-tier and multi-tier NTN networks \cite{10045716}. It is observed that existing SAGIN surveys predominantly concentrate on communication aspects, leaving a gap in the understanding of the impact of computing functions of SAPs on SAGIN's performance.

In the context of SAGIN, challenges encountered in pure communication issues predominantly revolve around the establishment and maintenance of reliable and efficient links between network nodes. Conversely, JCC issues encompass not only communication aspects but also extend to encompass processing and computational tasks within the network. The principal distinctions between the challenges in pure communication issues and JCC problems in SAGIN are as follows:

\begin{itemize}
	\item \textit{Network and Computing Architectures}: JCC challenges necessitate the formulation of architectures capable of efficiently supporting both communication and computational tasks. This encompasses the development of network nodes equipped with integrated processing capabilities and the design of distributed computing systems that can function in tandem with non-terrestrial communication networks.

\item \textit{Task Offloading}: JCC problems involve making decisions regarding the offloading of computational tasks from one node to another. This demands consideration of various factors, including the computational capabilities, communication delays, and resource availability at different nodes. However, pure communication problems do not involve such offloading decisions. 

\item \textit{Edge Computing}: In JCC scenarios, edge computing becomes a pivotal element, facilitating processing tasks closer to the end-users. This reduces latency and minimizes the need for extensive data transmission. Developing appropriate edge computing architectures, protocols, and algorithms is essential in JCC problems but is not typically addressed in pure communication issues.

\item \textit{Quality of Service (QoS)}: In pure communication problems, QoS focuses on metrics related to communication performance such as latency, throughput, and reliability. However, in JCC problems, QoS also needs to account for computation-centric metrics, like computing latency, energy consumption, and task completion rates.

\item \textit{Resource Management}: While pure communication issues mainly concentrate on allocating resources like bandwidth, power, and time slots to optimize communication performance, JCC challenges require the allocation of resources for computational tasks, including processing, storage, and energy. This introduces added layers of complexity in terms of resource management and optimization.
\end{itemize}


\subsubsection{Surveys on the Integration of Communication with Other Technologies}\label{subsec:survey_JCC}
Scholars are increasingly exploring the fusion of communication with other technologies like computing, sensing, and caching to enhance overall performance. Table \ref{tab:existing_survey_CC} summarizes existing representative surveys on this integration. 

Recent advancements in integrated sensing and communication (ISAC) have highlighted its potential to optimize the use of spatial, temporal, frequency, and power resources. This optimization comes alongside reductions in hardware and software complexity within wireless communications and radar sensing domains. The core of ISAC's effectiveness lies in its signal processing capabilities, which are pivotal for enhancing both sensing and communication functions.
Wei \textit{et al.} \cite{wei2023integrated} conducted a thorough examination of ISAC signal methodologies in the context of 5G-Advanced and 6G networks, addressing signal design, processing, and optimization, alongside methods for radar signal processing and interference management strategies. Utilizing the distinct benefits of multiple-input multiple-output (MIMO) technology, Fang \textit{et al.} \cite{fang2022joint} explored advancements in ISAC via spatial beamforming and waveform shaping, incorporating novel MIMO models with cloud radio access networks, UAVs, and reconfigurable intelligent surfaces (RISs).  
Luong \textit{et al.} \cite{9393464} presented an extensive survey on the resource management challenges inherent in ISAC systems. Concurrently, Ma \textit{et al.} \cite{8944276} reviewed the integration of IoT sensing, computing, and communication, with a particular focus on advancements in energy harvesting (EH) for IoT systems.

The increasing computational demands in recent years have prompted significant updates and integrations of communication and computation within wireless networks. Mach \textit{et al.} \cite{7879258} addressed critical aspects of MEC in cellular networks, including computing offloading decisions, computing resource allocation, and mobility management. Shirin Abkenar \textit{et al.} \cite{9910565} categorized MEC nodes by their environments— aerial, ground vehicular, spatial, and maritime—examining their network architectures, operational methods, challenges, and integrated solutions. Lastly, Wang \textit{et al.} \cite{8060515} expanded the discussion to encompass networking, caching, and computing integration, addressing key challenges and tradeoffs.


\begin{table*}[!h]
	\centering
	\caption{Existing Surveys Related to the integration of communication and other technologies}\label{tab:existing_survey_CC}
	\begin{tabular}{|m{0.6cm}<{\centering}|m{0.6cm}<{\centering}|c|m{1.8cm}<{\centering}|m{10cm}<{\centering}|} \hline
		Ref. & Year & Network &  Technique & Contribution \\ \hline \hline
		\cite{7879258} &2017 & Cellular network & Communication and computing &  This survey represents the pioneering work in the field of MEC, addressing key decision-making aspects such as computing offloading, allocation of computing resources, and mobility management within cellular networks. \\ \hline
		\cite{8060515} & 2018 & Cellular network & Networking, caching, and computing & A comprehensive discussion on the research challenges associated with the integration of networking, caching, and computing. This includes critical considerations like latency requirements, interface design, mobility management, and the tradeoffs between resources and architecture, as well as the aspects of convergence.  \\ \hline
		\cite{8944276} & 2020 & IoT & Sensing, computing and communication &An overview of the latest developments in EH for IoT, focusing on the integration of IoT sensing, computing, and communications. This encompasses commercial development aspects, hardware innovations, checkpointing and timekeeping techniques, applications of artificial intelligence (AI), packet loss, and backscatter communication technologies. \\ \hline
            \cite{zhang2021enabling}& 2021 & Mobile network & Sensing and communication& A comprehensive review that explores the progression of perceptive mobile network (PMN) and the potential issues in implementing ISAC technology. This includes performance bounds, waveform optimization, antenna array design, clutter suppression, sensing parameter estimation, resolving sensing ambiguities, pattern analysis, networked sensing under cellular topology, and sensing-assisted communications.\\ \hline
		\cite{9393464} & 2021 & Cellular network & Sensing and communication & An exhaustive review of literature pertaining to resource management in systems that integrate radar and communication. This includes an analysis of performance metrics, spectrum sharing strategies, power allocation methods, interference management, and security considerations. \\ \hline
           \cite{9910565}	& 2022 & Mobile Network&  Communication and computing & A detailed examination of MEC nodes, covering a range of environments such as spatial, aerial, vehicular, and maritime. \\ \hline
		\cite{fang2022joint}& 2023 & IoT & Sensing and communication& A survey specializing in ISAC applications using MIMO technology. This review discusses fundamental models combined with cloud-radio access networks (C-RANs), UAVs, and RISs, highlighting the potential and challenges of ISAC MIMO designs. \\ \hline
		\cite{wei2023integrated} & 2023 & Mobile network & Sensing and communication& An analysis of ISAC signals, encompassing ISAC signal design, signal processing techniques, and optimization strategies in ISAC applications. \\ \hline
 \textbf{Ours} & 2024 & SAGIN & Networking, communication and computing & A survey on the theoretical and engineering advancements in network architecture, application areas, and optimization methods within the context of JCC-SAGIN. \\ \hline
	\end{tabular}
\end{table*}

Compared to JCC challenges in TN, JCC issues in SAGIN present additional complexities:
\begin{itemize}

\item \textit{Mobility and Propagation Delay}: The mobility of SAPs is significantly higher than base stations (BSs) in TN. NTN also experience greater propagation delays due to the extended distances involved, impacting the efficacy of communication protocols (particularly those requiring prompt feedback, like transmission control protocol (TCP)), adaptive routing, handover management, and resource allocation strategies.

\item \textit{Line-of-Sight (LoS) and Channel Conditions}: NTN predominantly depend on LoS communication links, which are vulnerable to obstructions by buildings, terrain, or weather conditions, unlike TN. NTN are also more prone to severe channel impairments like signal fading, atmospheric attenuation, and interference from other systems, necessitating robust link management strategies, and advanced error correction and signal processing techniques to ensure consistent and reliable connectivity.

\item \textit{Scalability and Integration}: NTN must accommodate extensive geographic coverage and a large user base, presenting scalability challenges in routing and network management. They also require seamless integration with other communication systems, alongside the development of interoperable standards, protocols, and architectures.

\item \textit{Security}: The distinctive characteristics of NTN, such as their remote and distributed nature, expose them to different security threats than TN. This necessitates specialized security mechanisms and protocols to ensure secure communication and computing.

\item \textit{Resource Management}: NTN nodes, particularly those in space, often have constrained resources in terms of power and computational capabilities. This demands the creation of energy-efficient communication and computing solutions, as well as the implementation of onboard processing to meet QoS requirements. 
	
\end{itemize}
	
\subsubsection{Surveys on Resource Management}
\begin{figure}[h]
	\centering
	\includegraphics[width = 0.4\textwidth]{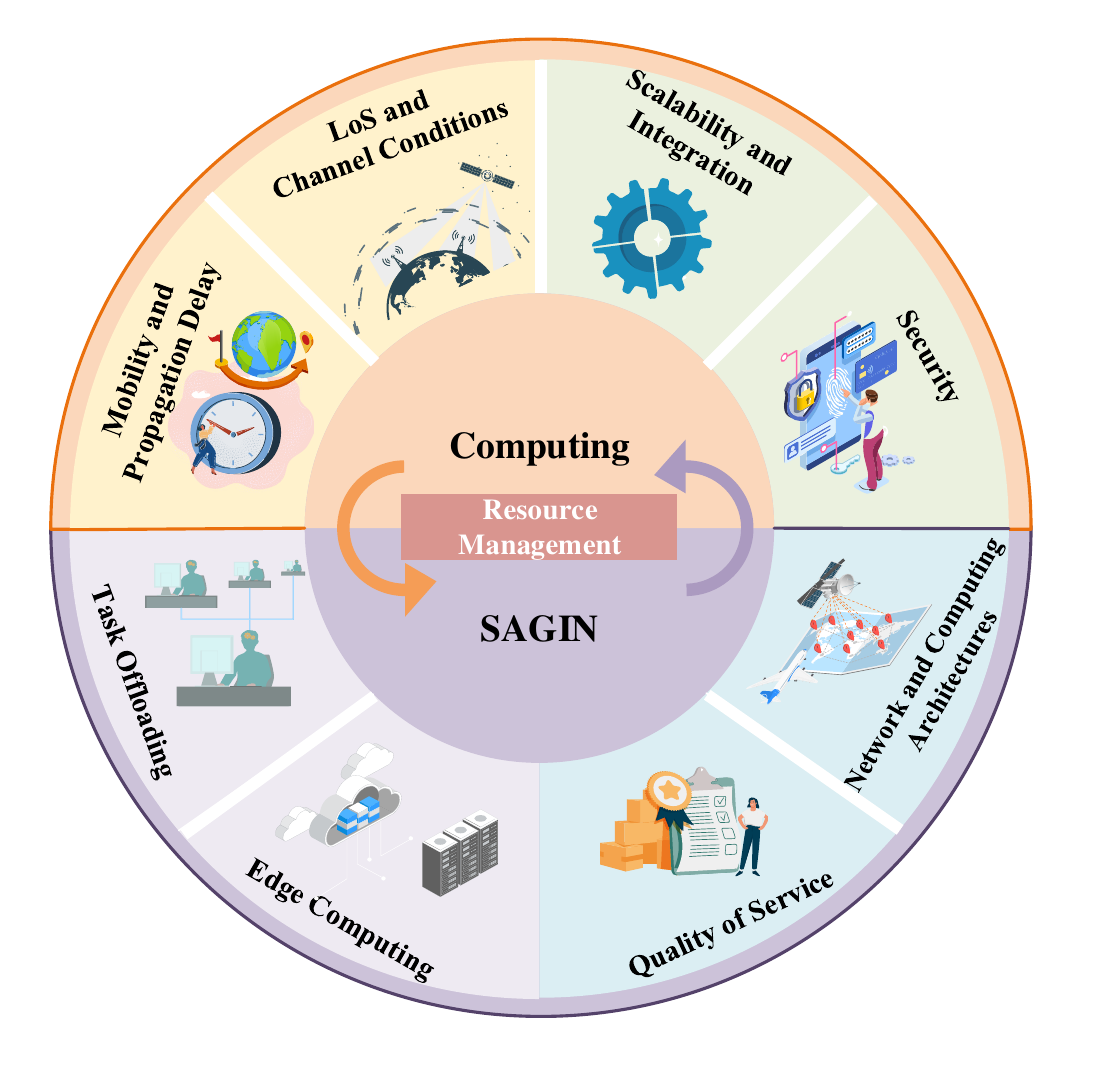}
	\caption{How computing impacts SAGIN, and how SAGIN influences computing. \label{fig:Challenge_JCCSAGIN}}
\end{figure}

Fig. \ref{fig:Challenge_JCCSAGIN} illustrates the interplay between computing and SAGIN, highlighting their reciprocal influence. This paper focuses on the pivotal challenges of resource management in JCC-SAGIN.
Previous literature has thoroughly examined radio resource management (RRM) in the context of 5G and beyond \cite{9139384}, heterogeneous networks (HetNets) \cite{9896125}, IoT networks \cite{8951180}, MEC systems \cite{9858872, 8847416}, and big data networks \cite{9478917}. These studies have advanced our understanding of resource management across various domains.
A summary of the major contributions from these surveys is presented in Table \ref{tab:existing_survey_RM}, which assesses the discussion on four critical aspects: the hardware used in communication and computing, application scenarios, traditional optimization methods, and the application of AI for intelligent decision-making strategies.
However, \textit{no existing study has comprehensively reviewed resource management within SAGIN that encompasses the aforementioned critical aspects}. This oversight underscores the novelty and importance of this paper.

\begin{table*}[!h]
	\centering
	\caption{Existing Surveys Related to Resource management}\label{tab:existing_survey_RM}
	\begin{tabular}{|m{0.6cm}<{\centering}|m{0.5cm}<{\centering}|m{1.1cm}<{\centering}|c|c|m{1.1cm}<{\centering}|c|m{7.5cm}<{\centering}|}
\hline 
Ref. & Year & Network &  Hardware & Applications & Traditional Alg. & AI Alg. & Contribution    \\ \hline \hline
\cite{9896125} & 2022 & 5G HetNets & \ding{55} & \ding{55} & \ding{51} & \ding{51} & A discussion on the critical role of RRM in enhancing 5G HetNets through addressing challenges such as interference and resource allocation, presenting a survey of RRM schemes, and outlining future challenges and opportunities.  \\ \hline
\cite{9858872} & 2022 & MEC system & \ding{55} & \ding{55} & \ding{55} &  \ding{51} & An overview of machine learning and deep learning applications for resource allocation in MEC systems, detailing their use in task offloading, scheduling, and joint resource allocation, and discussing future challenges and directions in this area. \\ \hline 
\cite{9478917} & 2021 & Big data networks &  \ding{51} & \ding{51} & \ding{55} & \ding{55} & A comprehensive review of big data deployment architectures, introducing a taxonomy for classifying these models based on their communication systems, as well as the implications, benefits, and challenges of modern big data environments compared to traditional dedicated clusters. \\ \hline
\cite{9139384} & 2020 & 5G and beyond & \ding{55} & \ding{55} & \ding{51} &  \ding{55} & A survey paper reviews research and standardization efforts aimed at addressing cross-link interference in dynamic time division duplex (D-TDD) systems for 5G New Radio, categorizing mitigation approaches, discussing signaling requirements, and presenting performance analysis in various environments. \\ \hline
\cite{8951180} & 2020 & Cellular and IoT networks & \ding{55} & \ding{55} & \ding{51} & \ding{51} & An in-depth survey of machine learning and deep learning based resource management mechanisms in cellular wireless and IoT networks, addressing the challenges of resource management, reviewing traditional and learning-based techniques, and identifying future research directions for enhancing IoT networks' efficiency and intelligence. \\ \hline
\cite{8847416} & 2020 &  MEC system & \ding{55} &  \ding{51} &  \ding{55} & \ding{51} & A survey presents the application of machine learning algorithms in MEC systems to address high-dimensional challenges in task offloading, resource allocation, and inter-server communication for 5G and IoT networks. \\ \hline
\textbf{Ours} & 2024 & SAGIN & \ding{51} & \ding{51} & \ding{51} & \ding{51} & We provide a comprehensive survey of computing hardware, JCC applications, traditional and AI algorithms of resource management optimization in JCC-SAGIN. \\ \hline
\end{tabular}
\end{table*}

\subsubsection{Contributions of This Paper}

This study distinguishes itself from existing surveys by focusing on several critical distinctions. Unlike other surveys on NTN, this research highlights the onboard computing capabilities of SAPs and addresses the complex resource management challenges inherent in coordinating SAGIN's multiple segments. Moreover, given the unique and additional challenges SAGIN present, our work diverges from prior works that primarily concentrate on JCC techniques or resource management.
Our objective is to fill the gaps identified in the literature by providing an extensive survey on resource management within JCC-SAGIN, contributing to both theoretical research and practical system design advancements.
To the best of our knowledge, \textit{this paper represents the first exhaustive review to detail the evolution of network integration and computing hardware in SAGIN. It discusses practical engineering applications and reviews theoretical resource management methods in JCC-SAGIN.}
The main contributions of this paper are outlined as follows:
\begin{itemize}
    \item We introduce the innovative concept of ``joint communication and computing-embedded SAGIN", showcasing non-terrestrial platforms equipped with onboard processing and data forwarding capabilities. This paper marks the first comprehensive exploration of  reviewing the evolution of computing hardware and network integration in SAGIN.

    \item We identify and analyze the enabling technologies in JCC-SAGIN. Following this, we discuss various supporting applications within JCC-SAGIN, illustrating their practical implementation through relevant engineering projects.

    \item We conduct an in-depth literature review on resource management within JCC-SAGIN, emphasizing modeling and optimization strategies. This encompasses traditional optimization techniques as well as learning-based intelligent decision-making methods. Additionally, we propose potential directions for future research that cover aspects from the physical layer to the application layer.
\end{itemize}

\begin{figure*}[!h]
	\centering
	\includegraphics[width = 0.6\textwidth]{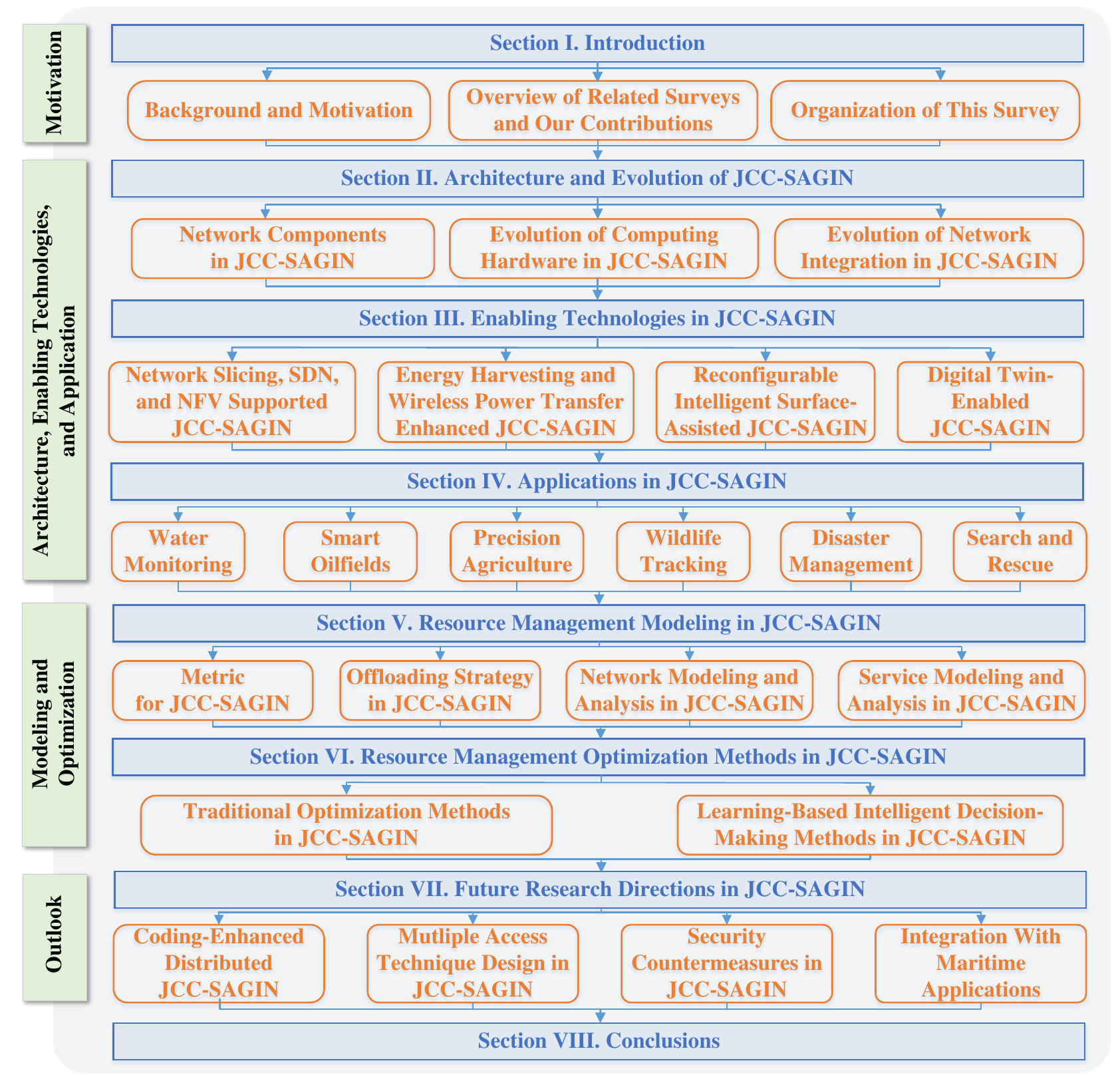}
	\caption{Organization of this survey and the logical relationship between different sections. \label{fig:thesis_organization}}
\end{figure*}
\subsection{Organization of This Survey}
For readability, Table \ref{tab:abbreviation} summarizes the primary acronyms. 
Fig. \ref{fig:thesis_organization} illustrates the structural organization of this survey. Following the introductory section, which illustrates the context and reviews pertinent existing surveys, a thorough examination of JCC-SAGIN is undertaken from Section \ref{sec:architecture_Evolution} to Section \ref{sec:future}.
Specifically, Section \ref{sec:architecture_Evolution} summarizes the architecture and evolution of JCC-SAGIN, tracing the network components and the evolution of network integration and computing hardware in JCC-SAGIN. 
Section \ref{sec:enabling_tech} discusses the enabling technologies, encompassing network slicing, software-defined networking (SDN), network function virtualization (NFV), EH and wireless power transfer (WPT), and RIS within JCC-SAGIN. Building upon the foundational architecture and promising techniques, Section \ref{sec:application} explores the typical applications facilitated by JCC-SAGIN and highlights engineering cases that have been successfully implemented.
The survey progresses from Section \ref{sec:offloading} to Section \ref{sec:RM_optimization_method}, where we cover resource management modeling and optimization methods in JCC-SAGIN. In Section \ref{sec:future}, we identify future research opportunities, paving the way for further exploration and innovation in this field. Our work is summarized in Section \ref{sec:conclusion}.

\begin{table*}[!ht]
	\centering
	\caption{LIST OF MAIN ABBREVIATIONs}\label{tab:abbreviation}
	\begin{tabular}{|p{1.5cm}<{\centering}|p{6cm}<{\centering}|p{1.5cm}<{\centering}|p{6cm}<{\centering}|}
		\hline
		{Acronyms} & {Description} & {Acronyms}	& {Description}  \\ \hline \hline
  AC	&	Active Communication		&	MATD3	&	Multi-Agent TD3		\\ \hline
AI	&	Artificial Intelligence		&	MDP	&	Markov Decision Process		\\ \hline
AP	&	Access Point		&	MEC	&	Mobile Edge Computing		\\ \hline
BCD	&	Block Coordinate Descending		&	MEO	&	Medium Earth Orbit		\\ \hline
BPP	&	Binomial Point Process		&	MINLP	&	Mixed Integer Nonlinear Programming		\\ \hline
BS	&	Base Station		&	ML	&	Machine Learning 		\\ \hline
CA	&	Civil Airplane		&	NFV	&	Network Function Virtualization		\\ \hline
CDC	&	Coded Distributed Computing		&	NPPP	&	Nonhomogeneous Poisson Point Process		\\ \hline
CSI	&	Channel State Information		&	NTN	&	Non-Terrestrial Networks		\\ \hline
CTCE	&	Centralized Training and Centralized Execution		&	PCP	&	Poisson Cluster Process		\\ \hline
CTDE	&	Centralized Training and Decentralized Execution		&	PLS	&	Physical Layer Security		\\ \hline
D3QN	&	Dueling Double DQN		&	PPP	&	Poisson Point Process		\\ \hline
DC	&	Difference of Convex		&	QKD	&	Quantum Key Distribution		\\ \hline
DDNN	&	Distributed DNN		&	QL	&	Q-Learning		\\ \hline
DDPG	&	Deep Deterministic Policy Gradient		&	QoE	&	Quality of Experience		\\ \hline
DDQN	&	Double DQN		&	QoS	&	Quality of Service		\\ \hline
DL	&	Deep Learning		&	QT	&	Quadratic Transform 		\\ \hline
DNN	&	Deep Neural Network		&	RIS	&	Reconfigurable Intelligent Surface		\\ \hline
DQL	&	Deep Q-Learning		&	RL	&	Reinforcement Learning		\\ \hline
DQN	&	Deep Q-Network 		&	SAGIN	&	Space-Air-Ground Integrated Networks		\\ \hline
DRL	&	Deep Reinforcement Learning		&	SAP	&	Space-Air Platform		\\ \hline
DT	&	Digital Twin		&	SCA	&	Successive Convex Approximation		\\ \hline
DTDE	&	Decentralized Execution		&	SDN	&	Software-Defined Networking		\\ \hline
EE	&	Energy Efficiency		&	SDR	&	Semi-Definite Relaxation		\\ \hline
EH	&	Enengy Harvesting		&	SR	&	Shadowed-Rician		\\ \hline
FL	&	Federated Learning		&	STIN	&	Satellite-Terrestrial Integration Networks		\\ \hline
FP	&	Fractional Programming		&	TD	& 	Temporal Difference		\\ \hline
GEO	&	Geostationary		&	TL	&	Transfer Learning		\\ \hline
GS	&	Ground Station		&	TN	&	Terrestrial Network		\\ \hline
HAP	&	High Altitude Platform		&	UAV	&	Unmanned Aerial Vehicle		\\ \hline
IoT	&	Internet of Things		&	URLLC	&	Ultra-Reliable and Low Latency Communications		\\ \hline
ISAC	&	Integrated Sensing and Communication		&	VNF	&	Virtual Network Function		\\ \hline
JCC-SAGIN	&	Joint Communication and Computing-Embedded SAGIN 		&	VR	&	Virtual Reality		\\ \hline
LEO	&	Low Earth Orbit		&	VSAT	&	Very Small Aperture Terminal		\\ \hline
MADDPG	&	Multi-Agent DDPG		&	WPT	&	Wireless Power Transfer 		\\ \hline

	\end{tabular}
\end{table*}

\section{Architecture and Evolution of JCC-SAGIN}\label{sec:architecture_Evolution}
This section is dedicated to the foundational aspects and architectural framework of JCC-SAGIN. Initially, we detail the network components spanning various segments, providing a structural overview. Subsequently, the development of computing hardware specifically designed for JCC-SAGIN applications is examined, highlighting technological advancements and capabilities. Finally, an exploration of network integration evolution is discussed, showcasing the advent of multiple emerging network frameworks and their significance in the broader context of JCC-SAGIN.
\subsection{Network Components in JCC-SAGIN}
\subsubsection{Space Segment}
The space segment of SAGIN primarily comprises satellites. The satellites are categorized based on their orbital heights into three types: geostationary (GEO), medium earth orbit (MEO), and low earth orbit (LEO). Table \ref{tab:satellite_COMPARATIVE} delineates the distinct characteristics of satellites operating at varying orbital altitudes, enabling the formation of hierarchical multi-layer satellite networks.

\begin{table*}[!ht]
	\centering
	\caption{Comparison of GEO, MEO, and LEO satellites}\label{tab:satellite_COMPARATIVE}
	\begin{tabular}{|m{2cm}<{\centering}|m{5cm}<{\centering}|m{4cm}<{\centering}|m{4cm}<{\centering}|}
		\hline
		Aspects & GEO & MEO	& LEO  \\ \hline \hline
            Orbit altitude	& 35786 km	& 7000-25000 km &	300-1500 km\\ \hline 
            Coverage	& Global	& Regional/Partial global	& Regional/Local\\ \hline 
            Orbit resources &  Limited	 & Moderate	 & Abundant \\  \hline 
            Propagation delay & High & Moderate & Low\\ \hline 
            Signal attenuation	& High &	Moderate & Low\\ \hline 
            Data rate	& High	 & Not as high as GEO &	High, often match to or even exceed those of GEO satellites\\ \hline 
		Advantages & Global visibility, fixed coverage area, and long-term stable communication. & Relatively lower cost & Cost-effective, rapid deployment, and suitable for large-scale constellations.\\ \hline 
		Disadvantages & High cost and complexity & Limited bandwidth & Large Doppler shift. Possible interruptions in communication.\\ \hline 
 	Suitable use cases & Television broadcasting, broadband internet, high-capacity communication services, and applications where low latency is crucial within the satellite’s coverage area. & Internet services, remote sensing, and applications where low latency and global coverage are essential.  &  Real-time communication, remote sensing, and applications where low latency and global coverage are essential.\\ \hline 
	\end{tabular}
\end{table*}

Previously, GEO satellite constellations, such as Immasat positioned at 35,786 km, were staple in satellite communications. Their stationary relative position to terrestrial users minimizes movement-related issues. A single GEO satellite, due to its high altitude, can cover a vast area, and three GEO satellites strategically positioned can facilitate global communication coverage. However, the significant altitude of GEO satellites introduces challenges such as substantial propagation delays and severe signal attenuation. Additionally, the synchronous orbit resources for GEO satellites are becoming increasingly scarce with the progression of satellite network development.

MEO satellite constellations, such as Inmarsat-P, Odyssey, MAGSS-14, O3b, operate at altitudes ranging from 7,000 to 25,000 km \cite{3GPP.21.917}. They primarily offer services like international search and rescue, positioning, navigation, and timing. Positioned between GEO and LEO satellites, MEO satellites amalgamate the benefits of both, exhibiting reduced propagation delays and broader coverage than GEO and LEO satellites, respectively. Moreover, MEO orbit resources are relatively abundant, given their less stringent altitude constraints.

The advancements in aerospace technology and electronic information have significantly reduced the costs associated with satellite design, manufacturing, and launch. Concurrently, the burgeoning demand for massive Internet and IoT services, particularly in areas devoid of cellular coverage, has heightened interest in mega LEO satellite constellations. LEO satellites, situated at altitudes between 300 km and 1500 km, experience considerably lower propagation delays and signal attenuation compared to their GEO and MEO counterparts. However, the rapid relative movement of low-orbit satellites results in substantial Doppler frequency shifts, posing challenges to inter-satellite links. 

\begin{table*}[!ht]
	\centering
	\caption{Typical LEO satellite constellations}\label{tab:satellite_constellation}
	\begin{tabular}{|c|m{3cm}<{\centering}|m{3cm}<{\centering}|m{3cm}<{\centering}|m{3.5cm}<{\centering}|}
		\hline
		Satellite system & Oneweb & Telesat & Project Kuiper & Starlink \\ \hline \hline
		Country & UK & Canada & USA & USA \\ \hline
  Availability & Northern Hemisphere  & Americas, Europe, Africa, the Middle East, and Asia-Pacific  & Unknown & Parts of the US, Canada, UK, Germany, New Zealand, Australia, France, and several other countries at higher latitudes. \\ \hline
		Number of satellites & 254 launched (total planned 648) to be completed till 2022 & 15 GEO + 298 LEO planned by 2023 & Total 3,236 in planning, 50\% by 2026 & Active 1000+ (total 42,000 by mid 2027) \\ \hline
		Orbit height (km) & 1200 km & 1100-1248 km & 630 km & 1110-1325 km / 540-570 km \\ \hline
		ISL & \ding{55} &  \ding{51}& \multirow{4}{*}{Unknown} & \ding{51} \\  \cline{1-3} \cline{5-5}
		Satellite function & Transparent forward & Onboard processing &    &Onboard processing \\ \cline{1-3} \cline{5-5}
		Beam pattern & Fixed multiple beams & Scanning spot beams &  & Scanning spot beams \\ \cline{1-3} \cline{5-5}
		Frequency band & Ku/Ka & Ka &   & Ku/Ka/V \\ \hline
	\end{tabular}
\end{table*}

Table \ref{tab:satellite_constellation} provides a comprehensive summary of notable LEO satellite constellations. A significant evolution in satellite functionality is observed, transitioning from transparent forwarding to onboard processing capabilities.
For instance, Starlink Internet project of SpaceX involves the deployment of 32,000 Linux computers and over 6,000 microcontrollers in the space\cite{Space-Linux}. This advancement underscores the necessity to consider the coexistence of satellites with varying functionalities within the space segment, particularly when developing communication strategies.
Moreover, the comparatively shorter distance between LEO satellites and mobile terminals as opposed to GEO and MEO satellites facilitates direct communication. This has spurred various satellite constellation programs aimed at providing global service coverage directly to mobile phones. Key developments in this area are listed as follows:
\begin{itemize}
	\item \textit{AST SpaceMobile}: This project plans to create a constellation of 243 satellites. Till now, two satellites had been successfully launched into orbit. The system is capable of impressive data transmission speeds up to 35 Mbps, indicating the potential for broadband-like instant communication services, enabling high-speed internet access akin to terrestrial broadband.
	
	\item \textit{Lynk Global}: This project aims to deploy over 1,000 satellites, with seven currently operational. Their system primarily supports text messaging services, indicating a focus on basic communication, likely targeting regions with inadequate existing services.
	
	\item \textit{Starlink V2}: As an extension of the Starlink initiative, this project envisages a vast network of over 2,000 satellites, with 21 already deployed. The system's data rate ranges between 2-4 Mbps, which, though lower than AST SpaceMobile's rates, is adequate for text messaging and fundamental internet services.
\end{itemize}

This advancement represents a revolutionary step in the mobile satellite services industry, enabling direct connectivity between existing, unmodified smartphones and satellites. The realization of these projects not only proves the feasibility but also highlights the practical engineering significance of satellite edge computing.

\subsubsection{Air Segment}
The air segment primarily comprises UAVs, civil airplanes (CAs), and HAPs, with typical flight altitudes ranging from 8 to 50 kilometers.

UAV networks represent a crucial subset of aerial access platforms (APs). Their growing significance in civil, commercial, and military communications is attributable to their flexible deployment, programmability, and networking capabilities \cite{7470933}. Evolving to meet computation-intensive service demands, UAVs have transitioned from functioning merely as relay nodes to serving as intelligent onboard computing platforms \cite{8786076}. This evolution in UAV roles enhances communication channel capacity and positioning accuracy. UAVs are instrumental in various applications, including data collection, processing, distribution, and emergency response, particularly in scenarios involving real-time video delivery. The relatively low flight altitude of UAVs reduces the launch requirements for mobile terminals and UAVs themselves, and also minimizes propagation delays. Swarm formations of multiple UAVs are employed to extend service coverage. However, the limited energy capacity of UAVs presents a significant challenge in scaling their use for large-scale service provision.

Furthermore, the tens of thousands of CAs traversing the sky offer a type of promising aerial platforms. Since CAs operate on existing flight routes, they do not incur additional launching costs. In comparison to UAVs, CAs face fewer constraints regarding load and energy capacity, and a single CA can provide substantial coverage.  Modern developments enable passengers to access internet services on board through air-to-ground connectivity, facilitated by communication links with LEO satellites or ground gateways. Several commercial and trial networks are operational, such as Gogo's network in the USA, Inmarsat's in Europe, and CMCC's trial network in China, as referenced in \cite{3GPP.draft}. Standards pertinent to these networks are poised for future implementation.Nonetheless, the flight trajectories of CAs are inherently less flexible than UAVs, being dependent on predefined flight routes. Additionally, CA trajectories are less stable compared to satellites, as they can be influenced by environmental factors and other variables.

\subsubsection{Ground Segment}
The ground segment encompasses terrestrial communication networks, support facilities for satellites and aerial platforms, as well as terrestrial terminals.

Terrestrial communication networks are categorized into cellular networks, mobile ad hoc networks (MONET), and wireless local area networks (WLAN). BSs within these networks are interconnected with the core network through optical cables, offering robust computing capabilities that often surpass those of satellites and aerial platforms.

Support facilities such as satellite GSs and CA gateways function as the control units for NTN. They are responsible for sending and receiving measurement and control signals, as well as data transmitted from SAPs. These facilities are also connected to the core network using high-speed optical cables, ensuring efficient data transfer and communication.

Terrestrial Terminals include very small aperture terminals (VSATs), mobile handheld terminals, and IoT devices. 
\begin{itemize}
    \item \textit{VSATs:} Often mounted on vehicles, ships, and other equipment, VSATs support multiple protocol stacks and can ideally track service beams by employing various antenna types.

    \item \textit{Mobile Handheld Terminals and IoT Devices}: These devices are typically equipped with omnidirectional or directional antennas, depending on the scenario and requirements. Due to their compact size, these terminals have limited transmitting capacity and energy consumption, which in turn restricts the achievable data rate.

\end{itemize}

These terrestrial terminals intermittently generate tasks with diverse QoS requirements. Visible SAPs can provide tailored access, transmission, offloading, and backhaul communication services to users, aligned with the specific QoS requirements of these tasks.

\subsection{Evolution of Computing Hardware in JCC-SAGIN}
In 2022, global computing devices achieved a total computing power of 906 EFLOPS, comprising 440 EFLOPS of basic computing power, 451 EFLOPS of intelligent computing power, and 16 EFLOPS of supercomputing power \cite{caict}. Projections suggest that by 2025, the total computing capacity of global computing equipment will surpass 3 ZFLOPS. IDC forecasts indicate that the number of global IoT devices will exceed 40 billion by 2025, generating nearly 80 ZB of data. Over half of this data will require processing by the computing capabilities of terminal or edge devices \cite{cic}. Cloud and edge computing continue to dominate as primary application scenarios for basic computing power.

The escalation in computing power is closely linked to advancements in chip manufacturing technology. Recent years have seen rapid growth in the development of AI chips and aerospace chips, providing the necessary hardware support for processing computationally intensive tasks. Notably, the swift progress in aerospace electronics technology and the extensive adoption of commercial off-the-shelf (COTS) products have significantly enhanced onboard computing capabilities. In the future, satellite communication is poised to move beyond the traditional ``repeater" mode, aiming for an efficient amalgamation of communication and computing \cite{8610431}.

This subsection will explore the development of high-performance computing processors that support the execution of computing tasks in SAGIN, focusing on onboard computing chips and AI chips.

\subsubsection{High-Performance Spaceflight Computing (HPSC) Processors}
In the aerospace sector, chip reliability is paramount, even more so than performance. Only chips with radiation resistance can guarantee the normal functioning of spacecraft. 
Numerous suppliers are producing highly integrated and modular onboard computing systems tailored for small spacecraft. Table \ref{tab:HPSC} outlines some of the latest advancements in aerospace-grade highly integrated airborne computing products \cite{NASA}. The vehicle column in the table correlates each onboard unit with a specific spacecraft classification, where ``general satellite" refers to larger SmallSat platforms, exceeding the size of CubeSats. The technology readiness level of a chip indicates its maturity for space applications, which can vary based on specific payload and mission requirements.

\begin{table*}[!h]
	\centering
	\caption{Sample of Highly Integrated Onboard Computing Systems}\label{tab:HPSC}
	\begin{tabular}{|m{2cm}<{\centering}|m{3cm}<{\centering}|m{4cm}<{\centering}|m{1.5cm}<{\centering}|m{2.5cm}<{\centering}|m{1cm}<{\centering}|}
		\hline
           Manufacturer& Product & Processor & Pedigree & Vehicle	& TRL \\ \hline \hline
           Space Micro	& CSP & 	AMD-Xilinx Zynq-7020 Dual-core ARM Cortex-A & \multirow{7}{*}{COTS} &  \multirow{4}{*}{CubeSat} & 	Ukn \\ \cline{1-3} \cline{6-6}
           GomSpace &	Nanomind A3200& 	Atmel AT32UC3C MCU	&  	& 	&Ukn\\ \cline{1-3} \cline{6-6}
           ISISPACE & iOBC	& ARM 9	&  	& 	& 9    \\ \cline{1-3} \cline{6-6}
           Pumpkin	& PPM A1 &	TI MSP430F1612 &	 	&   &	9 \\ \cline{1-3} \cline{5-6}
            Novo Space & 	GPU001AF	&  NVIDIA Jetson TX2i	& 	& General Satellite & 	Ukn \\ \cline{1-3} \cline{5-6}
           AAC Clyde Space	& Sirius OBC 	& SmartFusion Cortex-M3 &	  & SmallSat &	9 \\ \hline
           MOOG & 	G-Series Steppe Eagle & 	AMD G-Series compatible 	&  Rad Hard by design & General Satellite	& Ukn \\ \hline
          BAE	&  RAD750 & 	RAD750 	& Rad Hard 	& General Satellite & 	9 \\ \hline
	\end{tabular}
\end{table*}

\begin{figure}[h]
	\centering
	\includegraphics[width = 0.5 \textwidth]{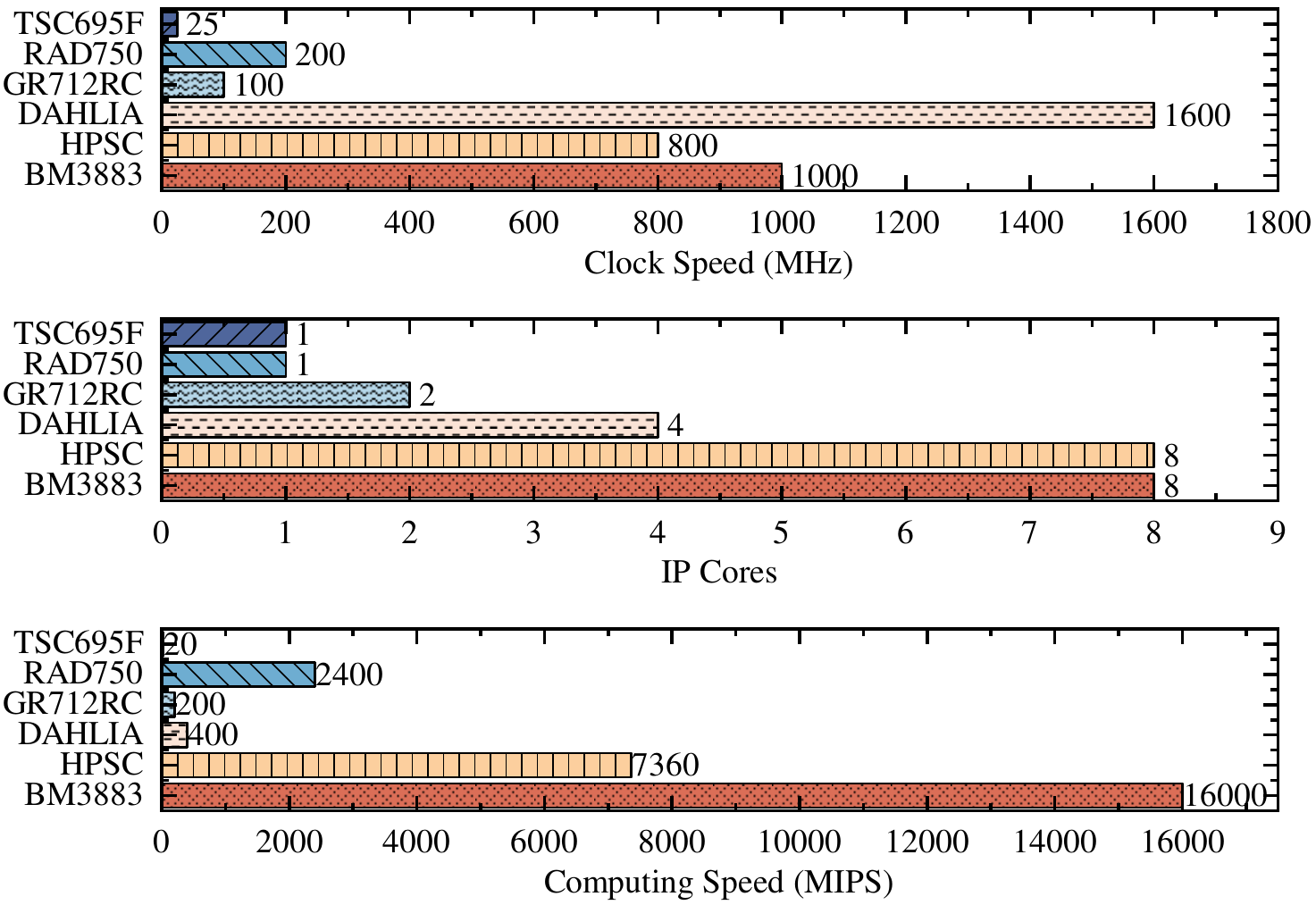}
	\caption{The performance of various aerospace-grade CPU chips is compared, highlighting chips like the TSC695F, GR712RC, and DAHLIA from Europe, RAD750 and HPSC from the USA, and BM3883 from China. \label{fig:Comparison_CPU}}
\end{figure}

With the advent of HPSC processors, the onboard processing capabilities of satellite networks have seen significant enhancement. Table \ref{tab:Typical  Satellite} provides a comprehensive summary of typical LEO satellite networks equipped with onboard computing capabilities, supported by advanced aerospace-grade entral processing unit (CPU) chips \cite{LEOCPU}.
Fig. \ref{fig:Comparison_CPU} presents a comparative analysis of the computing power of chips utilized for onboard data processing. The metrics used for this comparison are million instructions per second (MIPS) and giga instructions per second (GIPS), which measure the computing power of a CPU by indicating the number of instructions it can execute per second. This comparison provides valuable insights into the processing capabilities of various chips, highlighting their suitability for onboard data processing tasks in satellite networks.

 \begin{table}[!h]
	\centering
	\caption{Typical LEO Satellite Networks with onboard processing}\label{tab:Typical  Satellite}
	\begin{tabular}{|m{1cm}<{\centering}|m{0.6cm}<{\centering}|m{1cm}<{\centering}|m{1.4cm}<{\centering}|m{2.5cm}<{\centering}|m{3cm}<{\centering}|}
		\hline
           Satellite System & Year & Country & Networking	& Onboard Processing Function \\ \hline \hline 
           DARPA Blackjack &	2018 & \multirow{2}{*}{America}		& \ding{51} & Autonomous control and decision-making \\  \cline{1-2} \cline{4-5} 
           SDA NDSA &	2020 &   	& \ding{51}	& Distributed management and control\\ \hline
           Tianzhi Satellite &	2019 & \multirow{3}{*}{China}		& \ding{55}	& Cloud detection and satellite control \\  \cline{1-2} \cline{4-5} 
           Tianxian Constellation  &	2021 &  	& \ding{55}	& On-orbit processing of synthetic-aperture radar (SAR) images \\  \cline{1-2} \cline{4-5} 
           Tiansuan Satellite &	2021 &  		& \ding{55}	& Target recognition and model training \\ \hline
	\end{tabular}
\end{table}

 \subsubsection{AI Processors for Space, Air and Ground Segments}
 The advancement of industries such as AI has led to traditional chips falling short in meeting the performance and computing power requirements. AI chips broadly encompass hardware designed to accelerate AI applications, particularly in neural network-based deep learning (DL). The four primary chip types in AI computing are CPU, graphics processing unit (GPU), field programmable gate array (FPGA), and application specific integrated circuit (ASIC), each differing in computational efficiency, energy consumption, and flexibility.

\begin{itemize}
    \item \textit{CPU}: Operating under the von Neumann architecture, CPUs follow a Fetch-Decode-Execute-Memory Access-Write Back cycle. CPUs emphasize control and decision-making, leading to certain inefficiencies in parallel computing due to the need to fetch data into random access memory (RAM), decode instructions, perform calculations in the arithmetic logic unit (ALU), and then return results to RAM.

	\item \textit{GPU}: Originally utilized in image processing, GPUs have fewer data prefetching and decision modules and more computing units (i.e., ALUs), giving them an edge in parallel computing. Commonly used for tasks like vertex and pixel rendering, GPUs possess computing power far surpassing CPUs, making them the primary processors in general-purpose computers and supercomputers.
	
	\item \textit{FPGA}: FPGAs are ``reconfigurable" chips with a modular architecture, consisting of programmable logic modules and on-chip memory. They can achieve GFLOPS-level computational power with lower power consumption, presenting an efficient option for parallel artificial neural network implementations.
	
	\item \textit{ASIC}: ASICs are custom-designed for specific user and system requirements, like neural network processing unit (NPU) or tensor processing unit (TPU). Customization allows ASICs to outperform GPUs and CPUs in specific fields. They represent a major focus for AI chip design companies globally, with a growing presence in the market.
\end{itemize}

\begin{figure*}[!h]
	\centering
	\includegraphics[width = 0.7 \textwidth]{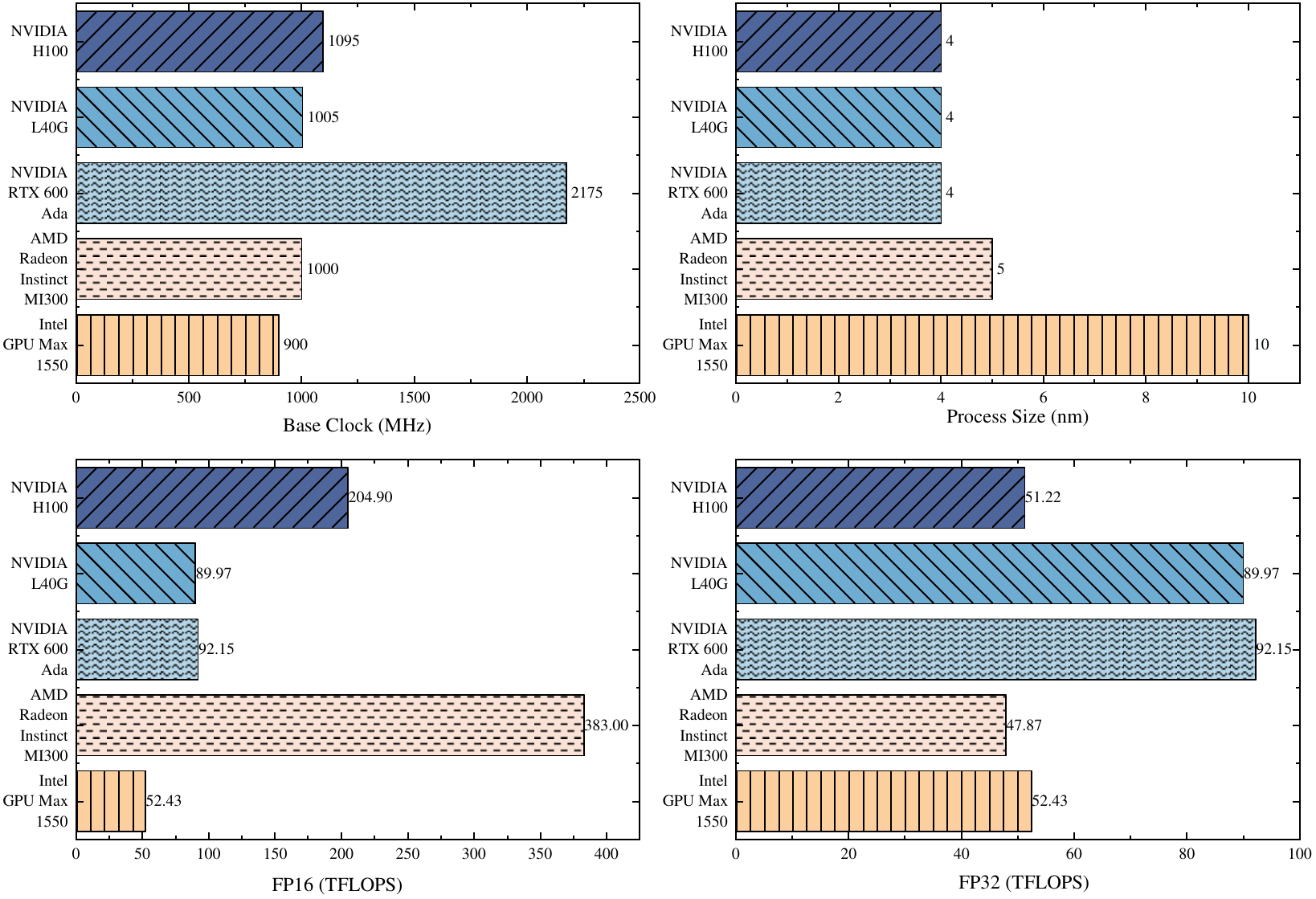}
	\caption{The latest advancements in GPU processing performance are exemplified by models such as the NVIDIA H100, NVIDIA L40G, and NVIDIA RTX 6000 Ada, all released in 2022, alongside the AMD Radeon Instinct MI300 and Intel Data Center GPU Max 1550, introduced in 2023. \label{fig:Comparison_GPU}}
\end{figure*}

GPU performance is often gauged by floating-point computing power, using semi-precision (16-bit) for applications like machine learning (ML), single precision (32-bit) for multimedia and graphics. In the manufacturing industry, leaders like NVIDIA, Intel, and American Semiconductor have launched high-performance processors. For instance, in 2022, Intel introduced the Intel Data Center GPU Max 1550 for enhanced AI computing. In January 2023, AMD released the MI300A chip, boasting a computing power of 47.87 TFLOPS for AI inference and training. NVIDIA launched the H100 chip with a Transformer engine in March 2022, achieving 51.22 TFLOPS \cite{TOPCPU}. Fig. \ref{fig:Comparison_GPU} compares the performance of current high-quality GPU processors. 

\textbf{\textit{- Takeaways:}}
The semiconductor industry has witnessed a remarkable surge in the number of transistors in computing chips, thanks to the ongoing refinement of semiconductor processes and technologies. This increase has catalyzed significant enhancements in the computing performance of CPUs, data processing units (DPUs), and GPUs. Concurrently, there have been substantial breakthroughs in the performance of aerospace-grade CPUs, establishing a robust hardware foundation for the integration of communication and computing in SAGIN. A notable development in SAGIN is the integration of generative AI, which has revolutionized system design and optimization. This form of AI incorporates complex models such as variational autoencoders (VAEs), generative adversarial networks (GANs), generative diffusion models (GDMs), and transformer-based models (TBMs), with GANs and TBMs being particularly prominent\cite{enwiki:1186331682}. Looking ahead, AI computing is set to become a ubiquitous element, significantly impacting various technological spheres.

\subsection{Evolution of Network Integration in JCC-SAGIN}
As network intelligence evolves, upcoming communication networks are set to experience an influx of computationally demanding applications. These include image processing, computer vision, video encoding/decoding, DL, and IoT data processing. These applications require significantly enhanced computing power, memory, and battery life, often surpassing the capabilities of local execution \cite{7925196}. MEC addresses these needs by shifting data processing closer to the network edge, aiming to decrease latency and improve service quality. In this context, resource-constrained devices can offload computational tasks to satellites, UAVs, and ground servers through a process known as computation task offloading, a topic that will be further elaborated in Section \ref{sec:offloading}. The scholarly exploration of computing task offloading within SAGIN has been robust, investigating various MEC architectures that include satellite-ground networks \cite{8610431}, UAV-ground networks \cite{8642376,8961914}, and SAGIN \cite{9013393}, reflecting the field's dynamic evolution and the growing complexity of integrated systems.

The application of RIS and EH technology further promotes the integration of different segments in SAGIN.
RIS addresses communication network challenges, such as signal occlusion, by facilitating RIS-assisted satellite and aerial communications. Furthermore, RIS is an active participant in computational task offloading, establishing a collaborative MEC framework with its assistance. Despite efforts to minimize energy consumption, the limited battery capacity of terminal devices poses an ongoing challenge. Offloading tasks to HAPs can exacerbate power consumption, thus reducing the operational lifespan of battery-dependent devices. Conversely, EH and WPT technologies empower devices to harvest energy from their environment, ensuring continuous communication and computing. Consequently, EH and RIS are viewed as promising technologies for the future advancement of networks. A more detailed exploration of these technologies and their applications in JCC-SAGIN is presented in Sections \ref{subsec:EHWPT} and \ref{subsec:RIS-JCC}.

In the era of rapid development powered by AI and computing hardware, the intelligent framework has emerged as a critical element in JCC-SAGIN. Digital twin (DT) technology, bridging the physical and virtual worlds, significantly enhances network monitoring, prediction, and intelligent control. Existing works have primarily explored its integration with MEC and AI to improve system performance and user experience, while also investigating its application in NTN for efficient resource allocation and learning. The application and implications of DT technology within JCC-SAGIN are further elaborated in Section \ref{subsec:digital_twin}.
Furthermore, learning-based intelligent decision-making techniques are proving to be instrumental in addressing complex, high-dimensional optimization challenges. Beyond centralized learning methods such as Markov decision processes (MDP), DL, and reinforcement learning (RL), decentralized approaches like federated learning (FL) and transfer learning (TL) offer solutions to alleviate the load on feeder links and ground cloud centers. A comprehensive discussion on these intelligent optimization strategies will be provided in Section \ref{sec:RM_optimization_method}.

\begin{figure}[h]
	\centering
	\includegraphics[width = 0.5\textwidth]{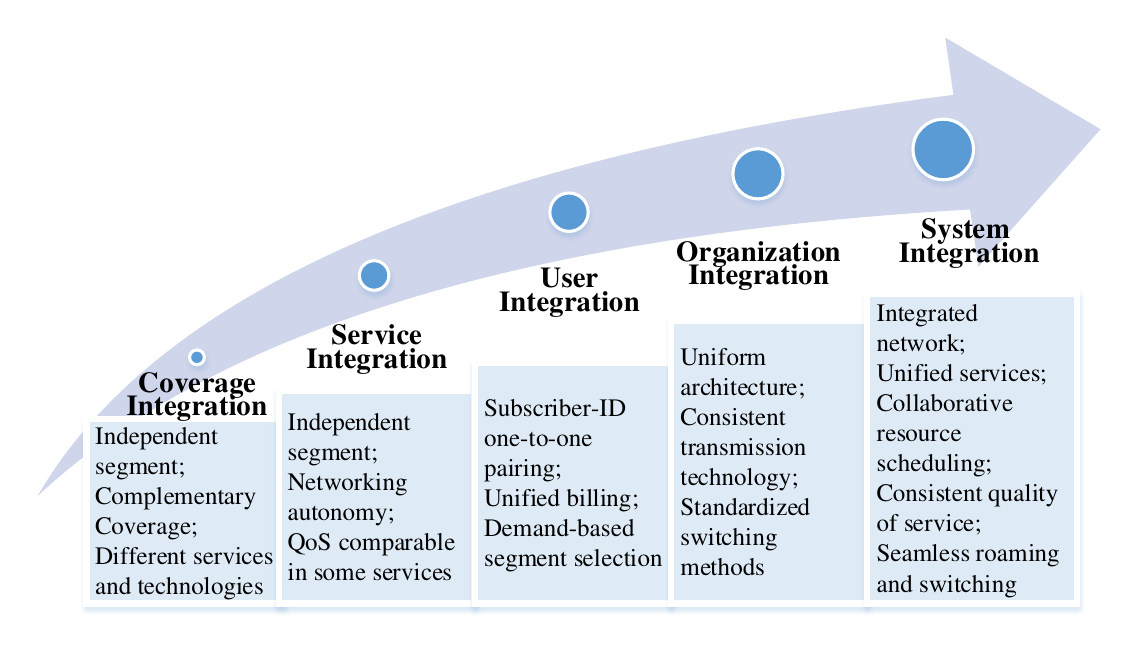}
	\caption{The integration process of SAGIN encompasses a series of progressive steps, commencing with initial coverage integration and advancing through service integration, user integration, organization integration, and culminating in system integration. \label{fig:integration_evolution_SAGIN}}
\end{figure}

The integration of heterogeneous networks is essential to harness the distinct advantages of each network segment, thereby enhancing overall performance. 
Zhongxing Telecom Equipment (ZTE) has pioneered the conceptualization of this evolutionary process in heterogeneous networks \cite{ZTE_Tian}, as depicted in Fig. \ref{fig:integration_evolution_SAGIN}. This process encompasses a progression from initial coverage integration to service, user, organization, and ultimately, system integration.

\begin{figure*}[h]
	\centering
	\includegraphics[width = \textwidth]{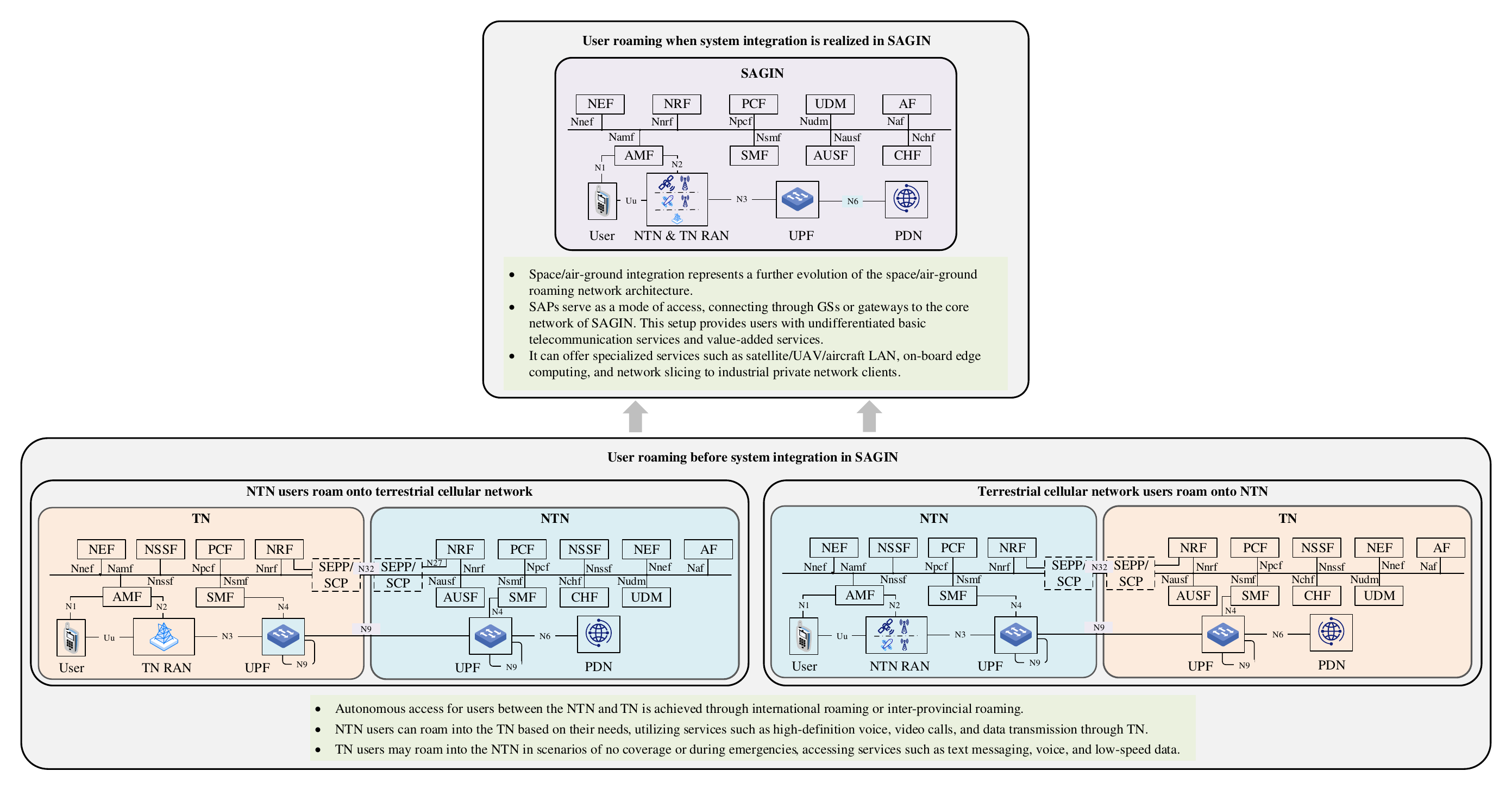}
	\caption{User roaming dynamics within SAGIN vary based on the degree of integration between NTN and TN. In scenarios where system integration is yet to be fully realized, user roaming typically involves NTN users transitioning onto TN, and users of terrestrial cellular networks moving onto NTN. This stage of integration is illustrated in the bottom two figures of the referenced diagram. Conversely, the top figure in the diagram represents the user roaming process within the framework of complete system integration in SAGIN, where seamless and unified network operations are achieved, facilitating smooth user transitions between NTN and TN. \label{fig:user roaming}}
\end{figure*}

The evolution of user roaming procedures in SAGIN, both before and after system integration, is illustrated in Fig. \ref{fig:user roaming}. Post-system integration, SAGIN is anticipated to facilitate a profound amalgamation of NTN with TN. This amalgamation will yield flexible and resilient networking capabilities, support adaptable network node function segmentation, and enable dynamic routing across multiple layers in SAGIN. Such an architecture will support the seamless deployment and transition of integrated systems. An intelligent, unified network management system will be instrumental in collaboratively scheduling network resources and optimizing spectrum utilization. Terminal users, including those with direct-to-vehicle and direct-to-mobile-phone connections, will benefit from imperceptible access to the most appropriate network node and seamless transitions across network types. This integration represents a significant advancement over traditional network structures where NTN and TN operate independently, positioning SAGIN to deliver ubiquitous, seamless, and superior communication services.

Recent advancements in network integration, reflecting contributions from both academia and industry, are summarized in Table \ref{tab:integration_SAGIN}. International organizations such as 3rd Generation Partnership Project (3GPP) and European Telecommunications Standards Institute (ETSI) have proposed standards for integrating various SAGIN segments, extending from physical to application layers \cite{3GPP_38811,3GPP_38821,ETSI_103611,ETSI_136133}. Additionally, entities like China Unicom \cite{chinaunicom} and China Future Mobile Communication Forum \cite{FutureCom} have published white papers highlighting the characteristics of network components, typical applications, and the impact of technology on SAGIN.
Beyond academic research, several corporations have initiated trials in network integration. For instance, EdgeSAT \cite{EdgeSAT} and Viasat \cite{Viasat} have concentrated on integrating satellite and TN, while Facebook \cite{Facebook} and Nokia \cite{Nokia} have developed air-terrestrial integration networks (ATIN).

\textbf{\textit{- Takeaways:}}
The evolution of 5G networks and the exponential growth of communication and computing applications have placed higher demands on the performance of SAGIN. Various architectures and technologies have emerged to enhance network performance. The conventional cloud-centered computing architecture is gradually transitioning to MEC to mitigate latency issues. Deploying RIS on satellites and drones to assist in communication and computing tasks offloading represents a promising avenue for future 6G networks. The integration of EH technology into forthcoming communication networks is anticipated to address the energy supply challenges of terminals. Moreover, the fusion of DT and AI with the network has the potential to significantly elevate the network's intelligence level. The synergistic integration of these architectures and technologies with SAGIN will continually drive the network evolution toward 6G.

\begin{table*}[!h]
	\centering
	\caption{ Academic and Industrial Trials on the network integration}\label{tab:integration_SAGIN}
	\begin{tabular}{|m{1.5cm}<{\centering}|m{1.5cm}<{\centering}|m{1cm}<{\centering}|m{1.2cm}<{\centering}|m{10cm}<{\centering}|} \hline
		Type & Institution &  Date  & Framework & Main Content \\ \hline \hline
\multirow{4}{*}{Standard}	& 3GPP \cite{3GPP_38811} & 2020-09  &  NTN	& Establish the deployment scenarios and pertinent system parameters for NTN, modify existing channel models, and highlight areas critically influencing new radio (NR) interfaces. \\ \cline{2-5}
	&	3GPP  \cite{3GPP_38821} & 2023-03  & NTN	& Conduct simulations at both the radio link and cell system levels to assess NR performance. Develop and define Layer 2 and Layer 3 solutions pertinent to NR, as well as interface protocols and solutions related to radio access network (RAN) architecture. \\ \cline{2-5}
 &	ETSI \cite{ETSI_103611} & 2020-06  & SAGIN	& Identify practical examples and necessary standards for integrating satellite and HAP communication systems into the 5G network. \\ \cline{2-5}
 &	ETSI \cite{ETSI_136133} &2023-05  & STIN & Detail the radio resource management requirements for frequency division duplexing (FDD) and TDD modes in E-UTRA, including measurement requirements for UTRAN and users, and the demand for dynamic node behavior and interaction, such as instant latency and response characteristics.  \\ \hline
  \multirow{2}{*}{White Paper} & China Unicom \cite{chinaunicom} & 2020-06 & SAGIN & Enumerate the benefits of different types of access platforms, introduce MEC and blockchain-enabled SAGIN, and identify typical applications within SAGIN. \\ \cline{2-5}
 & China Future Mobile Communication Forum \cite{FutureCom} & 2020-11 & SAGIN & Discuss the communication system of SAGIN from four angles: development drivers and vision, requirements and challenges, the architecture of the three-dimensional integrated network, and potential key technologies. \\ \hline
 \multirow{4}{*}{Industry Trial} & EdgeSAT \cite{EdgeSAT} & 2022-03 & STIN &  Illustrate the potential evolution of satellite networks to address specific use cases, focusing on seamless integration with terrestrial access networks and backhaul links, and managing resources of edge nodes to enhance scalability, especially in scenarios involving numerous edge nodes.  \\ \cline{2-5}
 & Viasat \cite{Viasat} & 2023-10 & STIN & Examine the current state of space-based cybersecurity, including recent advancements, unique challenges, and Viasat's core principles for protecting space networks. \\ \cline{2-5}
 & Facebook \cite{Facebook} & 2015-06& ATIN & Complete the first full-scale prototype of the solar-powered Aquila aircraft and achieve laser communications between aircrafts, offering data rates in the tens of Gbps. \\ \cline{2-5}
 & Nokia \cite{Nokia} & 2023-05 & ATIN & Provide the first fully certified automated drone solution connected through 4G/LTE and 5G networks. This solution includes advanced dual gimbal camera drones, a docking station, and edge data processing capabilities. It also supports third-party application integration via an open API framework, allowing for customization and enhanced functionalities. \\ \hline
	\end{tabular}
\end{table*}

\subsection{Summary Remarks}

This section provides a detailed examination of JCC-SAGIN, covering its network components, computing hardware evolution, and network integration processes. Various aerospace platforms in NTN, such as satellites, UAVs, and civil aviation aircraft, create a three-dimensional framework that promises ubiquitous connectivity and extensive coverage for future communication networks. The advent of aerospace-grade CPUs and AI acceleration chips offers vital hardware support, enabling the efficient execution of computation-heavy tasks within the network.
As networks continue to advance, there is a convergence of technologies and architectures aimed at improving network efficiency. This includes the use of RIS for computing task offloading, the integration of EH techniques, the application of DT technology, and the development of learning-based network architectures. However, the effective deployment of these technologies within SAGIN is met with numerous challenges, such as dealing with Doppler frequency shifts, adapting to rapidly changing network topologies, and accommodating the distinctive features of non-ground channels. These factors are crucial in the design and implementation of innovative network solutions.

\section{Enabling Technologies of JCC-SAGIN}\label{sec:enabling_tech}
Driven by the upgraded components and integrated networks, more flexible and functional techniques gradually emerged to enhance the performance of JCC-SAGIN.
In what follows, some promising technologies are discussed, enriching the framework of JCC-SAGIN.

\subsection{Network Slicing, SDN, and NFV Supported JCC-SAGIN}
One of the objectives of 6G technology is to actualize ultra-reliable low-latency communications (URLLC). Given that SAGIN encompass heterogeneous communication networks, and users engage in a variety of applications, it is imperative to allocate resources precisely according to user requirements to prevent resource wastage. 

Network slicing emerges as a pivotal technology in this new era of intelligent networking. It involves partitioning a physical network into multiple virtual networks, each precisely tailored to meet the demands of different services, as elucidated in \cite{MGA21-TNSM}. This approach significantly enhances overall resource utilization efficiency within SAGIN.
To ensure resource slices are appropriately aligned with the coverage and mobility of the SAPs, SAGIN employs three distinct types of resource slicing strategies: full slicing, partial slicing, and separated slicing \cite{JJJ22-WC}.

\begin{itemize}
	\item \textit{Full Slicing}: In this strategy, each SAP is dedicated to servicing a single area in collaboration with terrestrial facilities. The space, air, and ground layers function cohesively, catering to IoT applications within a defined topological range. The close interconnection between different SAGIN layers facilitates the implementation of full slicing, aided by resource slicing orchestrators and the cooperation among software-defined (SD) space/air/ground controllers.
	
	\item \textit{Partial Slicing}: Here, several SAPs are positioned at the boundaries of geographical regions governed by different SD controllers. These SAPs can service two adjacent areas, with resources provided by a specific SAP being partitioned to offer supplementary coverage to the ground as needed. This scenario necessitates the design of a rapid-response agreement scheme among various SD controllers, given the shared nature of typical SAP resources.
	
	\item \textit{Separated Slicing}: Considering the constant movement of SAPs and the consequent frequent changes in their regional affiliations, fast handovers can disrupt the continuity of service provided by resource slices. To address this challenge, resources of different layers are allocated to their respective slices, with terrestrial resources providing primary and reliable services and non-terrestrial layers offering complementary coverage. This approach significantly mitigates the impact of SAP dynamics on service continuity.
\end{itemize}

\begin{figure*}[ht]
	\centering
	\includegraphics[width =0.7 \textwidth]{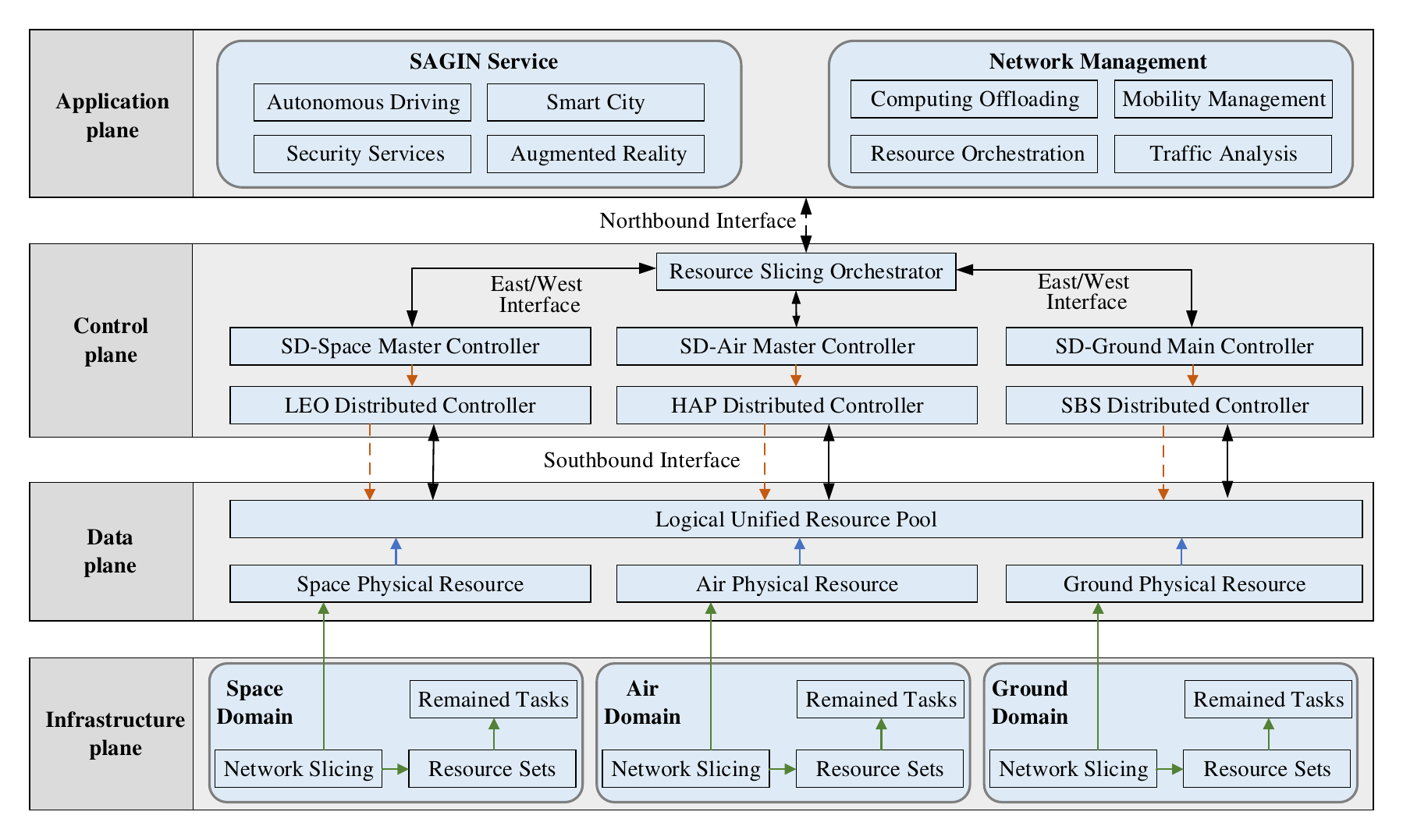}
	\caption{The logical structure of the SD-SAGIN, which consists of infrastructure plane, data plane, control plane and application plane. \label{fig:SDN_SAGIN}}
\end{figure*}

To achieve flexible and efficient network slicing in SAGIN, SDN and NFV are pivotal technologies, as highlighted in \cite{8967040}.
Fig. \ref{fig:SDN_SAGIN} illustrates the logical structure of SD-enabled SAGIN (SD-SAGIN). SDN's core concept involves intelligent network control achieved by decoupling the control and data planes through software programming, as detailed in \cite{8360850}. This architecture allows the control module, which has access to global network information, to logically allocate network slices based on task attributes. The control unit can be centralized in a single server or distributed across multiple platforms \cite{TK22-WC}, facilitating the realization of SAGIN services and network management. Previous works \cite{JJJ22-WC} have elaborated on the functions of each SDN plane, and these are not reiterated here.
For SDN applications, authors have proposed slicing-based architectures for SD-UAV \cite{JJJ22-WC} and SDN-based space-terrestrial integrated networks (SD-STIN) with MEC \cite{YGS19-NETWORK}. Additionally, Cao \textit{et al.} adopted a multi-level distributed SDN control framework in their SAGIN-Internet of vehicle (IoV) architecture, integrating both SDN and NFV technologies \cite{BJX22-IOTJ}.

NFV allows traditional network functions to be virtualized as software components on general-purpose hardware. This virtualization enables the abstraction of physical network resources into virtual ones, which can be allocated to various virtual network functions (VNFs) for different applications \cite{8344453}. The flexibility and reconfigurability of VNFs, which can be shared among multiple service requests, enhance network slicing and efficient allocation of communication and computing resources.
While many studies have addressed VNF scheduling issues in static scenarios \cite{9060940,9075271,9906071}, the field of online VNF mapping and scheduling in SAGIN is still in its nascent stages. Li \textit{et al.} developed an SDN/NFV-enabled framework for SAGIN to support the Internet of Vehicles \cite{9351537}. Their approach utilizes VNF live migration, reinstantiation, and rescheduling, maximizing service acceptance ratios and service provider profits through online Tabu search-based algorithms.

\textbf{\textit{- Takeaways:}} 
Both academic and industrial sectors are vigorously advancing SDN and NFV technologies to navigate resource management challenges in future SAGIN, ensuring QoS for varied services. While network slicing with SDN and NFV offers flexibility and scalability, its implementation in SAGIN presents numerous challenges, such as effective placement and management of network functions within and between slices. Additionally, the cross-national/regional coverage of satellite platforms demands heightened data privacy considerations in network slicing. Future research will be directed towards addressing these challenges, with a focus on efficient network slicing and management to provide tailored QoS services.

\subsection{Energy Harvesting and Wireless Power Transfer Enhanced JCC-SAGIN}\label{subsec:EHWPT}
In JCC-SAGIN, IoT devices face significant challenges due to their limited battery and computing capabilities. While computation offloading to nearby edge platforms can meet task delay requirements, the long transmission distances to SAPs often lead to substantial signal attenuation and consequently high transmit power requirements for IoT devices. Regularly charging these devices in remote areas without grid electricity is impractical, and frequent battery replacement is not feasible for mobile IoT devices \cite{YJK16-JSAC}. High-capacity batteries could alleviate this issue but at a higher hardware cost.

Most IoT devices are exposed to renewable energy sources like solar, wind, vibration, and radio frequency (RF), and installing solar panels/wind turbines near IoT sensors is a convenient solution. EH \cite{9215008} and WPT \cite{8444683} emerge as promising techniques to extend battery life and ensure seamless network operation. SAPs can simultaneously broadcast RF energy to terrestrial devices and support data offloading, combining MEC and WPT to enhance the computing capacity of energy-harvesting IoT devices.

Wireless powered-MEC (WP-MEC) in UAV systems is extensively explored. UAVs serve as edge servers, providing wireless energy and facilitating computing offloading from terrestrial IoT devices \cite{ZSY22-TWC,9215008,8937793,8854903,8434285,8422277}. For instance, a joint multi-task charging-offloading scheme in a UAV-assisted Beyond 5G (B5G) system minimizes task execution delay by optimizing offloading decisions and resource allocation \cite{9215008}. Other studies focus on minimizing UAV energy consumption \cite{8937793,8422277}, reducing packet loss \cite{8854903}, and maximizing user equipment computation rates \cite{8434285}.

Path loss between UAVs and terrestrial users often hampers the efficiency of RF energy harvesting. Energy beamforming is used to enhance energy transfer efficiency. For example, energy-efficient laser-beamed WPT significantly improves transmission efficiency in distributed UAV networks \cite{8703821}. An energy-efficient joint-WPT-MEC scheme is investigated in WP-MEC scenarios with time-varying channels \cite{8960510}, where maximum ratio transmission energy beamforming optimizes WPT towards users.

Considering that offloaded data transmission to MEC platforms via active communication (AC) components like oscillators and converters is power-intensive, backscatter communications (BackCom) are proposed as a low-power alternative. BackCom achieves higher energy efficiency (EE) than AC but with a lower offloading rate, allowing IoT nodes to modulate and reflect incident signals for task offloading \cite{9866050}. A hybrid active-passive communication scheme combines the advantages of both AC and BackCom, as investigated by Li \textit{et al.} in a UAV-enabled WP-MEC network \cite{QLZ22-IOTJ}. Simulation results demonstrate the superiority of this hybrid approach in terms of the weighted sum of computation bits compared to solely using AC or BackCom.

\textbf{\textit{- Takeaways:}}
EH offers a significant advantage for ground terminal devices, reducing the need for frequent battery replacements and lessening reliance on conventional power networks. While current research often focuses on drones as energy providers, other SAPs like satellites and HAPs should not be overlooked. HAPs equipped with large solar panels can sustain flight, communication, computing, and other functions using solar energy. This makes HAPs, especially larger ones, more promising platforms for powering IoT terminals compared to drones.

\subsection{Reconfigurable Intelligent Surface-Assisted JCC-SAGIN}\label{subsec:RIS-JCC}
RIS is an emerging technology in the field of telecommunications, characterized by its thin, flexible artificial metasurface comprised of numerous passive reflective components \cite{8910627}. RIS can intelligently alter the amplitude and phase of incoming signals, thereby customizing the electromagnetic response on its surface \cite{8811733}. This capability offers advantages in coverage expansion, spectrum and EE, and enhanced security, making RIS a key innovation for 6G networks.

RIS technology can be utilized in two primary ways: aiding existing communication systems and initiating new forms of communication \cite{SWY21-CM}. In RIS-aided communications, RIS serves as a relay to strengthen received signals at the receiver end. On the other hand, RIS-initiated communications treat RIS as an energy-efficient transmitter where signals are encoded and transmitted via phase shifting, leveraging an adjacent RF source.
While RIS has been widely explored in TN, where it is often deployed on building facades or indoor surfaces, this approach has limitations \cite{HYS21-TWC}. These include challenges in determining installation locations, high deployment costs for large-scale RIS, and the need for signals to undergo multiple reflections in complex environments. Consequently, integrating RIS within SAGIN is gaining interest.

When an SAP is equipped only with RIS (without edge servers), its primary role is to facilitate communication. In this setup, RIS can extend coverage to blind spots by reflecting signals from the transmitter. The mobility and rapid deployment capabilities of space-aerial platforms allow for efficient establishment of LoS paths, ensuring reliable uplink transmission.

Alternatively, RIS can function as an active transceiver, modulating and demodulating signals directly. In such cases, the aerospace platform typically hosts both edge servers and RIS. Stable communication links are established using RIS, followed by the offloading of computing tasks to either the aerospace platform or a ground BS. This process involves the user initiating the offloading request and MEC selection, while resource allocation and RIS configuration are managed on the MEC platforms \cite{9520318}.
The workflow in a RIS-assisted cooperative MEC framework involves several steps:
\begin{itemize}
    \item \textit{{MEC Selection}}: Initially, user devices attempt to process computing tasks locally. If local resources prove insufficient or unavailable, users must send an offloading request to an MEC platform. This step involves choosing an appropriate MEC platform for offloading the computational tasks.

    \item \textit{{Computation}}: Upon receiving offloading requests, MEC platforms share this information among themselves. They then engage in optimizing the offloading scheduling, allocating necessary resources, and reconfiguring the RIS. Once optimized, the MEC platforms communicate the results back to the user devices.

    \item \textit{{RIS Control}}: The MEC platform takes charge of controlling the RIS. This control is based on the optimal RIS configuration and resource allocation determined in the previous step. The aim is to assist in the efficient offloading of computing tasks from user devices.

    \item \textit{{RIS-Assisted Offloading}}: User devices, upon receiving feedback from the MEC platform, offload their computing tasks to the chosen space-aerial platform. This offloading is facilitated through a communication link established and maintained by the RIS.

    \item \textit{{Task Processing and Feedback}}: The selected computing platform, whether it's on a satellite, an aerial vehicle, or a terrestrial station, processes the offloaded computing tasks. Upon completion, the computing results are transmitted back to the user devices through the RIS-assisted communication link.
\end{itemize}

\subsubsection{RIS-Assisted Satellite Communications}
Future satellite communications are poised to evolve towards higher frequencies and larger-scale connectivity. Higher frequencies enable smaller antennas with more focused beams, enhancing the likelihood of LoS alignment. However, this development requires the deployment of multiple antennas and active relay components, which can conflict with the power and size constraints inherent in large-scale connectivity. This constraint necessitates the use of low-power, small-size satellites for comprehensive coverage.

RIS can play a pivotal role in addressing these challenges. Due to their passive nature, RIS units mounted on satellites can serve as effective passive reflectors. Furthermore, the simplistic hardware structure of RIS makes their deployment on satellites feasible and practical \cite{2020arXiv201200968T}. Integrating RIS with satellites can thus potentially mitigate issues related to limited computing capabilities and high transmission power requirements \cite{LLH22-NETWORK}.

Several studies have examined the performance of RIS-assisted satellite networks, focusing on aspects such as secrecy rate \cite{HZK22-TVT,9994309,9893807}, EE \cite{9539541,9726800}, and capacity \cite{9814615,9860805}. Notably, few studies have explored scenarios where RIS units are directly installed on satellites \cite{9539541,9954397}, with most considering ground-based RIS units. Additionally, the placement of the RIS controller—whether on the GS or the satellite—presents a dilemma. A terrestrial controller avoids interference by using different uplink and downlink links for tracking telemetry and command (TTC) channels and communications. However, a satellite-based controller could increase the computational burden on satellites, which typically have limited processing capacity.

The integration of RIS into satellite frameworks brings additional operational challenges. Satellites often require dynamic adjustments based on mission requirements, necessitating adaptable network configurations for onboard RIS, which could introduce significant overhead. Moreover, the reliability and longevity of hardware are critical for regular satellite operations. The durability and effectiveness of space-based RIS are yet to be fully tested, raising concerns about potential failures and the inability to update or replace RIS units in space as can be done on Earth.

\subsubsection{RIS-Assisted Aerial Communications}
The application of RIS in aerial platforms, as proposed by Alfattani et al. \cite{SWY21-CM}, offers an innovative approach to overcome the limitations in TN. RIS can be flexibly integrated with aerial platforms, either by coating their outer surface or as a separate entity. This flexibility, combined with the higher altitude of these platforms, enhances the likelihood of establishing LoS links. Aerial platforms equipped with RIS can provide extensive and adaptable coverage for wireless networks.

However, integrating RIS with aerial platforms like UAVs presents unique challenges. Frequent mobility can lead to increased overhead due to constant channel estimation and RIS reconfiguration. Additionally, the limited load capacity of these platforms restricts the size of RIS that can be deployed, potentially limiting the full exploitation of RIS benefits.

\subsubsection{RIS-Assisted Cooperative MEC framework in JCC-SAGIN}
Integrating RIS with MEC systems offers a dual advantage in improving communication links and computational capabilities. By aiding MEC servers with RIS, an additional degree of freedom is introduced, allowing for intentional adjustment of channel conditions to enhance the performance of MEC-enabled SAGIN. While research in this area is still nascent, emerging frameworks are exploring this integration.
One such framework is the RIS-assisted cooperative MEC framework in space information networks, which was outlined by XBC et al. \cite{XBC21-NETWORK}. This framework comprehensively covers aspects from MEC server selection to task processing. It underscores the potential benefits, challenges, and applications of integrating RIS and MEC in SAGIN.

Key issues in aerial RIS-assisted MEC include resource management, trajectory planning, and beamforming design. 
Shang et al. \cite{9771338} have highlighted these challenges, along with the need to balance excessive path loss in non-LoS (NLoS) conditions. 
To fully exploit the advantages of MEC and RIS, a max-min computation capacity problem is formulated under the orthogonal multiple access (OMA) scheme \cite{9785633}. The simulation results verified that the proposed scheme could achieve an 8.08 Mb computation capacity higher than the case without RIS. 
The authors extended their model and discussed the computation capacity maximization problem in the RIS-aided UAV-MEC system with  non-orthogonal multiple access (NOMA). The simulation results showed that compared to the OMA scheme, the NOMA scheme could achieve higher computation capacity \cite{9796538}.
Zhai \textit{et al.} \cite{9891794} and Mei \textit{et al.} \cite{9410588} investigated the EE issues in RIS-assisted UAV networks.

{\textbf{\textit{- Takeaways:}}
Research to date suggests that RIS can markedly improve communication quality while simultaneously lowering energy consumption, presenting a cost-effective approach. The integration of RIS with MEC in SAGIN promises to yield dual benefits in both communication and computational efficacy. Yet, the application of RIS as a transceiver on satellites and UAVs, especially to facilitate computation task offloading, is still in its early phases. The rapid movement of these aerial platforms necessitates frequent RIS reconfigurations, a crucial challenge that needs addressing. Considering these factors, future research could concentrate on enhancing both communication and computing efficiency by promoting cooperation among various network elements equipped with RIS in JCC-SAGIN.}

\subsection{Digital Twin-Enabled JCC-SAGIN}\label{subsec:digital_twin}
As an emerging technology in the 6G era, DT can establish connections between the physical and virtual worlds and realize the resource sharing between them \cite{FJQ18-IJAMT}. 
With the aid of DT technology, virtual models are created to represent physical entities and the entire network status can be monitored. After collecting the data from real objects, the dynamic DT can learn and update working conditions intelligently, predict future network events, and implement intelligent control to change the state of the physical entities \cite{9863238}.
The consecutive interaction with the physical world builds two-way closed-loop information feedback, making real-time intelligent decision-making possible \cite{9145588}.

Existing works on DT mainly focused on combining it with MEC and AI technology. 
Aided by DT, the offloading unit does not always need to interact with the physical environment and store the real-time status of each platform. Therefore, compared to the MEC network without DT, the system performance and user experience in the DT edge network can improve significantly \cite{BYL22-TVT}.
Up to now, the research on DT edge networks in SAGIN is still in the early stages. 
DT is first used to train the proactive deep reinforcement learning (DRL) scheme offline in a UAV-assisted MEC system to tackle the asynchronous training of deep recurrent Q-network (DRQN) \cite{9128981}.
To capture the time-varying features of SAGIN, dynamic DT is integrated with FL to maximize the clients' utilities, where the DT of the UAV acts as the leader of the formulated Stackelberg game \cite{9311405}. This is the first work to consider the unpredictable and mobile network topology during the FL procedure in the air-ground network.
Different from previous works, which assume that the mobility and task requirements can be predetermined, Sun \textit{et al.} proposed a DT-driven two-stage incentive scheme to jointly maximize the users' satisfaction and overall EE considering the limitations of UAVs \cite{9351542}. Compared to the centralized resource allocation algorithm, the proposed decentralized method can decrease the computing burden of vehicles while ensuring system performance.
The authors extended their research by considering the design of lightweight DT architecture in air-ground networks \cite{9931961}. The integration of DT and FL with a distributed incentive mechanism is formulated to maximize learning efficiency.

Utilizing augmented reality (AR) and VR, DT technology transforms collected data into virtual twin models, offering a significant advancement over traditional network simulators. DT provides a more accurate reconstruction of physical entities, enabling a more effective evaluation of the impacts of novel algorithms or configurations on the network. When combined with AI, DT is particularly proper for capturing the time-varying nature of network topology and forecasting dynamic changes, making it exceptionally suited for the variable MEC environments encountered in SAGIN.
It's crucial to acknowledge the computing limitations of SAPs, where the computational complexity of AI solutions becomes a pivotal concern. In light of this, there's a preference for lightweight AI models due to their ease of integration and deployment, to ensure that the systems remain efficient and responsive within the constrained computational environment in JCC-SAGIN.

\subsection{Summary Remarks}
This section examines the enabling technologies driving improvements in network communication and computing in JCC-SAGIN, focusing on network slicing, EH, RIS, and DT technologies. Table \ref{tab:technologies_capabilities} summarizes the capabilities of these enabling technologies. SAGIN's ability to cater to a wide variety of service demands overcomes the limitations of traditional physical network architectures in delivering on-demand services for diverse scenarios. Network slicing emerges as a vital solution, creating multiple virtual logical subnetworks within the same physical network infrastructure, thus addressing the need for customizable and flexible network services.
EH and WPT are pivotal in promoting sustainable network operations, enabling devices to collect energy from their environment or the network platforms, thereby alleviating the need for frequent recharging and battery replacements. Moreover, applying RIS significantly enhances future communication networks by supporting computing task offloading, improving network computing efficiency, and boosting communication capabilities. 
Furthermore, adopting DT technology in SAGIN facilitates intelligent decision-making and effective resource management, significantly enhancing network reliability and operational efficiency. The integration of enabling technologies underscores a strategic approach to overcoming SAGIN's inherent challenges, setting a foundation for the next generation of network communication and computing performance.

\begin{table*}[!h]
	\centering
	\caption{Capabilities of the Enabling Technologies in JCC-SAGIN}\label{tab:technologies_capabilities}
	\begin{tabular}{|m{2cm}<{\centering}|m{14cm}<{\centering}|}
		\hline 
		Enabling technologies & Ability and contribution to improve the performance of the JCC-SAGIN  \\ \hline  \hline
            Network Slicing, SDN, and NFV& 
Provide on-demand services, facilitate swift deployment and adaptable adjustment capabilities, enhance resource utilization efficiency and resource management efficacy, and achieve heightened flexibility and scalability.	\\ \hline 
            EH and WPT	& Ensure continuous communication and computing, enhance network sustainability, facilitate more flexible deployment and layout, and reduce energy consumption and costs.	\\ \hline 
            RIS &  Assist in communication and computing task offloading, facilitate signal and coverage enhancement, improve spectral and energy efficiency. Offer high flexibility, adaptability, and reconfigurability.	 \\  \hline 
            DT &  Beneficial for achieving efficient resource management and scheduling, intelligent decision-making, fault diagnosis, and risk assessment, thereby enhancing network reliability and stability.	 \\  \hline 
	\end{tabular}
\end{table*}

\section{Applications in JCC-SAGIN}\label{sec:application}
\begin{figure*}[ht]
	\centering
	\includegraphics[width = 0.7\textwidth]{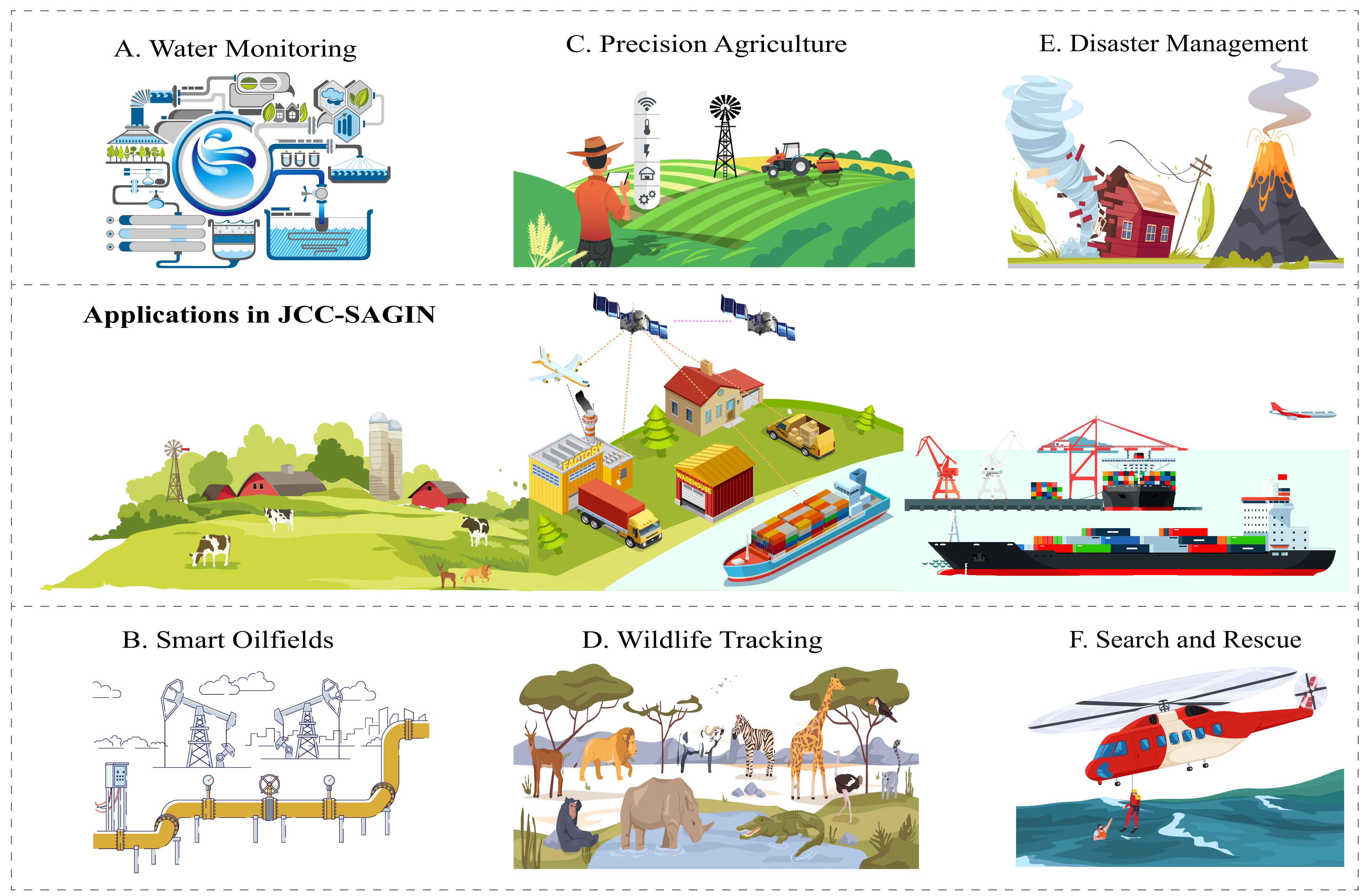}
	\caption{With onboard processing, JCC-SAGIN can intelligently support a broader range of applications, offering rapid response and reducing the reliance on terrestrial infrastructure. \label{fig:JCC_Application}}
\end{figure*}

Upon equipping SAPs with processing capabilities, computing functions traditionally executed by remote cloud servers (such as GSs) can be transferred to closer-edge platforms like SAPs. This shift enables service support in remote areas lacking cellular networks and enhances the provision of computation-intensive IoT services in urban and suburban areas, especially those with stringent latency requirements. Fig. \ref{fig:JCC_Application} illustrates the typical applications in JCC-SAGIN, which are detailed in this section.

\subsection{Water Monitoring}
The smart water resource monitoring platform solution leverages data from water quality monitoring devices for a variety of business applications, including monitoring water quality, controlling water pollution, observing water levels, and surveilling suspicious activities. This solution enables unified data management from these applications through a three-tier system comprising sensing devices, edge devices, and a central platform. This arrangement facilitates collaborative analysis and rapid edge response, alleviating heavy processing loads on backend systems. It also supports comprehensive monitoring, encompassing water quality, weather conditions, and environmental intrusions, thereby ensuring holistic data collection across varied scenarios.

The Gravity Recovery and Climate Experiment (GRACE) satellite, a collaborative endeavor between National Aeronautics and Space Administration (NASA) and the German Aerospace Center, exemplifies the integration of satellite technology in water monitoring. Launched in 2002, GRACE observes changes in Earth's gravity field, enabling scientists to assess water reserves in watersheds and track water movement across oceans, land, and the atmosphere. This satellite has revolutionized global gravity field observation and climate change experiments, becoming a critical tool for monitoring global environmental changes, such as melting glaciers, sea level and circulation changes, and variations in land water volumes.

\subsection{Smart Oilfields}
Numerous leading oil companies are currently embracing a strategy of ``reducing workforce and increasing efficiency," actively incorporating technologies such as AI, 5G, edge computing, and digitization into their entire business processes. In the oil industry, edge computing scenarios predominantly focus on numerous remote oil wells in isolated rural areas. Oil customers exhibit significant interest in combining ``edge computing and 5G" for applications like onsite oil well management, intelligent oilfield equipment management, and smart monitoring of oil reservoirs.

\subsubsection{Onsite Management of Oil Wells}
Video data captured by cameras at oil well sites are processed in real-time by an AI box. This device conducts extensive data processing at the edge, offloading a considerable amount of data and transmitting only critical analysis results. This method effectively reduces the bandwidth required for data transmission, enabling real-time monitoring of foreign object intrusions at oil wells and surveillance of worker behavior onsite.

\subsubsection{Intelligent Management of Oilfield Equipment}
Traditionally, equipment inspections at oilfields are conducted every three hours to ensure proper functioning. The deployment of an AI box at the site, equipped with relevant algorithmic models, allows for real-time analysis of operational data from connected devices and sensors. This system facilitates continuous equipment status monitoring and significantly reduces the frequency of manual inspections.

\subsubsection{Intelligent Monitoring of Oil Reservoirs}
Machine vision technology is utilized for assessing underground petroleum seepage conditions, transitioning from traditional to intelligent control methods. This technology enables intelligent calculations of oil well fluid volumes, autonomous interval pumping of pumping units, and smart regulation of water injection wells. 

In 2016, China's petroleum and petrochemical industry witnessed the first implementation of a Beidou-based oilfield exploration, development, and production satellite comprehensive application system. Integrating state-of-the-art technologies like satellite remote sensing, communication, and navigation, this system established a comprehensive service platform with functionalities including vehicle navigation monitoring, pipeline patrol, oil well condition monitoring, and oil and gas exploration and production monitoring. Additionally, the Landsat series of satellites, a joint project between NASA and the United States Geological Survey, plays a crucial role in monitoring, understanding, and managing the land and mineral resources essential for human sustenance through the Landsat program.

\subsection{Precision Agriculture}
With the global population projected to reach 9.1 billion by 2050, there is a pressing need for a substantial increase in food production \cite{food-data}. This necessitates a transition from traditional extensive farming methods to more intensive agricultural practices.

Precision agriculture (PA) represents an advanced integration of various technologies, encompassing aerospace, information, communication, and data analysis. The incorporation of emerging technologies like the IoT holds significant promise in PA. IoT-based PA systems automate critical farming operations such as crop growth monitoring, irrigation processes, fertilizer application, and disease detection. These systems utilize a diverse range of terrestrial, aquatic, and aerial sensors, alongside satellites and aerial platforms, to collect extensive geospatial data from multiple sources. This data is then processed to extract valuable insights, guiding decision-making and contributing to the enhancement of both the quantity and quality of agricultural products.

UAVs are increasingly employed in PA for various purposes, including weed identification and control, monitoring vegetation growth and yield estimation, detecting plant health issues and identifying diseases, managing irrigation, and crop spraying.
As farmers navigate through the seasons, from planting to growth monitoring and harvesting, they are progressively utilizing NASA Earth Science data to inform their decisions. For instance, the Landsat program, which has been monitoring Earth's surface changes since 1972, launched its ninth satellite in 2021. The Soil Moisture Active Passive (SMAP) satellite measures soil moisture, while the Moderate Resolution Imaging Spectroradiometer (MODIS)  and ECOsystem Spaceborne Thermal Radiometer Experiment on Space Station (ECOSTRESS) are utilized for assessing crop health.

\subsection{Wildlife Tracking}
NTN are increasingly being utilized in wildlife tracking and conservation efforts. These networks provide researchers with immediate data access, a broader geographical scope, and enhanced accuracy in monitoring animal movements, thereby enriching our understanding of wildlife behavior, habitat usage, and the various threats these species face. Here are some practical examples of how NTN support wildlife tracking.

\subsubsection{Migration Monitoring of Avian Species}
Researchers utilize satellite telemetry to monitor the migratory patterns of various bird species. By attaching small satellite transmitters to birds, scientists can track their movements and identify their migration routes. This data is crucial in understanding challenges faced by migratory birds, such as habitat degradation and climate change impacts, and in developing effective conservation strategies.

\subsubsection{Monitoring of Marine Fauna}
Satellite tags are used to track the movements of marine wildlife, including sharks, whales, and sea turtles. These tags transmit real-time positional data to researchers via satellite communication, providing insights into the migratory paths, foraging areas, and breeding sites of these species. This information is instrumental in establishing marine conservation measures, like creating marine protected areas and regulating fishing activities.

\subsubsection{Surveillance of Endangered Species}
Global Positioning System (GPS) or satellite collars are used for monitoring endangered species like elephants, tigers, and rhinos. These collars send geospatial data to researchers via satellite, allowing for effective tracking of the animals' territorial range, movement patterns, and habitat preferences. The data gathered is vital for developing strong conservation initiatives and anti-poaching strategies.


Argos, established in 1978, plays a key role in environmental research and protection, supporting programs like the Tropical Oceans Global Atmosphere (TOGA) and the World Ocean Circulation Experiment (WOCE). With over 22,000 active transmitters (8,000 of which are for animal tracking), Argos has been instrumental in tracking marine mammals and turtles since the late 1980s. The GPS tracking system, comprising a receiver and 24 GEO satellites, can pinpoint the location of animals when the receiver gets signals from at least three satellites.

\subsection{Disaster Management} 
NTN provide crucial information and communication services throughout various stages of natural and human-induced disasters, playing a significant role in reducing their impact on human lives and property. Here are some practical examples of NTN in disaster management:

\subsubsection{Early Warning Through Remote Sensing}
Remote sensing technologies are instrumental in identifying potential natural hazards like storms, floods, and wildfires. This data forms the foundation for early warning systems, allowing authorities to issue timely alerts and execute evacuation strategies, thereby minimizing risks to lives and properties.

\subsubsection{Emergency Response}
In scenarios where TN are compromised, NTN become vital in ensuring continuous communication. During disasters such as earthquakes or hurricanes, satellite phones and data links are essential for coordinating rescue and relief efforts, distributing medical supplies, and supporting affected communities.

\subsubsection{Earth Observation for Post-Disaster Recovery}
Post-disaster, satellite imagery aids in monitoring recovery efforts, providing insights into the reconstruction process and the effectiveness of relief measures. This information helps authorities identify areas needing further assistance, evaluate the long-term impacts of the disaster, and develop strategies for future risk reduction.

The Deep Space Climate Observatory (DSCOVR), launched in February 2015, plays a critical role in maintaining real-time solar wind monitoring capabilities, essential for the accuracy and timeliness of National Oceanic and Atmospheric Administration (NOAA) space weather warnings and forecasts.
The Global Precipitation Measurement Mission (GPM) provides advanced global rain and snow observation data, crucial for alerting potential natural disasters.

\subsection{Search and Rescue} 
\subsubsection{Damage Assessment and Relief Operations}
Drones, equipped with cameras and sensors, conduct aerial surveys of disaster zones, offering real-time insights into the extent of damage and survivor locations. This information is critical for emergency responders to prioritize actions and efficiently allocate resources. Drones also play a role in delivering essential supplies to inaccessible or hard-to-reach areas.

\subsubsection{Navigation for Search and Rescue Operations}
Satellite navigation systems, such as GPS or the European Galileo system, provide precise location information vital for search and rescue teams in disaster areas. This data facilitates navigation through damaged or unfamiliar terrains, aids in locating survivors, and enhances coordination among rescue teams.

The Cospas-Sarsat Global Satellite Search and Rescue system utilizes a network of LEO and GEO satellites, GSs, and control and coordination centers. Its mission is to deliver accurate distress signals and location data to assist in rescuing individuals in distress. In 2019, SpaceX began constructing the Starlink satellite system, becoming the world's largest commercial satellite operator by January 2020. This system is pivotal in establishing rapid disaster relief communication networks and assessing disaster scenarios when ground infrastructure is compromised.

\begin{table*}[!h]
	\centering
	\caption{Enhancements in Sector Applications Through JCC-SAGIN and Resource Management}\label{tab:benefits_application}
	\begin{tabular}{|m{2cm}<{\centering}|m{12cm}<{\centering}|}
		\hline 
		Applications & Impact of JCC-SAGIN and resource management on various applications  \\ \hline  \hline
 Water monitoring & JCC-SAGIN enables dynamic adjustments in water quality assessments and pollution control strategies, tailoring responses to environmental changes. Effective resource management optimizes water usage and conservation strategies, allowing for sustainable development and allocation of water resources across diverse ecosystems. \\ \hline
 Smart oilfields & By continuously monitoring conditions and automating responses, JCC-SAGIN minimizes downtime and maximizes oil production. Strategic resource management allows for precise allocation of drilling resources and maintenance efforts, ensuring operational efficiency while reducing environmental impact and operational costs. \\ \hline
 Precision agriculture & By creating a detailed view of crop health and soil conditions, JCC-SAGIN supports advanced predictive analytics for crop yields and resource use. Efficient resource management optimizes input usage like water and fertilizers, enhances land use efficiency, and reduces the ecological footprint of farming practices. \\ \hline
 Wildlife conservation & JCC-SAGIN boosts conservation efforts by providing detailed tracking of wildlife through satellite and ground-based monitoring networks. Resource management enable targeted conservation actions, optimal allocation of conservation funds, and quick mobilization of resources to protect threatened species and habitats. \\ \hline
 Disaster management & JCC-SAGIN streamlines communication between emergency teams and command centers, ensuring timely evacuations and effective deployment of resources. Through sophisticated resource management, JCC-SAGIN directs essential supplies and personnel efficiently, minimizing disaster impact and aiding in quicker recovery processes. \\ \hline
Search and rescue & JCC-SAGIN reduces search times dramatically and improves survival rates by enabling precise and swift delivery of emergency aid. Advanced resource management enhances logistical coordination, ensuring that rescue operations are well-equipped and responders can access critical resources promptly during emergencies. \\ \hline
	\end{tabular}
\end{table*}

\subsection{Summary Remarks}
This section underscores the wide range of application scenarios enabled by JCC-SAGIN, alongside listing engineering examples that illustrate the network architecture's adaptability. 
The impact of JCC-SAGIN and resource management on various applications have been listed in Table \ref{tab:benefits_application}.
Within the diverse use cases highlighted, components such as satellites and UAVs are tasked with not just fulfilling basic communication needs but also delivering computational services for complex and computation-intensive tasks like remote sensing data processing, image recognition, and ML applications. This dual requirement emphasizes the necessity and critical importance of addressing joint communication and computation challenges in SAGIN.
Effective resource management in JCC-SAGIN plays a crucial role in enhancing system performance, minimizing costs, and aligning with user demands more effectively. Resource management covers various aspects, such as user association policies, allocation of power and computing resources, beamforming, and UAV trajectory optimization. The strategic management of these resources is pivotal for the design and operation of future 6G SAGIN, a topic that is thoroughly explored in Section \ref{sec:RM_optimization_method}. This focus on resource management is integral to unlocking the full potential of JCC-SAGIN, ensuring it meets the evolving demands in the future.

\section{Resource Management Modeling in JCC-SAGIN}\label{sec:offloading}
Against the background of ubiquitous computing tasks, aerospace platforms equipped with edge servers can offer effective computing services. This section provides an overview of the fundamental modeling of resource management in JCC-SAGIN, encompassing metrics, offloading strategies, network modeling, and service modeling. These components form the basis for the subsequent optimization methods of resource management in JCC-SAGIN.

\subsection{Metric for JCC-SAGIN}
The primary rationale for utilizing SAPs as edge servers is to reduce latency, lower energy consumption, and alleviate the load on the core network. Therefore, the main metrics in current literature on JCC-SAGIN fall into three categories: task completion time \cite{FCL22-JSAC,JPB22-CC,YTJ21-TVT,TJB22-IOTJ}, energy consumption \cite{CJH22-TWC,LKC22-TMC,XKK19-TWC}, and the joint minimization of the weighted sum of delay and energy consumption \cite{MLJ22-CC,YCL22-TNSE,SSL22-IOTJ,NZY21-TWC}.
In the following paragraphs, we will first review existing research focused on these three key aspects, followed by an examination of other vital performance indicators.

\subsubsection{Task Completion Time}
In JCC-SAGIN, task completion time encompasses both the transmission time for communication and the computing time for executing tasks. Transmission time is influenced by the allocation of communication resources (like bandwidth and power) and the distance of communication, whereas processing time depends on the allocation of computing resources and task offloading strategies. Zhang \textit{et al.} examined the JCC issue in satellite communication networks and introduced a game theory-based approach to minimize total weighted delay for computationally intensive tasks \cite{9495368}. In another study, the authors considered the interplay between communication and computation, developing a broadband aggregation algorithm for over-the-air federated learning (AirFL) to reduce communication delays \cite{8870236}. Previous research in TN often assumes the computed results entail minimal data, overlooking the return time of these results. However, in JCC-SAGIN, the considerable propagation delay between satellites and the ground necessitates accounting for return time, highlighting a distinct difference in JCC problem research between SAGIN and TN.

\subsubsection{Energy Consumption}
Given the limited energy resources of satellites and HAPs, JCC-SAGIN research must consider energy consumption of MEC servers. This consumption relates to the allocation of power and computing resources, as well as offloading strategies. Many studies focus on resource allocation with the goal of minimizing system energy consumption. For instance, one study addressed the resource allocation problem in MEC scenarios to minimize total energy consumption while adhering to delay constraints \cite{8488502}. Another proposed a three-layer offloading architecture for a hybrid cloud and edge computing LEO satellite system, devising a low-complexity algorithm to minimize system energy use \cite{9344666}. A novel satellite-ground IoT MEC architecture was proposed in references \cite{9383778}, featuring an energy-saving computing offloading strategy and resource allocation algorithm that divides computing energy consumption between ground and space segments. Although energy consumption is a critical factor in JCC problems, the constraints on computing resources and energy in SAGIN's satellite and aerial platforms are more stringent compared to ground networks.

\subsubsection{Weighted Sum of Delay and Energy Consumption}
By considering a weighted sum of task completion time and energy consumption, a trade-off between these two metrics is achievable. Different weights can be assigned according to the specific requirements of various application scenarios, indicating the relative importance of delay or energy constraints. For instance, a study proposed a fast-converging game theory-based algorithm to minimize both task completion time and energy consumption in scenarios where LEO satellites provide computing services to users \cite{9861225}. Chen \textit{et al.} investigated offloading issues within a multi-tier hybrid computing framework in SAGIN \cite{9978924}, involving local computing by IoT devices, edge computing by SAPs, and cloud computing by GSs. They formulated an optimization problem to minimize the weighted sum of task completion latency and energy consumption under communication and computing resource constraints.

\subsubsection{Other Key Performance Indicators}
In addition to the primary metrics, various other performance indicators are employed in research, offering unique perspectives and insights.
For instance, studies have focused on maximizing the quality of experience (QoE) for all users under system constraints in UAV-assisted edge computing networks \cite{LJ22-TVT,LBN22-IOTJ}. In particular, the normalized QoE for mobile VR users is correlated with the downlink data rate \cite{LJ22-TVT}. Given the importance of latency for computation-intensive IoT applications, service latency is mapped to differing QoE levels, such as excellent, fair, poor, and dissatisfied \cite{LBN22-IOTJ}.
Maximizing the sum of computation task-input bits within given constraints is another focus \cite{XKY20-TWC}. 
To enhance the freshness of computed results, one study in air-ground collaborative MEC architecture focused on UAV task scheduling, computing resource allocation, and trajectory optimization, aiming to minimize the weighted age of information (AoI) of all users \cite{9930881}. Han \textit{et al.} developed a stochastic model for vehicle-to-vehicle communication reliability based on probability theory, introducing computational reliability as a new metric for computational offloading, and proposed a dynamic programming method to maximize this reliability \cite{9293144}.


In summary, these diverse metrics are utilized to assess the performance and efficiency of the JCC-SAGIN, providing a comprehensive evaluation of the actual system. The choice of specific metrics may vary depending on the research problem and system architecture, necessitating researchers to select metrics aligned with their research objectives.

\subsection{Offloading Strategy in JCC-SAGIN}
Computation offloading strategies can be broadly categorized into binary offloading and partial offloading. 
In binary offloading, all tasks are processed on a single platform. The decision-making process involves determining not only the type of platform but also identifying the specific platform on which to offload the tasks. This approach typically hinges on the suitability and capacity of the chosen platform to handle the entirety of the workload.
Partial offloading, on the other hand, utilizes a data-partition model where tasks generated by users are independent on a bit-wise level \cite{CMM19-TCOM}. This independence allows for the parallel execution of the task across multiple platforms in different segments. Enabling parallel processing, task completion latency is significantly reduced compared to binary offloading. In partial offloading, both the selection of offloading platforms and the task processing ratio at different platforms are critical considerations. 

From a broader perspective, binary offloading can be seen as a specific case of partial offloading. The flexibility and efficiency of partial offloading make it a more versatile and often more effective strategy, especially in scenarios where workload and latency are paramount.

Numerous works discuss the offloading issues in JCC-SAGIN under binary offloading \cite{XRA22-TNSM,XRA22-TNSE,LJ22-TVT,SXQ22-IOTJ,BJX22-IOTJ,FCL22-JSAC,JPB22-CC,MLJ22-CC,YCL22-TNSE,LN20-IOTJ,SSL22-IOTJ,YWT22-TON,QOG22-TWC,AJH22-JSAC,AGB21-TNSM,JQY21-WCL,JHJ19-CL} and partial offloading strategies\cite{CJH22-TWC,TJQ22-IOTJ,XZY21-WCL,ZYY21-IOTJ,NZY21-TWC,ZSY22-TWC,PGE21-TMC,YTY22-TCOM,CWX21-TWC,XKY20-TWC,EHC21-TCOM,SMC21-IOTJ}. Recent works have also considered both offloading strategies \cite{ZCL21-TCCN,YTJ21-TVT,WJN21-JSAC,QLZ22-IOTJ}. Specifically, the resource allocation process during computation offloading has been modeled as a double auction framework, where IoT devices act as buyers and MEC servers as sellers \cite{ZCL21-TCCN}. In this framework, binary and partial offloading are represented as buyer-choose-one seller and buyer-choose-multi-sellers scenarios, respectively.
Xu \textit{et al.} developed two effective alternating optimization algorithms for binary and partial offloading modes \cite{YTJ21-TVT}. Simulation results indicated that task completion time under binary offloading is typically longer than under partial offloading in comparable situations, highlighting the benefits of partial offloading. The superiority of partial over binary offloading in maximizing sum computation bits is also corroborated in studies like \cite{WJN21-JSAC,QLZ22-IOTJ}.

\subsection{Network Modeling and Analysis in JCC-SAGIN}
\subsubsection{Platform Distribution}
Prior research on network performance analysis has predominantly focused on TN \cite{SAS19-Access,QYQ19-CC,YQY18-Access,AB18-UEMCON,YL19-TVT}. Key studies have examined system capacity in ultra-dense networks, particularly in relation to interference analysis \cite{SAS19-Access,QYQ19-CC,YQY18-Access}. Akter \textit{et al.} developed a vehicle traffic model to evaluate transmission rates in vehicular ad-hoc networks (VANETs) under both sparse and dense traffic conditions \cite{SMM19-GCCE}. These studies commonly assume that the locations of terrestrial BSs follow a Poisson point process (PPP).

While the PPP model yields accurate results for TN performance analysis, it is less applicable to communication scenarios involving a limited number of SAPs dispersed over a given area \cite{SM10-TVT}. The Binomial point process (BPP) is a more suitable geometric method for modeling the spatial distribution of a finite number of SAPs \cite{CDW13-BOOK}. Table \ref{tab:performance_related works} presents several notable studies on the performance analysis of NTN. Chetlur \textit{et al.} investigated the coverage probability of downlink UAV networks, with UAVs modeled as a 3D BPP and assuming UAV-to-user links undergoing Nakagami-m fading \cite{VH17-TCOM}. Likewise, research on satellite networks has modeled the location distribution of LEO satellites as a 3D BPP \cite{DJWJ22-TCOM,NTDIR20-TCOM}. Okati \textit{et al.} explored satellite network performance under a nonhomogeneous PPP (NPPP) model, analyzing communication variances for users located at different latitudes \cite{NT20-PIMRC,NT22-TCOM}. This approach extends the understanding of NTN performance by considering the distinct spatial distributions and characteristics of SAPs, as opposed to traditional TN.

\begin{table*}[!h]
	\caption{Typical Related works of performance analysis for NTN} \label{tab:performance_related works}
	\begin{center}
		\begin{tabular}{|c|c|p{7cm}<{\centering}|c|c|}
			\hline
			Ref.& Type of network & Distribution model & Key metric & Small fading type            \\ \hline \hline
			\cite{VH17-TCOM}& UAV & Multiple UAVs with 3-D BPP and one terrestrial user & SIR &Nakagami-m \\ \hline
			\cite{DKYM22-TWC}& \multirow{6}{*}{Satellite}  & Single satellite and multiple users with PPP & \multirow{3}{*}{SNR} & SR \\ \cline{1-1} \cline{3-3}\cline{5-5}   
			\cite{NT20-PIMRC}&   & Multiple satellites with NPPP and one terrestrial user &   & Rician\\ \cline{1-1} \cline{3-3}\cline{5-5}  
			\cite{DJWJ22-TCOM}&   & Multiple satellites with 3-D BPP and one terrestrial user & & SR \\ \cline{1-1} \cline{3-5}  
			\cite{NTDIR20-TCOM}&   & Multiple satellites with 3-D BPP and one terrestrial user & SNR\&SINR &Non-fading \& Rayleigh\\ \cline{1-1} \cline{3-5} 
			\cite{NT22-TCOM}&   & Multiple satellites with NPPP and one terrestrial user & SINR &Nakagami-m \\ \cline{1-1} \cline{3-5}
			\cite{7981353} &   & Single satellite and multiple users & SNR  & SR, Nakagami-m \\ \hline
			\cite{10136779} & NTN & Multiple satellites and CAs with BPP and one terrestrial user & SINR & SR, Rician, Rayleigh \\ \hline
			\cite{JSBM20-TWC} & SAGIN & A GEO satellite, a moving HAP, a BS and a user & SNR & SR, Rician, Rayleigh \\ \hline
		\end{tabular}
	\end{center}
\end{table*}

\subsubsection{Performance Analysis}
Network performance analysis is crucial for the deployment of SAPs and system design. Current research on the performance analysis of NTN primarily focuses on coverage and capacity. These studies can be categorized based on their methodology into simulation analysis and numerical/theoretical analysis.

\textit{* Simulation Analysis}:
For satellite networks, simulation tools like Systems Tool Kit (STK) and OMNET++ are commonly used. Koukis utilized OMNET++ to simulate a satellite constellation-assisted IoT device scenario with satellites at an orbital altitude of 600 km \cite{Koukis_thesis}, and evaluated the impact of constellation design (such as the number of satellites and aircraft) and inter-satellite links on latency and packet loss rates. A large number of experiments have shown that increasing the number of satellites or the existence of inter-satellite links can significantly reduce network delay and packet loss rate. In addition, the authors mentioned that OMNET++ is a tool suitable for network topology simulation, network protocols and the implementation and evaluation of novel paradigms, and its practicality in space work is also considerable. Bi employed OPNET to simulate LEO satellite network constellations, demonstrating that the network capacity increases with the expansion of the network scale. However, this increase is not linear, as the gain in capacity gradually diminishes with the growth of the constellation scale. \cite{BYY_thesis}. In CA network, simulation data often derives from public aviation data. For instance, Li \textit{et al.} leveraged FlightRadar24 data to analyze the coverage performance of civil aircraft-assisted networks (CAAN) \cite{9693468}. The simulation results demonstrated that civil aircraft assistance can significantly reduce the number of satellites in civil aircraft-assisted SAGIN (CAA-SAGIN) at the same data rate. Sun \textit{et al.} examined network characteristics like transmission path and clustering coefficients using VariFlight data \cite{9708632}.

\textit{* Numerical and Theoretical Analysis}:
Coverage in SAGIN is typically measured by signal outage probability, defined as the likelihood of received signal-to-noise ratio (SNR) or signal-to-interference-plus-noise ratio (SINR) exceeding a threshold. Capacity analysis often relies on ergodic capacity derived from Shannon's formula. Shadowed-Rician (SR) fading has been identified as a more appropriate model for small-scale fading and shadowing in SAP-to-ground channels compared to traditional Nakagami-m fading \cite{C85-TVT}. Huang \textit{et al.} formulated a constrained optimization problem to maximize system capacity, derived closed-form expressions for the outage probability and ergodic capacity in hybrid satellite-terrestrial systems under SR fading \cite{QMW20-AES}, and proposed two multi-user scheduling schemes (best user scheduling and user fairness scheduling) to improve system performance. Bhatnagar \textit{et al.}  investigated the maximum ratio combining (MRC) scheme in the land mobile satellite channel under SR fading channels \cite{MA14-CL}. This study derived approximate closed-form expressions for the bit error rate, outage probability, and capacity of the considered scheme. The resulting bit error rate expression is particularly valuable for assessing the coding gain of the MRC schem. However, system-level analysis for SR fading channels is limited. Na \textit{et al.} and Jung \textit{et al.} analyzed outage probability and coverage performance in satellite networks under SR fading, but did not fully address co-channel interference \cite{DKYM22-TWC, DJWJ22-TCOM}.

Addressing this gap, Chen \textit{et al.} analyzed the coverage performance of SAPs, including satellites and CAs with sectorized beams \cite{10136779}. They provided outage probabilities for scenarios with and without interference using the Laplace transform of interference in SR fading channels. This approach also applies to Rayleigh and Rician fading scenarios, offering a comprehensive method for evaluating network performance under various fading conditions. This research contributes significantly to understanding the impact of fading and interference on the performance of complex SAGIN systems.

\subsection{Service Modeling and Analysis in JCC-SAGIN}
\subsubsection{Service Attributes}
To effectively differentiate between various computation-intensive IoT applications, parameter tuples are commonly employed to describe task attributes.

Task characterization often involves specifying the input data size and the required number of CPU cycles per bit of input data \cite{XZY21-WCL,ZHX22-SJ}. These parameters are foundational in defining the tasks. Additionally, the tolerable delay for task completion is introduced as a critical attribute for delay-sensitive services, reflecting the urgency and time sensitivity of each task \cite{SYJ22-TITS,YYN21-CL}. The combination of task size and required CPU cycles, along with a maximum completion latency, is also considered in various studies to further detail task parameters \cite{GCM22-TVT,ZYX21-TMC}. To enhance fairness and resource utilization efficiency, the required CPU cycles on different platforms are optimized based on the allocated tasks, rather than being fixed.

After processing by the platforms, the results of applications, such as environmental monitoring data or surveillance videos, are transmitted back to the local devices. The ratio of the output data size to the input size is frequently represented by a proportional factor \cite{SGY21-CC}, which can be viewed as a compression ratio in these scenarios. Given that this factor is typically limited, some studies simplify the analysis by omitting the return transmission process from the platforms back to the users \cite{ZYS20-IOTJ,QYG19-IOTJ}.

For a more general approach, the attributes of a task $ f $ generated by user $ k $ can be represented by a tuple $ f_k = \left\{a_k,L_k,{\mathbf{C}}_k, T_k^{\max} \right\} $ Here,  $ a_k $ indicates the service type (either data transmission or a computing task), $ L_k $ specifies the input data size, ${\mathbf{C}}_k $ is a vector denoting the processing density, which includes the required CPU cycles for processing a unit task at local, edge, and cloud platforms, and $ T_k^{\max} $ represents the task's latency tolerance. This comprehensive characterization aids in tailoring the network's response to the specific demands of each task, ensuring efficient and effective service delivery.

\subsubsection{Service Analysis}
In JCC-SAGIN, the communication requirements of a service significantly influence resource block occupancy by users, which in turn affects network interference. This interference impacts the performance of communication links between users and SAPs, thereby affecting service processing. Given the tightly intertwined relationship between service attributes and network performance, a thorough modeling and analysis of service characteristics and network performance is essential for effective system design and construction.

\begin{figure*}[h]
	\centering
	\includegraphics[width = 0.85 \textwidth]{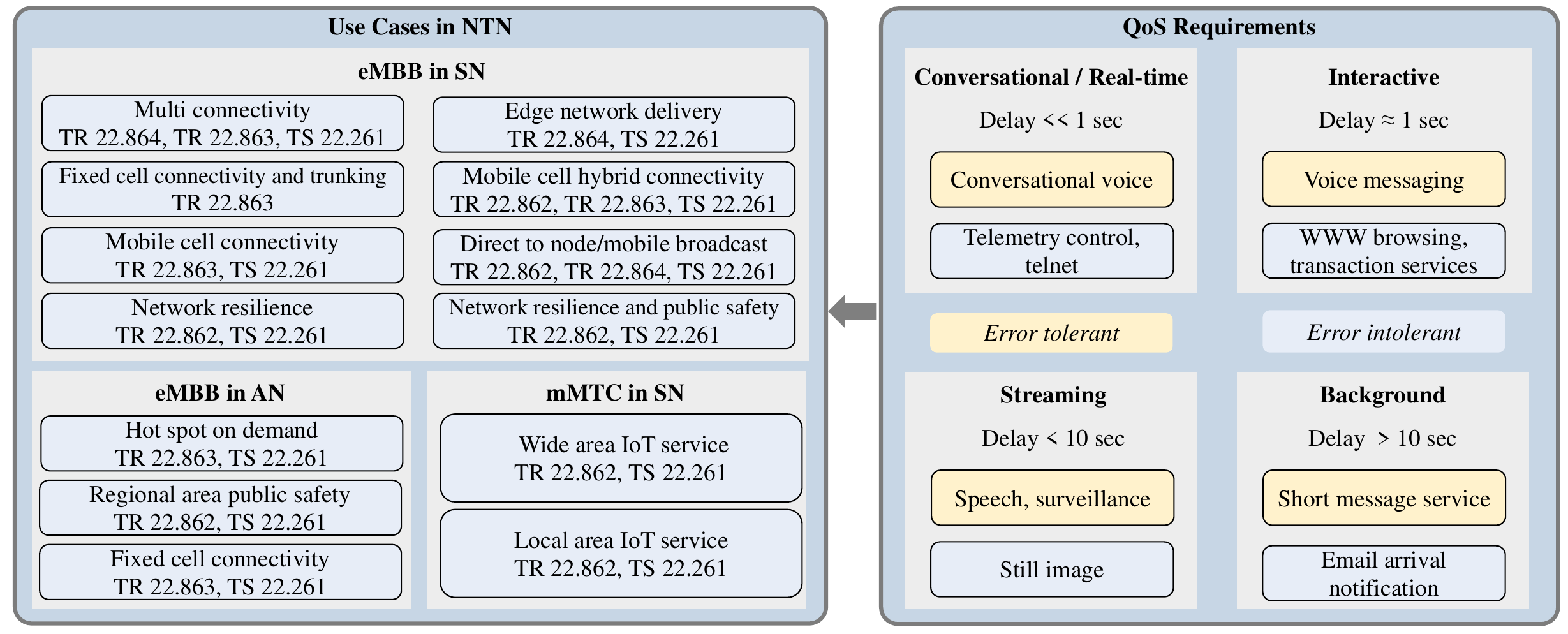}
	\caption{Use cases supported by NTN. In the left part, ``TR" and ``TS" denote Technical Report and Technical Specification, respectively, and the following numbers are the specification numbers of 3GPP.  \label{fig:use_case_QoS}}
\end{figure*}

Space and aerial networks are particularly advantageous in offering services to remote areas where deploying cellular networks is challenging, such as deserts, oceans, and forests. They are also crucial in providing emergency communications when TN are compromised. The use cases in NTN, as per 3GPP standards, encompass a wide range of scenarios, as outlined in left part of Fig. \ref{fig:use_case_QoS}. They provide detailed specifications for these use cases.

Advancements in mobile terminal technology, including handsets and IoT devices, have led to an increase in their processing capabilities. The intersection of 6G with disciplines like big data, AI, and advanced computing will foster the integration of communication with sensing, computing, and control. This cross-pollination is pivotal in supporting advanced communication services like immersive cloud extended reality (XR), holographic communication, DT, and comprehensive area coverage. These services, often sensitive to delays and requiring high data transmission rates, are categorized in the 3GPP standard \cite{3GPP.22.105} and depicted in right part of Fig. \ref{fig:use_case_QoS}.

Service distribution studies, such as those examining temporal and spatial service hotspots throughout a day \cite{7248881}, reveal distinct patterns in service usage. To accurately represent the varied data traffic in SAGIN, both temporal and spatial characteristics of service distribution must be considered. This dual consideration is crucial in effectively modeling and analyzing services within the SAGIN framework, ensuring that network design and resource allocation are optimized to meet the diverse and evolving demands of users.


The spatial distribution of services often correlates with the geographic distribution of users. Theoretical analysis in this domain largely employs stochastic geometric models \cite{8168355,9829241,8267238}.
Wang \textit{et al.} used the NPPP to model the spatial distribution of interfering users in cellular networks, exploring the meta-distribution of signal-to-interference ratio (SIR) under fractional power control \cite{8168355}. Gu \textit{et al.} investigated service offloading processes in Poisson edge computing networks, deriving the distribution of service offloading success probabilities \cite{9829241}. Afshang \textit{et al.} focused on service access strategies in heterogeneous cellular networks, where micro BSs and users followed a Poisson cluster process (PCP) in densely populated regions \cite{8267238}. These studies reflect the complexity of modeling service distribution in various network types.

Temporal service distribution analysis includes modeling packet arrival processes and dynamic processing. Studies often utilize queuing theory for this purpose. The discrete time-slot ALOHA systems with multiple terminals have been explored in \cite{8303680} and \cite{9765641}. Pappas \textit{et al.} modeled packet arrivals as a Bernoulli process, studying throughput and average delay in wireless caching systems \cite{8281002}. Most of these studies assume infinite storage capacity for task queues, although practical IoT nodes often face storage limitations.

Considering limited storage resources and the need for prioritizing grant-based services, effective access control schemes based on service attributes are crucial. These can be modeled as an M/G/1 queuing system, where "M" denotes exponentially distributed service arrival times, "G" represents generally distributed service times, and "1" indicates a single server \cite{CHX19-WCNC,DR92-Datanetworks}. Teng \textit{et al.} and Musumpuka \textit{et al.} investigated the impacts of correlated service delays on communication link performance under different queuing models \cite{WMK19-TVT,RTF15-NOF}.

Analyzing spatiotemporal characteristics requires integrating stochastic geometry with queuing theory, increasing the complexity of service characterization. Jiang \textit{et al.} proposed a service-aware spatiotemporal model combining these theories to address the dynamic nature of task queuing in IoT devices and the static nature of the physical layer network \cite{8408843}. Zhong \textit{et al.} and Wang \textit{et al.} further extended these models to study service time characteristics under different user distributions and scheduling schemes, delving into aspects like queue stability and delay \cite{7886285,GYR20-TCOM}.

In summary, service analysis in JCC-SAGIN requires a multifaceted approach, considering both spatial and temporal dimensions of service distribution. The integration of stochastic geometry and queuing theory in these analyses reflects the intricate nature of service delivery and performance in such complex networks.

\subsection{Summary Remarks}
In this section, we reviewed resource management modeling in JCC-SAGIN from four aspects: metrics, offloading strategies, network modeling, and service modeling. Table \ref{tab:offloading_JCCSAGIN} summarizes the typical related works on offloading in JCC-SAGIN and analyzes them concerning scenario, computation framework, computation strategy, and optimization objective. 

We discussed the metrics of JCC-SAGIN, including task completion time, energy consumption, and the weighted sum of latency and energy consumption. In addition to the aforementioned mainstream metrics, other significant and novel metrics such as QoE, AoI, and computing reliability were also listed, providing insights for future research. 
Subsequently, current offloading strategies were discussed, including binary offloading and partial offloading. The offloading strategy is not selected independently, and it is necessary to consider whether the computation task supports being divided. Existing works on network modeling primarily concentrate on TN, assuming that the locations of ground BSs follow a PPP distribution. Compared to PPP models, BPP is a more suitable mathematical model for characterizing the distribution of spatial platforms. 
Service modeling includes service description and characterization of spatiotemporal distribution characteristics. Most work in service modeling relies on existing queue models and derives statistical results. Exploring and designing specific task queues for different service needs in the future would be valuable.

\begin{table*}[ht]
	\centering
	\caption{Summary of Related works on offloading in JCC-SAGIN} \label{tab:offloading_JCCSAGIN}
	\begin{tabular}{|c|p{3cm}<{\centering}|c|p{1.5cm}<{\centering}|p{7cm}<{\centering}|}
		\hline
		Ref.       & Scenario                                                       & Computation   Framework & Computation   Strategy & Objective                                         \\ \hline \hline
		\cite{ZFZ21-TNSE} & \multirow{5}{*}{\makecell*[c]{UAV-Terrestrial \\ Integrated Networks}}           & Local-Edge              & Binary                 & Computation efficiency                            \\ \cline{1-1} \cline{3-5} 
		\cite{QYG19-IOTJ} &                                                                & Local-Edge              & Partial                & Delay                                             \\ \cline{1-1} \cline{3-5} 
		\cite{ZYS20-IOTJ} &                                                                & Edge-Cloud              & Partial                & Delay \& energy                                     \\ \cline{1-1} \cline{3-5} 
		\cite{RJC19-TCOM} &                                                                & Edge-Cloud              & Partial                & Energy                                            \\ \cline{1-1} \cline{3-5} 
		\cite{9625737 }   &                                                                & Local-Edge              & Binary                 & Energy                                            \\ \hline
		\cite{8945402}    & \multirow{6}{*}{\makecell*[c]{Satellite-Terrestrial \\ Integrated Networks}}     & Local-Edge              & Partial                & Delay \& energy                                     \\ \cline{1-1} \cline{3-5} 
		\cite{ZYY21-IOTJ} &                                                                & Local-Edge              & Partial                & Energy                                            \\ \cline{1-1} \cline{3-5} 
		\cite{9043505}    &                                                                & Edge-Cloud              & Binary                 & Delay \& energy                                     \\ \cline{1-1} \cline{3-5} 
		\cite{SXQ22-IOTJ} &                                                                & Edge-Cloud              & Binary                 & Task completion time and satellite resource usage \\ \cline{1-1} \cline{3-5} 
		\cite{BJX22-IOTJ} &                                                                & Edge-Cloud              & Binary                 & Delay, resource utilization, security, energy     \\ \cline{1-1} \cline{3-5} 
		\cite{SGY21-CC}   &                                                                & Local-Edge-Cloud        & Binary                 & Delay                                             \\ \hline
		\cite{9013393}    & \multirow{3}{*}{\makecell*[c]{Satellite-UAV-Terrestrial \\ Integrated Networks}} & Edge-Cloud              & Partial                & Delay                                             \\ \cline{1-1} \cline{3-5} 
		\cite{NFW19-JSAC} &                                                                & Local-Edge-Cloud        & Binary                 & Delay                                             \\ \cline{1-1} \cline{3-5} 
		\cite{9351533}   &                                                                & Local-Edge-Cloud        & Partial                & Energy      \\ \hline     
		\cite{9978924} &  {\makecell*[c]{Satellite-CA-Terrestrial \\ Integrated Networks}} & Local-Edge-Cloud        & Partial                & Delay \& energy   \\ \hline
	\end{tabular}
\end{table*}

\section{Resource Management Optimization Methods in JCC-SAGIN}\label{sec:RM_optimization_method}
On the basis of resource management modeling in JCC-SAGIN, this section will review the optimization methods exhaustively, including traditional optimization methods and learning-based intelligent decision-making methods.

\subsection{Traditional Optimization Methods in JCC-SAGIN}
\begin{figure}[h]
	\centering
	\includegraphics[width =0.48\textwidth]{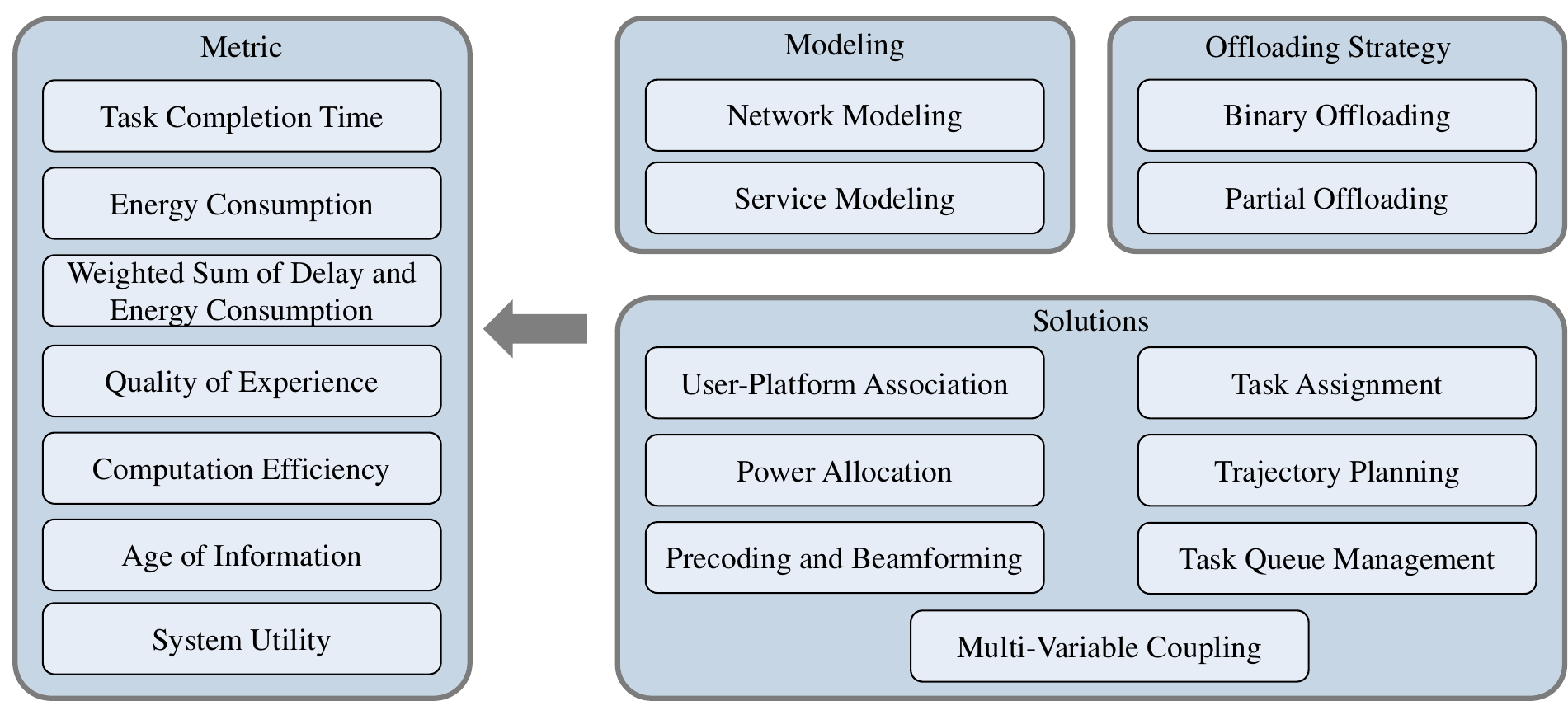}
	\caption{Metric, modeling, offloading strategy, and solutions of resource management in JCC-SAGIN. \label{fig:summary_JCC_SAGIN_resource_management}}
\end{figure}

Fig. \ref{fig:summary_JCC_SAGIN_resource_management} illustrates a comprehensive overview of resource management in JCC-SAGIN. On the left, the figure highlights various metrics pivotal to resource management, while the right side presents an array of offloading strategies, modeling techniques, and their principal optimization solutions aimed at achieving these metrics.
In the realm of JCC-SAGIN, optimization algorithms for tackling JCC problems fall into two primary categories: traditional optimization algorithms and AI-empowered algorithms. This section focuses on the former, while AI-enabled approaches are the subject of the following section.

\subsubsection{User-Platform Association}
Considering the limited computing capability of local devices, IoT devices often rely on nearby space or aerial platforms for traffic offloading to meet latency requirements of computation-intensive tasks. The user-platform association in these scenarios involves binary, rather than continuous, variables, transforming JCC problems into mixed integer nonlinear programming (MINLP) problems. 

Common integer programming methods are summarized in Table \ref{tab: Integer Programming} \cite{BB19-TPAMI}. Let's denote the association variable between user $k$ and SAP $i$ as $ \alpha_{k,i} \in \left\lbrace 0,1 \right\rbrace $.
Continuous methods typically handle binary constraints in two ways. The first is through relaxation-based methods, where binary constraints are relaxed into a continuous range, $ \alpha_{k,i} \in \left[0,1\right]$. These can be further categorized into approximate and exact methods. Approximate methods round continuous solutions to discrete values. For instance, Yu \textit{et al.} used a relaxation-based (approximate) method in discussing UAV-BS association optimization \cite{YTY22-TCOM}, omitting one constraint for easier subproblem reconstruction before recovering binary solutions. However, this approach doesn't guarantee local optimality due to separate optimization and rounding processes. Exact methods like Branch-and-Bound and cutting plane approaches avoid rounding but require repetitive linear relaxations, leading to longer runtimes.

The second continuous method is replacement-based, substituting the concerned space with a set of equivalent continuous variables. This often involves designing penalty functions and solving non-convex optimization problems iteratively, a process that can be complex and time-consuming. Wu \textit{et al.} proposed the $\ell _p$-Box alternating direction method of multipliers (ADMM) algorithm to bridge integer programming and continuous optimization, enhancing convergence speed and computational efficiency \cite{BB19-TPAMI}.

\begin{table*}[ht]
	\centering
	\renewcommand\arraystretch{1.2}
	\caption{Summary of Integer Programming Methods} \label{tab: Integer Programming}
	\begin{tabular}{|cccc|}
		\hline
		\multicolumn{1}{|c|}{\multirow{3}{*}{Continuous Methods}} & \multicolumn{1}{c|}{\multirow{2}{*}{Relaxation-Based}} & \multicolumn{1}{c|}{Approximated Approaches} & Linear, Spectral, SDP relaxation, etc. \\ \cline{3-4} 
		\multicolumn{1}{|c|}{}                                    & \multicolumn{1}{c|}{}                                  & \multicolumn{1}{c|}{Exact Approaches}        & Branch-and-Bound, Cutting plane, etc.  \\ \cline{2-4} 
		\multicolumn{1}{|c|}{}                                    & \multicolumn{1}{c|}{Replacement-Based}                 & \multicolumn{2}{c|}{Penalty methods, $\ell _p$-Box ADMM \cite{BB19-TPAMI}, etc.}                                       \\ \hline
		\multicolumn{1}{|c|}{Discrete Methods}                     & \multicolumn{3}{c|}{Potential game, Hungarian algorithm, etc.}                                                                                              \\ \hline
	\end{tabular}
\end{table*}

The potential game, a discrete approach in integer programming, is effectively utilized in multi-user offloading decision-making processes \cite{DL96-GEB}. Functioning as a non-cooperative game, it involves players (users) independently strategizing to maximize their individual payoffs until Nash equilibrium points are reached. At these points, no player can unilaterally change their strategy for a better outcome.
Optimizing the connection to SAPs for computation offloading can be framed as an exact potential game, leading to a Nash equilibrium \cite{MLJ22-CC,JQY21-WCL}. This approach is particularly pertinent when centralized optimization problems are NP-hard. Game-theoretic approaches enable the design of distributed computation offloading algorithms that converge to a Nash equilibrium, offering a practical solution to these complex optimization challenges \cite{XLW16-TON,BHJ22-TSC}.
The potential game method is also applied to address co-channel interference issues. Here, the resource block allocation subproblem is often resolved using graph coloring methods \cite{CFC17-TVT,AWA13-PIMRC}, which effectively manage resource allocation among multiple users with overlapping communication channels.
Furthermore, to facilitate transmission cooperation between UAVs and unmanned ground vehicles (UGVs), Wang \textit{et al.} transformed the task offloading challenge into a two-sided matching problem. This transformation enabled them to determine the optimal UAV-UGV association strategy \cite{YWT22-TON}, exemplifying the versatility of potential game theory in solving diverse offloading and resource allocation problems in JCC-SAGIN.

Overall, potential game theory offers a robust framework for decision-making in multi-user environments, especially where individual users' strategies interact and impact each other’s outcomes. Its application in JCC-SAGIN demonstrates its effectiveness in solving complex, interdependent optimization problems, particularly in scenarios involving multiple users and competing resources.

\subsubsection{Power Allocation}
In NTN, managing energy consumption is a more critical issue compared to TN. IoT devices, for example, require increased transmit power to counter severe free-space path loss and atmospheric loss. Additionally, the dynamic topology of NTN, characterized by varying distances between users and SAPs, necessitates real-time adjustment of transmit power by users with limited energy capacity. Furthermore, energy-constrained platforms like UAVs require regular charging, making the control of their transmitting and computing power imperative.

In scenarios where interference is not a factor, the SNR becomes the key metric. Under these conditions, the power allocation subproblem is a relatively straightforward convex problem, solvable using Karush-Kuhn-Tucker (KKT) conditions \cite{HXQ22-CC}.

When interference is considered, the SINR is the primary indicator. Since optimization variables appear in both the numerator and denominator, the non-convex power allocation subproblem requires transformation to become tractable. One effective approach is to rewrite the problem as minimizing the difference of two convex functions. Techniques like difference of convex (DC) programming and successive convex approximations (SCA) can then be applied to derive optimal power allocation solutions \cite{ZYY21-IOTJ,YHH22-JSAC}.
The quadratic transform (QT) technique addresses multiple-ratio concave-convex fractional programming (FP) problems, particularly suited for power control, EE, and beamforming \cite{KW18-TSP}. Tentu \textit{et al.} utilized a block-QT method for global EE optimization in UAV-enabled cell-free systems \cite{VED22-TCOM}. Ding \textit{et al.} applied QT-based FP and DC to solve the terrestrial user association subproblem \cite{CJH22-TWC}.

Other general methods have also been developed for power control. Lu \textit{et al.} used SCA in each iteration, obtaining optimal power through first-order Taylor expansion \cite{WYY22-TCOM}. Bilevel programming has been employed for energy minimization, deriving optimal closed-form expressions for power allocation \cite{FKZ21-TCOM}. Sequential FP, combined with first-order Taylor expansion and Dinkelbach's algorithm, has been utilized to address EE in terrestrial-satellite IoT systems, leading to optimal power allocation solutions \cite{ZYY21-IOTJ}.

\subsubsection{Precoding and Beamforming}
In the domain of computation offloading between SAPs and end devices, many studies have traditionally utilized a single antenna communication model \cite{BHL19-TWC,ZCK19-TWC,XKK19-TWC}. However, there is a growing interest in exploring more advanced precoding and beamforming techniques, such as massive MIMO transmission schemes, particularly in the context of LEO satellite networks \cite{LKJ20-JSAC} and UAV systems \cite{LSR19-TWC}.

Ding \textit{et al.} have leveraged the MIMO communication model to address the challenges of severe path loss and to enhance offloading efficiency in satellite and high-altitude platform-mounted edge computing networks \cite{CJH22-TWC}. They employed the QT based fractional programming (QTFP) approach to tackle the precoding subproblem. This approach involves transforming the subproblem into a convex problem using a weighted minimum mean-squared error method. Additionally, QT theory has been applied to optimize the beamforming matrix in multi-beam satellite communications, further enhancing communication efficiency and signal quality \cite{CYR22-CL}.

\subsubsection{Task Assignment}
After establishing a user association strategy in SAGIN, optimizing the task processing ratio across various computing platforms becomes crucial in the task offloading process. The task assignment subproblem is typically convex and can be resolved using methods like the interior point method, as applied in satellite-aerial integrated networks \cite{CJH22-TWC} and UAV-assisted computing systems \cite{SHJ21-TVT}. For instance, when UAV-ground links are modeled as NLoS channels, the SCA method is employed to determine task allocation within an alternating optimization algorithm \cite{YTJ21-TVT}. Additionally, considering user heterogeneity and the potential failure of UAV-mounted MEC servers, prospect theory can describe users' behavioral patterns, leading to task allocation modeled as a non-cooperative game where pure Nash equilibrium achieves the optimal solution \cite{PGE21-TMC}.

\subsubsection{Trajectory Planning}
UAVs serving as both computing and relay nodes aim to minimize average latency for IoT devices. The 3-D UAV placement challenge is often addressed using exhaustive search methods \cite{LN20-IOTJ, MAH18-WCL}. In UAV-aided wireless-powered MEC networks, placement optimization algorithms based on sequential unconstrained convex minimization, like the polyhedral annexation procedure, are proposed to determine the optimal 3-D locations of UAVs \cite{WJN21-JSAC}.
The SCA technique, which doesn't require the convexity of the JCC problem, is commonly used for UAV trajectory optimization. This method iteratively tightens the upper bound of the problem until convergence is achieved \cite{ZSY22-TWC,YTY22-TCOM,XKY20-TWC,LKC22-TMC,XKK19-TWC,QLZ22-IOTJ,SHJ21-TVT}.

\subsubsection{Task Queue Management}
In scenarios like event-driven environment sensing, computation tasks often arrive stochastically, and due to limited computing capacities, tasks may be processed at different times, leading to queues at IoT devices and SAPs. Balancing task queue stability and computation efficiency is therefore essential.

Lyapunov optimization theory is a valuable tool for addressing long-term optimization problems while stabilizing task queues. This approach involves decoupling the original problem into several real-time ones, using the drift-plus-penalty minimization approach to minimize the upper bound of the drift-plus-penalty expression at each time slot \cite{QSX22-IOTJ,TJQ22-IOTJ}.

\subsubsection{Addressing Multi-Variable Coupling}
The necessity to optimize multiple variables simultaneously often leads to complex coupling relationships among these variables in JCC-SAGIN. This complexity renders the original problems NP-hard, posing significant challenges to direct solution approaches. To effectively manage and solve these multi-variable JCC problems, one widely adopted method is block coordinate descent (BCD) \cite{YTD21-TCOM,LFZ22-TVT}. The core principle of BCD is to iteratively solve for specific variables while keeping others fixed \cite{LM00-ORL}, simplifying the optimization process by breaking down the problem into more manageable parts.

In scenarios where BCD is applied to multi-variable JCC-SAGIN optimization problems involving two blocks, the BCD algorithm essentially operates as an alternating optimization scheme. This scheme is a prevalent iterative framework in many related works \cite{CHZ21-IOTJ,JMQ19-TVT,ZYY21-IOTJ}, offering a structured approach to progressively refine solutions. By alternating between optimizing different variable sets, the algorithm can gradually converge to a solution that balances the objectives and constraints associated with each variable.

The application of BCD and alternating optimization schemes in tackling multi-variable JCC problems underscores the need for strategic problem decomposition in complex optimization scenarios. By focusing on one variable or a subset of variables at a time, these methods reduce the computational complexity and enhance the tractability of problems that would otherwise be too challenging to solve directly. This approach is particularly valuable in the context of SAGIN, where the interdependencies of communication, computing, and networking variables demand sophisticated optimization techniques to achieve efficient and effective resource management.

\subsection{Learning-Based Intelligent Decision-Making Methods in JCC-SAGIN}
Considering the limitations of the traditional optimization tools, researchers began to exploit the application of AI to resource allocation and scheduling in JCC-SAGIN.

AI technology can avoid time-consuming iterations by learning environmental features and determining optimization policies without certain information \cite{KWY19-CM}. The introduction of AI provides an opportunity for intelligent edge computing in SAGIN.
As a promising subset of AI, ML has been widely employed in wireless communication networks and enables SAGIN to monitor, learn and predict communication-related parameters \cite{QRF22-IOTJ}. 
ML can bring out the following benefits for SAGIN \cite{BFY21-NETWORK}: 
\begin{itemize}
	\item Predict the dynamic network environment and unexpected changes.
	
	\item Map the complex relationships among a large number of network factors.
	
	\item Reduce human interventions by automatic management. 
\end{itemize}

When discussing the resource management in JCC-SAGIN, the most widely used ML algorithms and their applications in JCC-SAGIN will be introduced in this subsection. 
\subsubsection{Markov Decision Process}
MDP comprises five parts: state space, action space, reward function, transition probability, and policy, which is often used to model the computation offloading process in SAGIN \cite{CZH22-TITS}. 

Considering the network dynamics, Cheng \textit{et al.} formulated the offloading decision-making as an MDP and proposed a centralized offloading strategy based on the collected global information \cite{NFW19-JSAC}. However, it will bring out additional overhead. Assuming that the flight trajectory of UAVs can be pre-determined, a CL-MADDPG scheme was proposed to learn the task offloading strategy \cite{ZHS21-WCSP}. However, this assumption is weak for the multi-UAV scenarios. An MDP framework characterizes the mapping relationships among task queues and computation procedures in the multi-layer satellite-terrestrial integrated networks \cite{TJQ22-IOTJ}. 

Based on the basic MDP framework, additional MDP models have been further developed. Song \textit{et al.} considered an optimization problem, which jointly minimizes the task latency and energy consumption and maximizes the number of tasks collected by UAVs simultaneously, which is modeled by multi-objective MDP \cite{FHX22-TMC}.
Different from the full observability mandated by the MDP model, for the partially observable Markov decision process (POMDP), the agent cannot directly observe the complete system state but can utilize the observations to form a belief state which can be given by the probability distribution over all possible states \cite{Spaan2012}. The multi-UAV sensing and computation problem is formulated as a POMDP in a UAV-enabled mobile crowdsensing system to maximize the overall utility \cite{TZY22-TNSE}.

\subsubsection{Deep Learning}
With the rapid development of hardware, DL refreshes the implementation of AI techniques and becomes a promising technique to improve the system performance of SAGIN \cite{NZF19-WC}. 

With multiple feature extraction and transformation layers, DL algorithms can address complex scenarios with intricate input-output mappings \cite{QRF22-IOTJ} and massive data \cite{QRF22-IOTJ}. 
These DL algorithms enable the intelligent functions including but not limited to user association strategy, resource allocation, routing, and congestion control \cite{FCX22-NETWORK,NZF19-WC}. 
Based on the historical UAV trajectory and request service rate, DL algorithms can predict the communication demand distribution when UAVs boost the edge intelligence \cite{FCX22-NETWORK}. To support the computing services of Internet of vehicles, Yu \textit{et al.} put forward a deep imitation learning-driven offloading and caching algorithm aiming to jointly minimize the task execution latency and satellite resource usage \cite{SXQ22-IOTJ}.

Deep neural networks (DNNs) are often applied for augmented reality applications, natural language processing, and computer vision \cite{YJ21-IWQOS}. Based on DNN, a two-layer optimization algorithm is presented to solve the nonlinear programming problem in a multi-UAV-enabled MEC system \cite{JPB22-CC}. The upper layer algorithm utilizes differential evolution to obtain the optimal location of UAVs, while the lower algorithm uses distributed DNN (DDNN) to address the offloading problem. 
Compared to a single DNN, the adopted multiple DNNs accelerate the convergence speed and lead to significant differences in the outer layer.

Due to the diversity of generated tasks, massive data are from non-Euclidean domains and are denoted by graphs. Also, DL approaches for graph data have emerged, while the complexity of graph data introduces tremendous challenges to the current ML algorithms. A comprehensive survey on graph neural networks (GNNs) was reviewed regarding data mining and ML \cite{ZSF21-TNNLS}. 
Considering the uncertainty of instantaneous network states in UAV networks, GNNs supervise the UAV's training in actor-critic (A2C) and exploit the hidden network states based on the feature correlation. The simulation results verify the convergence of the proposed GNN-A2C framework in the aerial Edge IoT system.

\subsubsection{Reinforcement Learning}
RL is a learning procedure based on MDP, where the agents observe the network states, make the decisions, and dynamically refine strategies to attain predefined objectives \cite{JHL20-WC}. With RL algorithms, the optimal solutions are obtained by sequential trial-and-error actions \cite{JTR21-NETWORK}. 
RL approaches as a model-free solution have been extensively used in decision-making, like user association strategy, resource allocation, and routing \cite{LDD18-TVT}. However, RL faces limitations in addressing problems characterized by extensive state space, continuous action space, and large-scale input samples\cite{BCH20-WC}.

\textit{* Deep Q-Network}: As one of the essential enabling approaches of RL, the agent of the Q-learning (QL) framework takes action from the current state based on the expected cumulative discount reward and establishes a Q-table step by step \cite{VKD15-NATURE}. QL demonstrates its advantages in learning the optimal policy for any finite MDP via a simple RL. By optimizing the secure offloading strategy in a multi-UAV-assisted MEC network, Lu \textit{et al.} propose a single-agent scheme based on QL and a multi-agent scheme based on Nash QL in order to maximize the system utility \cite{WYY22-TNSE}.

Deep QL (DQL) or Deep Q-network (DQN) is a model-free RL algorithm and consists of fixed and target neural networks. The fixed neural network estimates the Q value, while the target network aids in the convergence of DQL. Typically, DQN works with a discrete action space. 
Facing the challenge of high-scale state and action space in the scenario with multiple UAVs and multiple mobile devices, a two-layer hierarchical trajectory optimization and offloading optimization problem is formulated in \cite{TJB22-IOTJ}, where DQN is utilized to develop offloading scheduler and further to solve the offloading optimization problem.

Other learning algorithms are also developed based on DQN.
A dueling DQN is proposed to determine the node selection and block size decision in the UAV-assisted MEC system, which can calculate the state value and action reward. The adopted dueling DQN can also increase the estimating capability of the environmental state to achieve the optimization objective \cite{MFP20-NETWORK}. 
Since the DQN algorithm may result in overestimating the Q value, double DQN (DDQN) can overcome this disadvantage by separating the selection action from the evaluation one. 
In order to minimize energy consumption, a DDQN algorithm is proposed to jointly optimize the user association and UAV trajectory in the DT-assisted UAV-enabled MEC system \cite{BYL22-TVT}. An online DDQN framework is also adopted in air-ground integrated edge computing networks, where the two DQNs are used to generate the Q-factor and the post-decision Q-factor of the mobile device \cite{XCT22-JSAC}. 
However, the sampling from reply memory is random in the conventional double DQL, neglecting the timeliness and heterogeneity of different actions. This fact is improper for delay-sensitive applications. Aiming at improving the sampling efficiency, a latency-sensitive replay memory algorithm based on double DQL is presented to obtain the offloading strategy by utilizing the local and neighboring historical information \cite{FHN22-JSAC}. 
Furthermore, a DRL offloading method based on dueling double DQN (D3QN) addresses the MDP formulation. By combining the DDQN and the dueling framework, the D3QN-based offloading policy trains the approximating function and calculates the Q value of each action at each state \cite{TJQ22-IOTJ}. 

\textit{* Deep Reinforcement Learning}: DRL is a learning framework extended based on DL and RL, which aims to learn the optimal strategy according to the action feedback in a dynamic environment without prior knowledge \cite{AGS21-IOTJ}. 
In-depth research has been conducted on JCC problems in SAGIN using DRL techniques. 
A DRL-based algorithm in the UAV-mounted MEC network is proposed to maximize the long-term system reward by optimizing the UAV trajectory \cite{QLL20-TVT}. 
DRL techniques are also used in mobile crowdsourcing systems to enable model-free UAV control. To ensure the mobile unmanned charging station reaches the destination in time, the proposed DRL-based experience-driven control framework shows the effectiveness and robustness via simulations \cite{BCJ18-TII}. 
The two problems of multi-objective RL and multi-agent RL are investigated in heterogeneous satellite networks, where the advantages of DRL in resource allocation have been verified \cite{BCH20-WC}.

Although using DRL approaches in SAGIN can significantly make resource management more efficient, training DRL frameworks is always resource-consuming \cite{ZZM21-NETWORK}. Specifically, the training model based on local data is only helpful for the local environment, limiting the universality in other new scenarios. Moreover, DRL agents may spend a long time for DRL training due to collecting enough usable data, leading to a long convergence time.

Using multiple SAP servers in SAGIN promotes the development of multi-agent learning methods. There are three typical multi-agent DRL algorithms developed based on an actor-critic framework, i.e., centralized training and centralized execution (CTCE), centralized training and decentralized
execution (CTDE), and decentralized training and decentralized execution (DTDE) \cite{SHJ22-WC}. 

\begin{figure}[ht]
	\centering
	\subfigure[CTCE]{\includegraphics[width =0.25\textwidth]{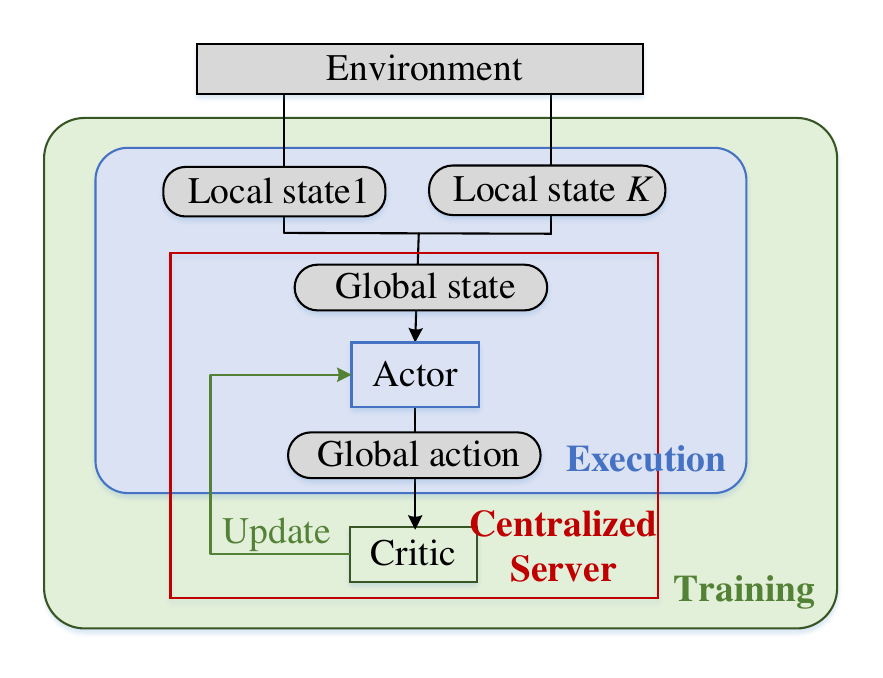}\label{fig:CTCE}}
	\subfigure[CTDE]{\includegraphics[width =0.29\textwidth]{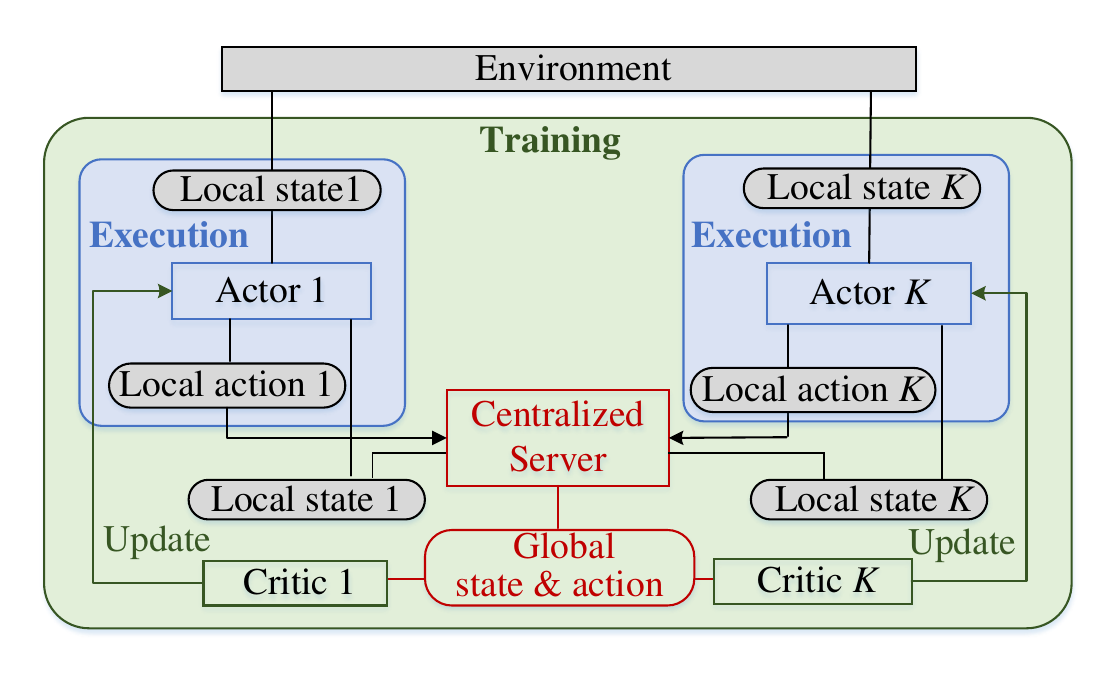}\label{fig:CTDE}}
	\caption{Comparison between CTCE and CTDE.} 
	\label{fig:CTCE_CTDE}  
\end{figure}

Note that the update of the Q-value uses temporal difference (TD)-error. Since the experience with high TD error usually means successful attempts, assigning different weights for samples is an efficient way to select the experience. To overcome the problem of divergence and vibration of the prioritized experience replay scheme, the importance-sampling weight is adopted to denote the importance of selected experience under CTCE in the UAV-assisted MEC system \cite{LKC22-TMC}. The process of CTCE algorithm is shown in Fig. \ref{fig:CTCE}.
However, the CTCE framework assumed that a single actor-critic pair is deployed on a central controller to coordinate the multiple UAVs. 
The CTDE framework requires the global state during centralized training. Then, each SAP is executed by an individual agent, which can make independent decisions according to local observations. The working procedure of CTDE algorithm is shown in Fig. \ref{fig:CTDE}.
Extensive research has been focused on CTDE algorithms for UAV network \cite{NZY21-TWC,AJH22-JSAC,AGB21-TNSM,TZY22-TNSE,HX21-JSAC} and HAP station-assisted MEC network \cite{QOG22-TWC}. 

\begin{figure*}[ht]
	\centering
	\includegraphics[width = 0.7\textwidth]{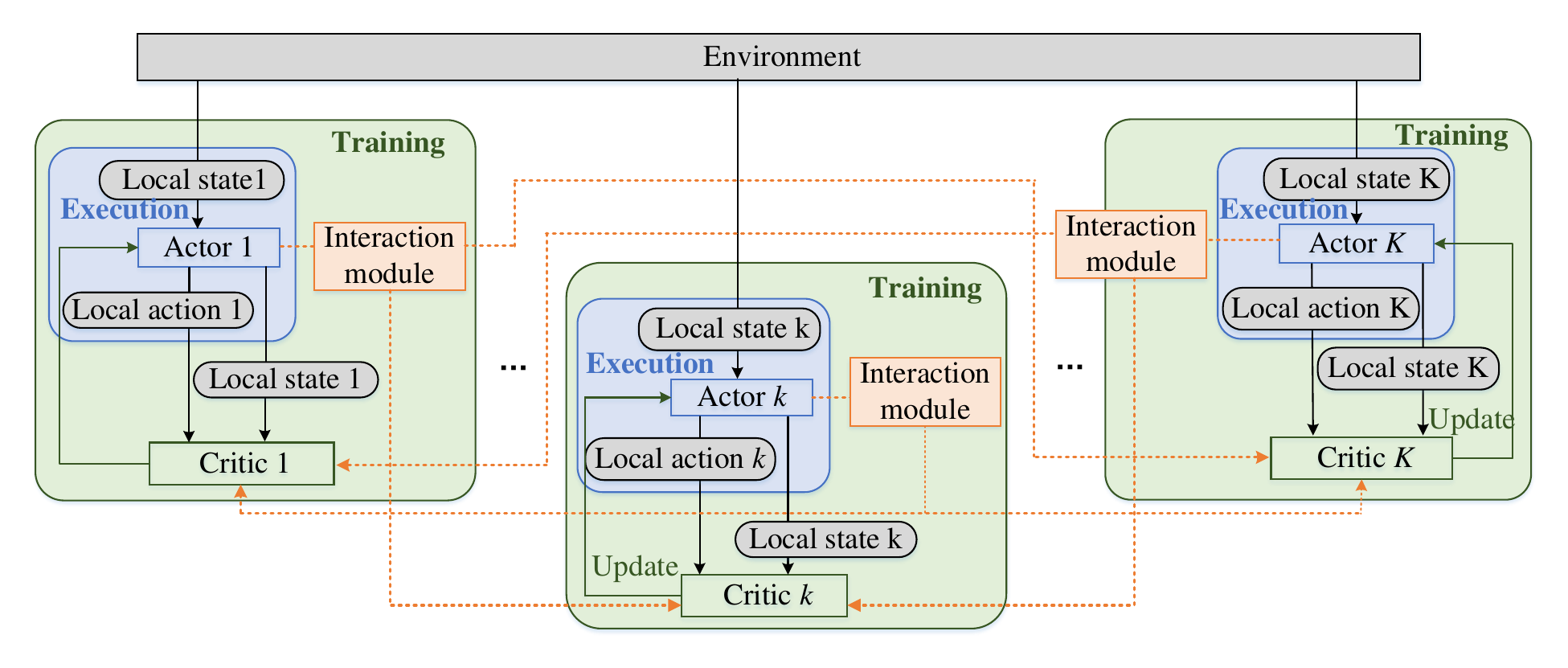}
	\caption{Procedure of DTDE. \label{fig:DTDE}}
\end{figure*}

Both CTCE and CTDE require a centralized coordinator for the training process of actor-critic DNN. 
However, this assumption is weak for future wide-area SAGIN without cellular coverage and centralized coordination units. Aiming to decrease reliance on centralized processing, Hwang \textit{et al.} proposed a DTDE framework to realize decentralized implementation \cite{SHJ22-WC}, and the procedure of DTDE is shown in Fig. \ref{fig:DTDE}. To compensate for insufficient network information, the DTDE algorithm introduces an efficient mechanism among agents for interacting messages related to decentralized operations.

\textit{* Deep Deterministic Policy Gradient (DDPG)}: 
DDPG is an offline learning method, which integrates the DRL and actor-critic framework \cite{NIPS1999_6449f44a}. 
There are three key components in DDPG: the primary network, the target network, and the replay memory \cite{JTR21-NETWORK}.
Recall that DQN is suitable for the optimization variables with discrete action space. Compared to DQN, the main advantage of DDPG is its capability to address the problem with continuous variables. 

The DDPG approaches are widely investigated in multi-UAV MEC system 
and satellite network \cite{SSL22-IOTJ}. 
Specifically, in the high-dynamic UAV-aided MEC system, it is unrealistic to apply conventional optimization methods to obtain the hybrid beaming matrices according to the real-time system state. A DDPG-based algorithm is designed for hybrid beamforming when the UAV and mobile devices roam, which is more suitable than the semi-definite relaxation (SDR) approach \cite{WJN21-JSAC}.  
Due to the continuity of the UAV's horizontal locations, an offline RL algorithm based on DDPG is proposed to perform the trajectory optimization problem \cite{TJB22-IOTJ}. 
In satellite networks, the DDPG approach also shows its advantages against DQN regarding energy consumption, network reward, dropped tasks, and risk \cite{SSL22-IOTJ}.
These single-agent DDPG algorithms are useful for optimization problems with small-scale variables. However, the network environment between SAPs and IoT devices is wide-ranging and changes unpredictably at each time slot. 

\begin{figure*}[ht]
	\centering
	\includegraphics[width = 0.7\textwidth]{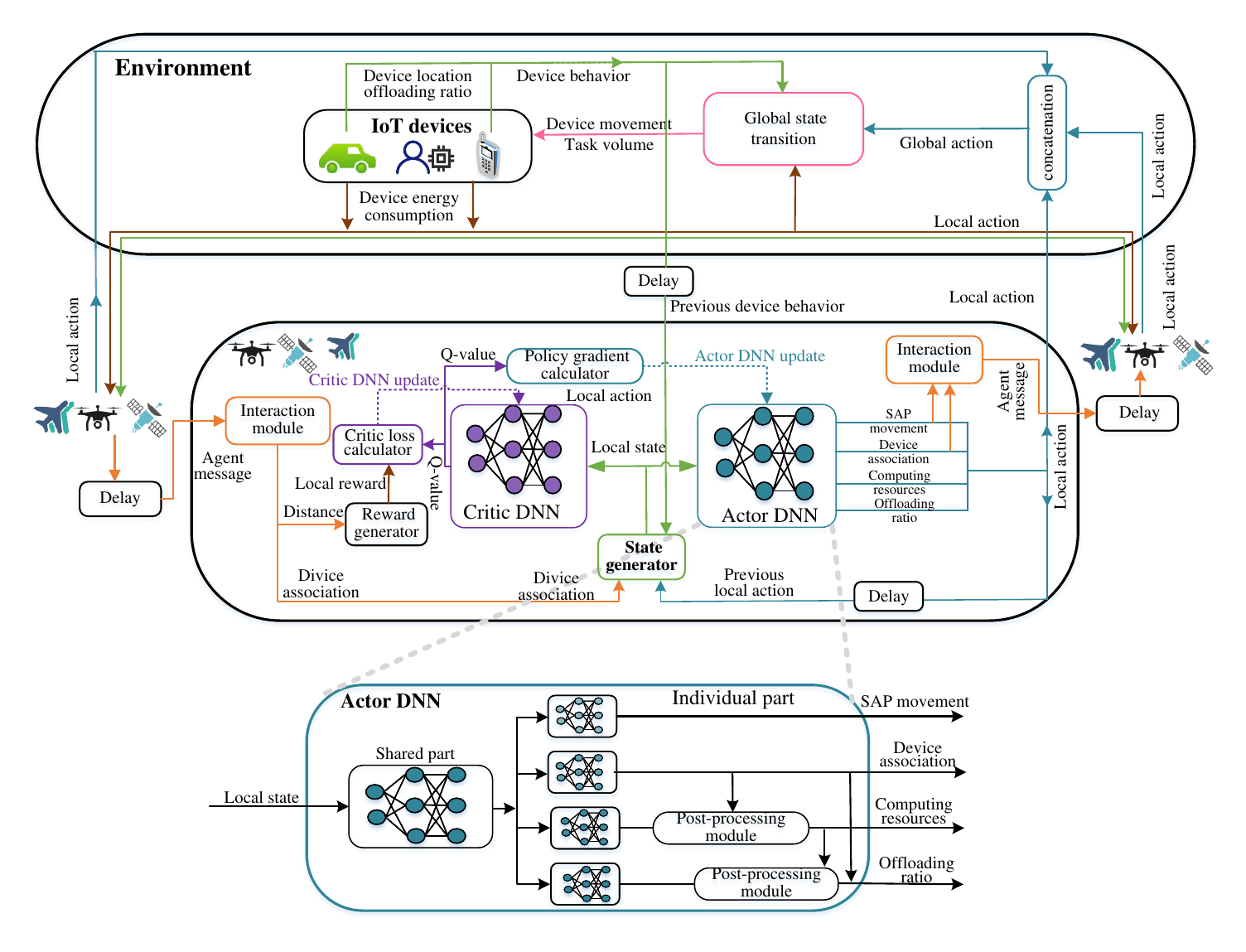}
	\caption{DTDE multi-agent DDPG in JCC-SAGIN. \label{fig:detail_DTDE}}
\end{figure*}

Generally, multiple agents exist in the dynamic SAGIN-enabled MEC systems. As one of the multi-agent DRL frameworks, the multi-agent DDPG (MADDPG)-based algorithm is proper for the multi-agent environment since the system performance gets improvement with agents learning cooperatively \cite{RYA17-ICNIPS}. 
In order to maximize the long-term reward in the aerial-enable MEC network, the MADDPG-based multi-UAV assisted IoT edge framework is proposed, which achieves the optimal strategy and decreases the training cost with the CTDE technique \cite{AGB21-TNSM}.
A multi-leader multi-follower Stackelberg game is formulated for the task offloading problem, and a model-free MADDPG algorithm is utilized to maximize long-term reward by optimizing the offloading decision in multi-UAV-assisted IoT networks \cite{AJH22-JSAC}. The simulation results show the advantages of MADDPG in delay, energy consumption, and latency compared to DDPG, asynchronous advantage actor-critic, and dueling DQN approaches. The offline-training MADDPG model obtains the optimal association and resource allocation strategies \cite{HX21-JSAC}. Compared to the single-agent DDPG-based and random schemes, the QoS satisfaction proportion of the proposed MADDPG-based resource management model is higher. 
A DTDE multi-agent DDPG framework is proposed in \cite{SHJ22-WC}. By extending the framework into JCC-SAGIN, the algorithm is shown in Fig. \ref{fig:detail_DTDE}.

As the number of SAPs increases, obtaining the optimal action under MADDPG would be hard, and the system performance would deteriorate. Considering the high-dimensional continuous action space, the multi-agent TD3 (MATD3) optimization algorithm is proposed to obtain UAV trajectory, offloading strategy, and resource allocation scheme in a dynamic MEC environment \cite{NZY21-TWC}. Compared to MADDPG, the proposed MATD3 scheme significantly outperforms in terms of total system cost.

\textit{* Summary Remarks}:
Plenty of works investigated the applications of RL methods in resource management due to its advantage of model free.
Table \ref{tab:Related_Work_RL} summarizes and compares the typical works which introduces RL method to provide intelligent and adaptive solutions for resource management in JCC-SAGIN.
Most works assume that there is a server for centralized training, which may lead to higher costs in SAGIN than in TN. 
In addition, the training and execution process will introduce additional latency and energy consumption. These facts will incur little performance loss on delay and increase the burden of energy-limited devices.
Therefore, it is necessary to design adpative DTDE algorithms, which consider different features of platforms or devices and their deployed resources in JCC-SAGIN.

\begin{table*}[h]
	\centering
	\caption{Typical works using Reinforcement Learning Method in JCC-SAGIN}\label{tab:Related_Work_RL}
\begin{tabular}{|m{0.9cm}<{\centering}|m{0.6cm}<{\centering}|m{1.5cm}<{\centering}|m{2cm}<{\centering}|m{5cm}<{\centering}|m{5cm}<{\centering}|}
\hline
       Method              & Ref.       & Scenario                                 & Algorithm & Key issues & Objective            \\ \hline \hline
\multirow{5}{*}{DQN}  & \cite{TJB22-IOTJ} & \multirow{3}{*}{ \makecell*[c]{UAV-assisted \\ MEC system}}                                             & DQN, HTO3       & High-dimensional state and action space       & Minimize the average task delay       \\ \cline{2-2} \cline{4-6}
                      & \cite{MFP20-NETWORK} & 
                                      & Dueling DQN       & Data transmission, security and reliability      & Ensure security, maximize data computation capacity and throughput                           \\ \cline{2-2} \cline{4-6}
                      & \cite{BYL22-TVT} &                                    & DDQN      & Intelligent task offloading      & Minimize energy consumption                          \\ \cline{2-6} 
                      & \cite{TJQ22-IOTJ} & STIN                                     & D3QN   & Offloading path selection and resource allocation      & Maximize the number of offloaded tasks and minimize the power consumption of LEO satellites           \\ \cline{2-6} 
                      & \cite{FHN22-JSAC} & SAGIN                                    & Delay-sensitive
replay memory algorithm 
      & High dynamic and complex optimization issues     & Reduce package loss and increase 
throughput, improving packet delay    \\ \hline
\multirow{4}{*}{DRL}  & \cite{QLL20-TVT} & \multirow{7}{*}{\makecell*[c]{UAV-assisted \\ MEC system}} & DRL      & UAV trajectory
      & Maximize the long-term system reward     \\ \cline{2-2} \cline{4-6}
                      & \cite{LKC22-TMC}  &                                          & CTCE       & Divergence and vibration     & Minimize energy consumption                          \\ \cline{2-2} \cline{4-6}
                      & \cite{HX21-JSAC} &                                          & CTDE     & Resource management      & Maximize the number of offloaded tasks                           \\ \cline{2-2} \cline{4-6}
                      & \cite{SHJ22-WC} &                                          & DTDE       & Time-varying offloading demands and mobility      & Minimize energy 
consumption                            \\ \cline{1-2} \cline{4-6} 
\multirow{4}{*}{DDPG} & \cite{WJN21-JSAC}  &                                          & DDPG     & hybrid beamforming design and resource allocation      & Maximize the sum computation rate    \\ \cline{2-2} \cline{4-6}
                      & \cite{NZY21-TWC} &                                          & MATD3     & UAV trajectory, offloading strategy and resource allocation     & Minimize the sum of execution delays and energy consumption                        \\ \cline{2-2} \cline{4-6}
                      & \cite{AGB21-TNSM} &                                          & MADDPG      & Task offloading and resource allocation   & Maximize the long-term reward                      \\ \cline{2-6} 
                      & \cite{SSL22-IOTJ} & Satellite networks                       & DDPG     & Security-aware computation offloading      & Minimize the time, energy,
and security cost    \\ \hline
\end{tabular}
\end{table*}

\subsubsection{Federated Learning}
Most of the above learning procedures applied for the JCC problem in SAGIN rely on centralized control, leading to high signaling and computation overhead in large-scale SAGIN. Integrating data and control information also results in severe control delay, energy consumption, and synchronization problems \cite{NBF20-WC}. Moreover, global information collection at central units is improper when the generated tasks are private and cannot be shared \cite{FCX22-NETWORK}.

As a distributed ML technique, FL enables the end devices to train the learning model locally and transmit the training parameters instead. With the data remaining at the devices, FL motivates the computation of many AI applications to the IoT devices without violating privacy \cite{MZW21-TWC}, like face detection, next-word prediction, and voice recognition \cite{HMZ22-WC}. 

After introducing FL into the SAGIN-enabled computation network, space/aerial aggregation becomes a promising application. The flying SAPs as servers aggregate the model in space/air, which can help the learning coverage of terrestrial FL-enabled networks. These SAPs can also act as FL users, train the AI models locally and transmit their updates to the cloud servers \cite{QRF22-IOTJ}.

The effects of introducing FL into a UAV-assisted MEC network are investigated in \cite{ZMX21-WC,CYY21-NETWORK}. 
FL is utilized to improve the training efficiency of DRL and DNN, and the simulation results verify that it outperforms centralized and separate learning schemes in terms of prediction accuracy \cite{ZMX21-WC}. 
To boost edge intelligence, UAVs join the model training of FL among multiple IoT devices, leading to the lower energy consumption of UAV \cite{CYY21-NETWORK}. This work also points out the necessity of deploying lightweight ML models for inference on UAVs.

Although federated RL (FRL) has been verified its superior performance in edge learning systems \cite{RYS21-NETWORK}, it is still tough to design task offloading schemes with FRL techniques in high dynamic SAGIN with heterogeneous features. Tang \textit{et al.} formulated the task offloading problem in SAGIN as an MDP and proposed the blockchain-based federated asynchronous advantage actor-critic (BFA3C) algorithm to ensure the security and improve system performance \cite{FCL22-JSAC}.
Since the practical actions simultaneously include discrete and continuous ones, the JCC optimization problem cannot be addressed only by leveraging DQN or DDPG alone. Based on the hybrid DRL framework, which combines DQN and DDPG algorithm \cite{2018arXiv181006394X}, the FL technique is introduced to avoid privacy leakage and reduce massive communication overhead in UAV-MEC system \cite{ZZM21-NETWORK}. Compared to the baseline algorithms, the proposed federated DRL (FDRL) method can achieve higher average reward and lower execution delay within limited iterations.
Liao \textit{et al.} investigated the security problems in computation offloading and the adverse impacts of electromagnetic interference (EMI) in SAGIN for the first time \cite{HZZ22-JSP}. Based on FDRL, the proposed federated deep actor-critic-based EMI-aware task offloading method guarantees model convergence and decreases the adverse effects of EMI via discarding the aberrant and inferior local models from the federated set.  
The learning methods based on FL are also developed in satellite and high-altitude networks. In order to satisfy the URLLC constraints in space-assisted vehicular networks, an asynchronous federated DQN-based algorithm is proposed to address the task offloading problem based on the environment observations \cite{CZH22-TITS}. 
Aiming to jointly minimize the energy consumption and latency, a support vector machine (SVM)-based FL algorithm is posed to obtain the user association strategy in high-altitude balloon networks \cite{SMC21-IOTJ}. 

The above works utilize FL to improve the computing efficiency of complex optimization algorithms. Recently, another emerging research branch is investigating the FL framework in satellite networks \cite{2023arXiv231101483L}, where multiple satellites and GSs collaboratively train a global model. They formulated an optimization problem to find the optimal model parameters aiming to minimize the global loss. Compared to terrestrial FL, one of the challenges when implementing FL to NTN is to investigate the impact of the SAP's movement on learning performance. The research in this area is still in the initial phase.

In summary, the dual aspects of how FL influences optimization processes and how optimization strategies impact FL constitute the primary research points when applying FL in JCC-SAGIN. It becomes imperative to consider the mobility patterns and trajectories of SAPs in designing and developing an efficient FL framework.

\subsubsection{Transfer Learning}
Unlike the traditional ML agents who learn a new task each time, the TL agents can use the prior knowledge from others who have performed similar tasks, resulting in higher learning efficiency and accelerated convergence rate \cite{ZKA20-ARXIV}.  
By utilizing these advantages, a TL-enabled aerial edge network is proposed, where the UAVs can share and use the information from neighbor ones \cite{KDW21-WC}. The authors also design the joint UAV deployment and resource allocation schemes to optimize the resource utility by leveraging TL \cite{DKF21-ICCT}.

\subsection{Summary Remarks}
\begin{figure}[h]
	\centering
	\includegraphics[width = 0.5\textwidth]{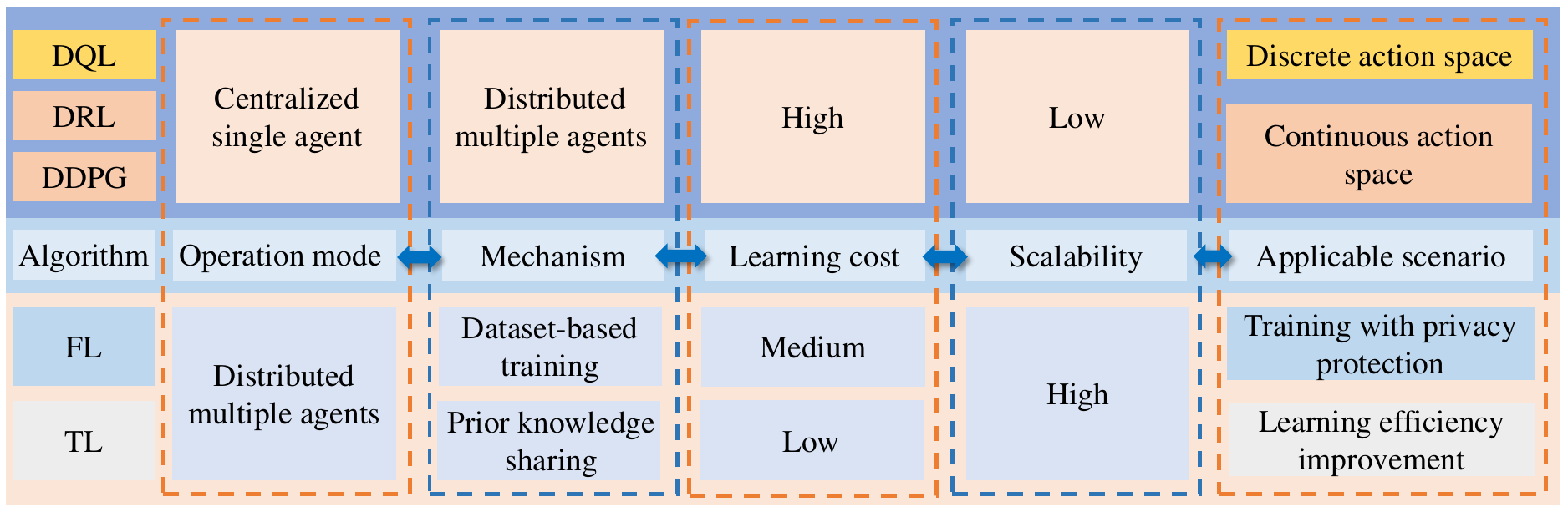}
	\caption{Comparison of typical ML algorithms for resource management in JCC-SAGIN. \label{fig:learning_method}}
\end{figure}

This section provides an in-depth analysis of resource management methods in JCC-SAGIN, covering both traditional optimization strategies and learning-based intelligent decision-making methods. The traditional optimization approaches for the JCC problem in SAGIN spans seven critical dimensions: association between users and platforms, power allocation, precoding and beamforming, task assignment, trajectory planning, task queue management, and the intricacies of multi-variable coupling. The application of potential game theory is highlighted for its efficacy in addressing complex, intertwined optimization challenges, where power allocation plays a pivotal role in minimizing interference and amplifying system EE.
With the evolution of network technologies, there's a burgeoning interest within the academic sphere in pioneering advanced precoding and beamforming methodologies to mitigate the significant path loss in NTN, especially in the context of task offloading. Furthermore, task assignement, UAV flight path planning, and the efficient management of task queues are focal points in the resource management for JCC-SAGIN. Addressing these optimization problems often entails numerous variables, introducing substantial complexity to the problem-solving process.
Transitioning to the learning-based intelligent decision-making methods can effectively address the limitations of traditional optimization techniques. Typical AI techniques are depicted in Fig. \ref{fig:learning_method}. Our work indicates that ML applied to SAGIN holds significant application value in enhancing network robustness, delineating complex mapping relationships, and facilitating rapid responses.

\section{Future Research Directions in JCC-SAGIN}\label{sec:future}
Implementing JCC in SAGIN still requires further design at different system levels, including but not limited to coding, multiple access techniques, security, and integration with maritime applications. This section will discuss these future research directions and challenges in JCC-SAGIN.

\subsection{Coding-Enhanced Distributed JCC-SAGIN}
Most computation-intensive tasks need to be retrieved after processing by the edge or cloud servers. However, uncertain disturbances, like frequent link errors and straggler nodes, will increase transmission failure and decrease computing efficiency.
With the fault tolerance requirement, If the returned copies from the straggling edge/cloud servers are fewer than the required recovery threshold, the devices that generate the tasks cannot retrieve the received result. This side effect will be enlarged when the tasks are executed by multiple servers parallelly.

To reduce packet losses and boost throughput through information fusion from multiple channels, coded storage-and-computation (CSC) has gradually been used in distributed SAGIN. The main idea of CSC to alleviate node failure is associating different clusters and nodes in a clustered distributed system (CDS) by information interleaving \cite{BKV18-SCIS}.
CSC techniques have gradually become a theoretical foundation that balances communication, caching, and computing loads. 
In addition, integrating CSC and AI can promote the multiple functions of ML algorithms, like traffic prediction, resource management, and trajectory planning. 
Aiming to enhance the service quality of SAGIN, the CSC technique integrated with AI is used in content distribution according to the workload and capacity of storage nodes \cite{SQW20-WC}. The lost and disrupted data stripes can be regenerated with data repair mechanisms. Also, the bandwidth can be reduced using an adequately coded structure, and data synchronization can be achieved. 
The authors further focus on the coding design to overcome the soft errors in outer space for satellite clustered storage systems. 
Considering the diversity of inter-cluster bandwidth, an asymmetric repair of regenerating codes (RCs) and generalized RCs is proposed to reduce the data repair cost \cite{9580435}. 

Besides fault tolerance, the unavoidable straggler is another challenge in  SAGIN-enabled distributed computing. The offloading delay depends on the slowest distributed computing nodes (workers). The randomness of the distributed parallel computing framework, like the time-varying task volumes of the workers and uncertain poor processing capabilities, will cause straggler effects, significantly increasing the communication and computation delay.

Adding redundancy to computation tasks is the key idea to deal with the straggler effect. The related methods mainly include task replication with redundant scheduling \cite{gardner2017redundancy,joshi2017efficient,amiri2019computation} and coded distributed computing (CDC) \cite{8002642}. 
The widely used CDC schemes contain maximum distance separable (MDS) coding \cite{reisizadeh2019coded}, gradient coding \cite{10043662}, rateless coding \cite{MCS22-CACM}, and multivariate polynomial coding \cite{9519610}.
Compared to simple replication, CDC can improve computing efficiency by using error correction codes to add redundancy. However, the task partition ratio to workers and the redundant load amount should be carefully designed.

To minimize the network cost and prevent the improper subscription of the resources, a two-phase stochastic coded offloading approach is proposed in a UAV-assisted computing network \cite{9709578}. The proposed coding scheme can overcome the uncertainty of weather, demand, and shortfall when applying the CDC technique to mitigate the stragglers. 
A novel framework with CDC is proposed to solve the delay-energy cost tradeoff problem when tasks are offloaded from multiple UAVs to ground edge servers \cite{9417305}. This is the first work to combine CDC and task offloading in UAV networks. 
A coded computation offloading strategy is designed to mitigate the terrestrial dense computation tasks to distributed satellite constellations. As a result, the computing stragglers with MDS and rateless codes can decrease \cite{9498862}. Also, the latency-energy tradeoff problem is investigated, and the optimal processing platforms and coding parameters are jointly obtained. 

The above works discussed the CDC techniques in space/air-terrestrial integrated networks. The coding caching approach consisting of content placement and coded transmission phases is proposed in SAGIN-IoV. The authors discuss single road section (RS) and multi-RS scenarios, and the optimal content placement, power allocation, and coverage deployment are jointly derived respectively \cite{9377456}. 

\subsection{Multiple Access Technique Design in JCC-SAGIN}
\subsubsection{Orthogonal Multiple Access}
Orthogonal frequency-division multiple access (OFDMA) and time-division multiple access (TDMA) are typical OMA techniques. 
Herein, TDMA requires a strict time sequence, which will cause high costs for information exchanges.
Since most computation tasks have stringent delay requirements, OFDMA is more widely used than TDMA.

OFDMA technique is widely used during the computation offloading from terrestrial users to SAPs. With the assumption that all the uplinks work on orthogonal channels, the interference among users can be omitted \cite{9453824,LJ22-TVT,NZY21-TWC,YTY22-TCOM,AJH22-JSAC,EHC21-TCOM,LKC22-TMC}, and the SNR is taken as the metric of interest.

Co-channel interference may occur when the offloading tasks exceed the number of orthogonal subchannels.
In a multi-UAV task execution network, multiple UAV members will offload their tasks to their coalition head, and the UAVs will cause co-channel interference when they select the same subchannel \cite{JQY21-WCL}.
The authors extended the framework into a hierarchical UAV-assisted MEC network \cite{9732351}. The member UAV can offload its task to the UAV in the coalition head layer for edge computing directly or forward to the central layer by the relay node. Mutual interference may exist when multiple UAVs transmit their tasks to the head UAVs.
Some works also adopt full frequency reuse, where all the users which transmit signals at the same time can cause interference \cite{YWT22-TON,PGE21-TMC,TZY22-TNSE,BHJ22-TSC,HX21-JSAC}.

\subsubsection{Non-Orthogonal Multiple Access}
NOMA has been regarded as a promising technique for 6G wireless networks. Unlike OMA, NOMA can enable massive computation tasks offloaded to SAPs simultaneously within the same given frequency/time resource block, and the signal can be decoded with successive interference cancellation (SIC) techniques. Therefore, the combination of NOMA and MEC can realize efficient spectrum utilization, increase throughput, and support ultra-high connectivity. NOMA holds several potential advantages in SAGIN, especially within the framework of joint communications and computing :
\begin{itemize}
   \item \textit{Enhancing Spectrum Efficiency}: NOMA allows multiple users to communicate simultaneously on the same time and frequency resources. By employing interference management techniques such as SIC, NOMA enhances spectrum efficiency. This is crucial for SAGIN systems with limited spectrum resources, enabling more effective utilization.
    \item \textit{Scalability for Massive Connectivity}: NOMA supports large-scale connectivity, facilitating multiple users to share the same spectrum resources simultaneously. This is particularly significant in SAGIN scenarios with a substantial number of ground terminals or IoT devices requiring communication and computation.
    \item \textit{Flexibility and Adaptability}: NOMA exhibits strong flexibility, adapting to diverse communication requirements and channel conditions among users. In SAGIN, where different environments such as air, ground, and satellite coexist, NOMA's adaptability contributes to optimizing communication performance.
    \item \textit{Improving Energy Efficiency}: NOMA achieves enhanced EE by allocating users to different power levels based on their channel conditions. In SAGIN, this aids in extending the battery life of ground terminals or IoT devices, especially in environments where recharging is challenging.
    \item \textit{Support for Multi-Task Offloading}: In JCC scenarios, ground terminals may need to concurrently handle both communication and computation tasks. NOMA's capability to support simultaneous offloading of multiple tasks contributes to overall system efficiency.
\end{itemize}

It is important to note that while NOMA possesses these potential advantages, its actual performance depends on system design, channel conditions, and specific application scenarios. Therefore, considering the adoption of NOMA in SAGIN necessitates detailed system analysis and performance evaluation. UAV-NOMA-MEC networks have confirmed their advantages in large-scale access networks with stringent spectrum constraints \cite{9940945,ZMX21-WC,WYY22-TCOM}.

\subsubsection{Combination of OMA and NOMA}
The above works discuss OMA and NOMA techniques separately. Recently, some works have considered combining them and utilizing their advantages simultaneously. 

To maximize secure computing capacity, TDMA and NOMA are adopted in a UAV-assisted MEC system \cite{YTD21-TCOM}. Compared to the TDMA scheme, the algorithm complexity is higher, while the security performance is superior under the NOMA scheme. 
Dai \textit{et al.} investigated the computation offloading problem in UAV-assisted marine networks \cite{MYL22-TNSE}. NOMA is used for acoustic transmission between multiple underwater sensor nodes and the unmanned surface vehicle (USV). FDMA is used for RF transmission between the USV and multiple UAVs. 
Cui \textit{et al.} \cite{8685130} and Mu \textit{ et al.} \cite{9447204} proposed different optimization algorithms for OMA and NOMA modes, respectively and showed their different advantage areas through simulation results.

To sum up, the multiple access techniques in SAGIN mainly adopt OMA or NOMA, which is a binary choice problem. Considering OMA has the advantages of no co-channel interference in the cell and low complexity while NOMA can further improve the spectral efficiency, especially for scenarios with multiple users, it can achieve more performance gains when adopting partial multiple access techniques according to task attribute. That is, combining OMA and NOMA simultaneously and designing the ratio of these two strategies.

\subsection{Security Countermeasures in JCC-SAGIN}

\subsubsection{Blockchain}
During the multi-hop transmission within and between different segments in large-scale SAGIN, the transmitted data is vulnerable, and intrinsic trust issues may appear due to decreasing human intervention. 
Although many works proved the benefits of introducing SAPs to IoT services, few discussed data security and privacy while ensuring communication and computing efficiency.

\begin{figure}[h]
	\centering
	\includegraphics[width = 0.5\textwidth]{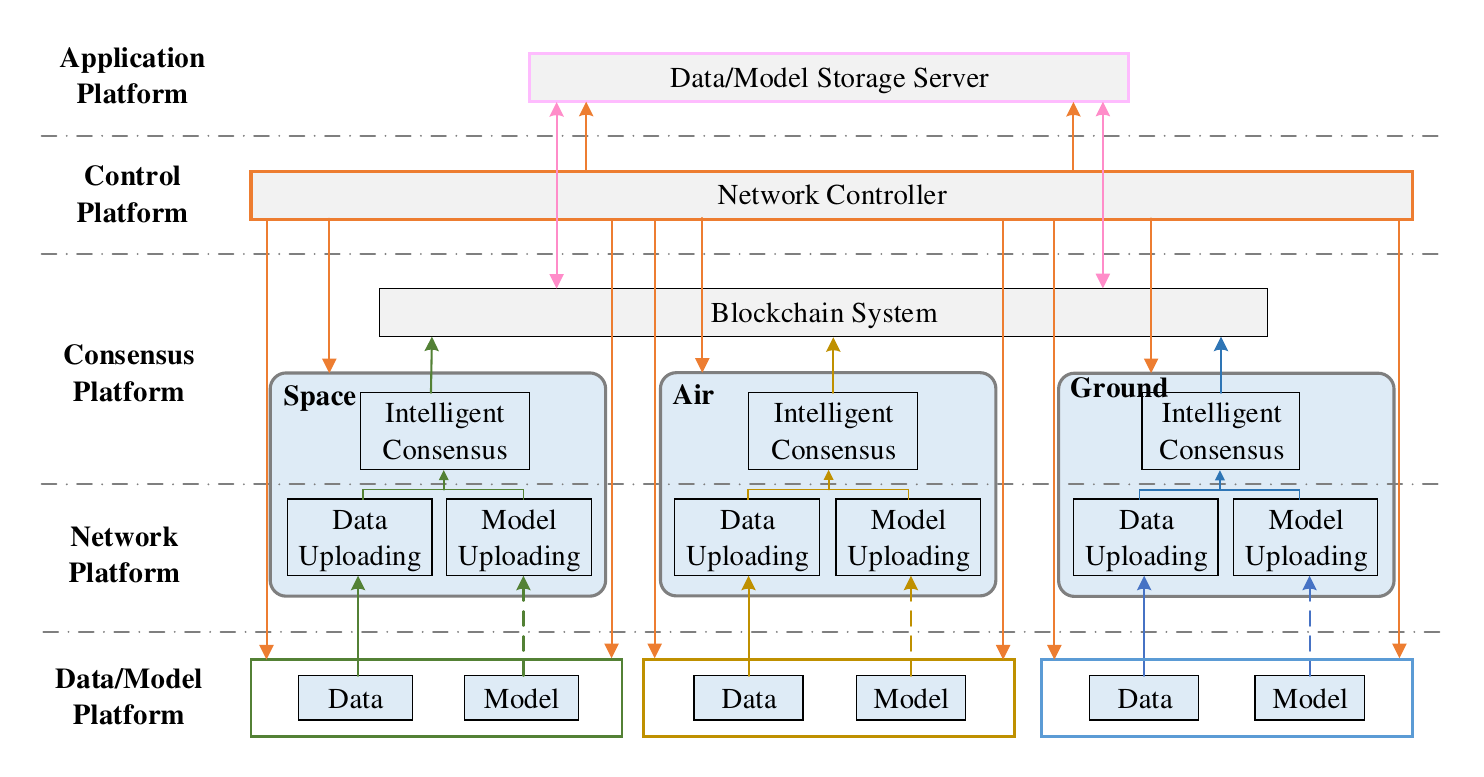}
	\caption{Logical architecture of blockchain-embedded JCC-SAGIN. \label{fig:logical_blockchain}}
\end{figure}


Blockchain is an emerging and promising technology to provide a feasible solution for security management in SAGIN. The logical architecture of blockchain-embedded JCC-SAGIN is shown in Fig. \ref{fig:logical_blockchain}.
With blockchain technology, any collected data is recorded by an immutable cryptographic signature through one-way hash functions and consensus mechanisms. 
Compared to traditional security solutions like encryption and intrusion detection, blockchain has the advantages of decentralization, transparency, immutability, traceability, and auditability \cite{9631953}.
By employing blockchain technology in the system, SAGIN can realize decentralized control without a central authority and avoid disrupting the system due to specific node destruction.

Despite the great potential of blockchain-empowered SAGIN, such a framework faces some fundamental challenges.
First, blockchain technology requires frequent sequential validation requests, bringing out intolerable latency. Second, duplicating the whole chain from source to destination will occupy numerous storage resources. 
Furthermore, computing power also increases the burden of IoT devices and SAPs.

Wang \textit{et al.} focused on the integration
of blockchain technologies for securing SAG-IoT applications, and reviewed the blockchain-based applications such as trustworthy resource management and resilient network design \cite{9631953}.
In order to realize efficient management of diverse resources, a collaborative blockchain architecture is proposed for SAGIN \cite{9316453}. In each segment of SAGIN, a specific blockchain is designed to manage its resources independently. Then, the interaction of different blockchains makes global information interoperability and authentication possible.
Compared to the centralized control protocol, the cost of storage resources under the proposed incentive-based collaborative scheme decreases significantly while ensuring the failure probability.

The integration of blockchain and other efficient technologies can achieve better system performance. 
Due to the same features of decentralization and interdependence, the combination of MEC and blockchain can bring out lower latency and higher EE.  
By adopting the Byzantine fault tolerance (PBFT) consensus protocol, UAVs as blockchain nodes verify the data and transmit the unchanged data to the blockchain system \cite{MFP20-NETWORK}. 
The security issues of FL can be addressed by storing and sharing models via blockchain. 
A blockchain-based FL framework is exploited to guarantee the secure sharing of topological and model information \cite{FCL22-JSAC}. The authors improved the traditional PBFT protocol and developed an enhanced algorithm based on the proposed node security evaluation mechanism.
Blockchain also integrates with multi-agent DRL by storing the learning model-related data to reduce the error probability \cite{8726067}.
To overcome the emergency and vulnerability problems faced by EH, Seid \textit{et al.} formulated a joint optimization problem of computing offloading and EH to minimize computation costs and maximize utilities in multi-UAV-assisted IoT network \cite{AJH22-JSAC}.
Liao \textit{et al.} investigated the adverse impacts of EMI and security issues in computation offloading in SAGIN \cite{HZZ22-JSP}. 
The consortium blockchain was also introduced in this work to reduce block creation delay by leveraging LEO satellites for broadcasting messages among blockchain nodes.

\subsubsection{Quantum-Based Approaches}
Due to its heightened security characteristics, quantum communication holds significant potential for widespread application in space-air integrated network (SAIN) \cite{9210567}.

In satellite networks, secure communication between satellites and between satellites and GSs can be realized with entanglement-based schemes.
Furthermore, quantum communication can serve as a valuable tool for enhancing the precision and stability of satellite navigation systems through precise positioning and clock synchronization processes.
A comprehensive design and performance analysis are investigated in an LEO satellite quantum communication system, and the potential applications in secure communication links over long distances are highlighted \cite{Bourgoin_2013}. 
The authors propose a novel architecture that utilizes a global quantum key distribution (QKD) network to establish secure communication links between the GS and provide a detailed description of the system design, including the satellite orbits, the satellite payload, the GS equipment, and the cryptographic protocols.
The article also presents a thorough performance analysis of the proposed system. It evaluates the system performance in terms of the key generation rate, the critical transmission rate, and the error rate. The results show that the proposed LEO satellite quantum communication system has the potential to provide high-speed, secure communication links over long distances.
The satellite-based QKD system is also discussed in \cite{chen2021integrated,liao2017satellite}. Specifically, a kilohertz key rate from the satellite to the ground can be realized over a distance of up to 1200 kilometers in \cite{liao2017satellite}.
By integrating the fiber and free-space QKD links, the communication distance between users can extend up to 4600 kilometers \cite{chen2021integrated}. 
These works mark significant achievements for long-distance quantum communication.

Similarly, secure communication can be achieved between aircraft and gateways by setting up quantum links to ensure the safety and reliability of air communication.
Xue \textit{et al.} provides an overview of the current state of research on airborne QKD systems, highlighting the technical challenges and potential applications of these systems \cite{XueChen21}. 
Due to the difficulty in adapting the traditional optomechanical structure to small UAV platforms, there needs to be more research on long-range optical quantum links for small-sized UAVs.
Tu \textit{et al.} design a new acquisition and tracking system for smaller and lighter airborne quantum systems \cite{9913625}. This article provides the design of the optical system, the electronics, and the software used to control the system. The experimental results prove the wide-range tracking, tracking accuracy, and fast response. 
Tian \textit{et al.} propose a new approach to set up a compact QKD system, which is mounted on a drone and is capable of establishing a secure communication link with a GS \cite{2023arXiv230214012T}. The drone-based QKD system is capable of generating secure keys at a high rate and over a range of several kilometers. 
Trinh \textit{et al.} discuss the advantages of using free-space optical (FSO) for QKD in satellite and UAV networks, which can enhance the secrecy performance and reduce implementation cost. The effects of atmospheric turbulence, beam pointing and tracking, and the need for stable and reliable hardware are also provided as technical challenges \cite{8597918}.

\subsection{Integration With Maritime Applications}
The marine network comprises a seafloor observation network and a self-organized underwater network \cite{9019858}. The former communicates by laying optical cables and underwater cables, mainly used for communication with shore-based networks and the related centers on the ground. 
Due to the high construction and maintenance costs caused by the harsh marine environment, the large-scale seafloor observation network is hard to deploy. 
The latter, including fixed sensor nodes and mobile ones like autonomous underwater vehicles (AUVs) and USVs, is an expansion of the underwater network in wireless communications. The self-organized underwater network uses acoustic waves for communication, sensing, navigation, and other fields. The signal attenuation rate with acoustic waves is 1/1000 of the electromagnetic wave, and the transmission distance is long. However, acoustic communication has the disadvantages of complex multi-path fading, high ambient noise, narrow bandwidth, and long transmission latency. 
Diamant \textit{et al.} proved that there is a strong correlation between the characteristics of the acoustic links and the reliability of optical communications, and the higher frequency measurements of the acoustics are more predictable to the optical links in most cases \cite{8053791}.

Upon completion of data transmission from subaqueous sensors to maritime vessels, the acquired information necessitates uploading to visible SAPs for subsequent exploitation. Although the integration of marine networks and SAIN can enhance maritime coverage effectively \cite{9344715}, the JCC issues of maritime applications are required to be solved urgently.
Unlike previous works with perfect channel state information (CSI) at all scales, only the position-related large-scale CSI is assumed to be available, and a minimum ergodic achievable rate maximization optimization problem is formulated in the hybrid satellite-UAV networks \cite{8960465}.
In the hierarchical satellite-UAV-terrestrial network, the joint link scheduling and rate adaptation problem of maritime communication networks is addressed in the same framework to minimize the total energy consumption while guaranteeing the QoS requirements \cite{9453860}. 
The works on computation offloading in UAV-assisted marine networks are still limited. After uploading the sensing data via NOMA from underwater sensor nodes to USV, the collected tasks are offloaded from the USV to multiple UAVs via OFDMA to avoid co-channel interference, and an energy consumption minimization problem is solved when ensuring secure performance.

\section{Conclusion}\label{sec:conclusion}
Against wide-coverage connectivity and an ever-increasing computing demand for services, SAGIN and MEC are at a pivotal stage in their development. Existing literature often focuses on isolated segments of SAGIN or the computing frameworks in terrestrial cellular networks, offering solutions that address these areas separately. Recognizing the gap in comprehensive research on computing functions of NTN, this paper aims to merge academic and industrial perspectives, offering a comprehensive overview of solutions and insights for JCC-SAGIN.

In this paper, we introduced the architecture, key enabling technologies, and applications of JCC-SAGIN, marking the transition from transparent forwarding to onboard processing. Then, an in-depth analysis of existing research on resource management in JCC-SAGIN was presented, especially in the areas of modeling and optimization for resource management. Subsequently, we envisioned the future research directions in JCC-SAGIN and outlined the existing and upcoming solutions to overcome these bottlenecks. 

This paper serves as an initial point for embedding MEC with SAGIN, aiming to advance research and exploration of JCC-SAGIN in the context of the 6G era. We hope this survey can contribute to the ubiquitous wireless communication and decentralized low-cost processing era.

	\ifCLASSOPTIONcaptionsoff
\newpage
\fi

\bibliographystyle{IEEEtran}
\bibliography{IEEEabrv,reference.bib}

\newpage
\begin{IEEEbiography}[{\includegraphics[width=1in,height=1.25in,clip,keepaspectratio]{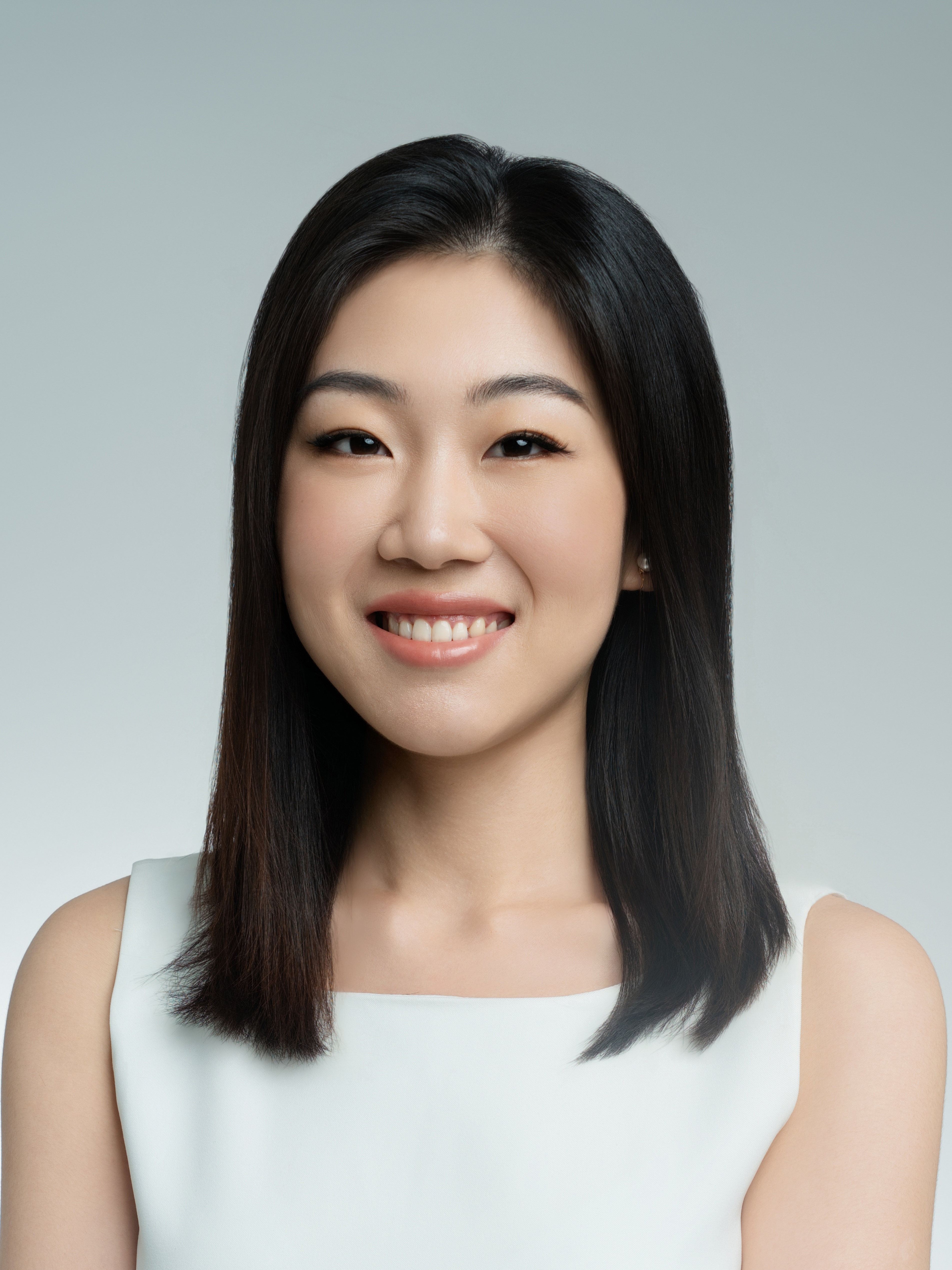}}]{Qian Chen}
	(Member, IEEE) received the B.E. and Ph.D. degrees in Information and Communication Engineering from the Harbin Institute of Technology (HIT), Harbin, China, in 2018 and 2023, respectively. She was a Visiting Ph.D. Student with the Information Systems Technology and Design (ISTD) Pillar, Singapore University of Technology and Design (SUTD), Singapore, from 2021 to 2022. 
She was the Student Chair of IEEE Communications Society Harbin Chapter from 2018 to 2023.
Her current research interests include edge intelligence and performance analysis in space-air-ground integrated networks.  
\end{IEEEbiography}

\vspace{-40pt}
\begin{IEEEbiography}[{\includegraphics[width=1in,height=1.25in,clip,keepaspectratio]{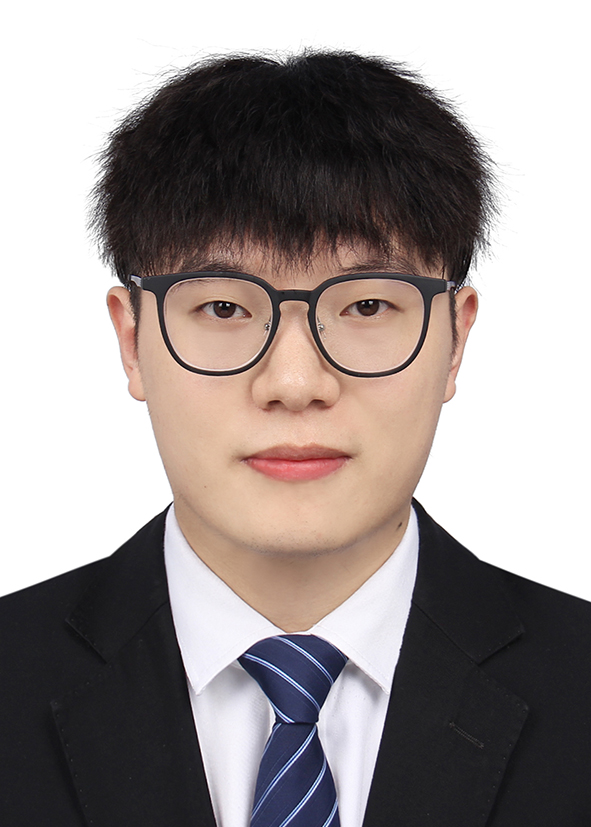}}]{Zheng Guo}
	(Graduate Student Member, IEEE) received the B.E. and M.E. degrees in information and communication engineering from the Harbin Institute of Technology (HIT), Harbin, China, in 2021 and 2023, respectively, where he is currently pursuing the Ph.D. degree. His current research interests include networking theory and resource management in space-air-ground integrated networks.
\end{IEEEbiography}

\vspace{-40pt}
\begin{IEEEbiography}[{\includegraphics[width=1in,height=1.25in,clip,keepaspectratio]{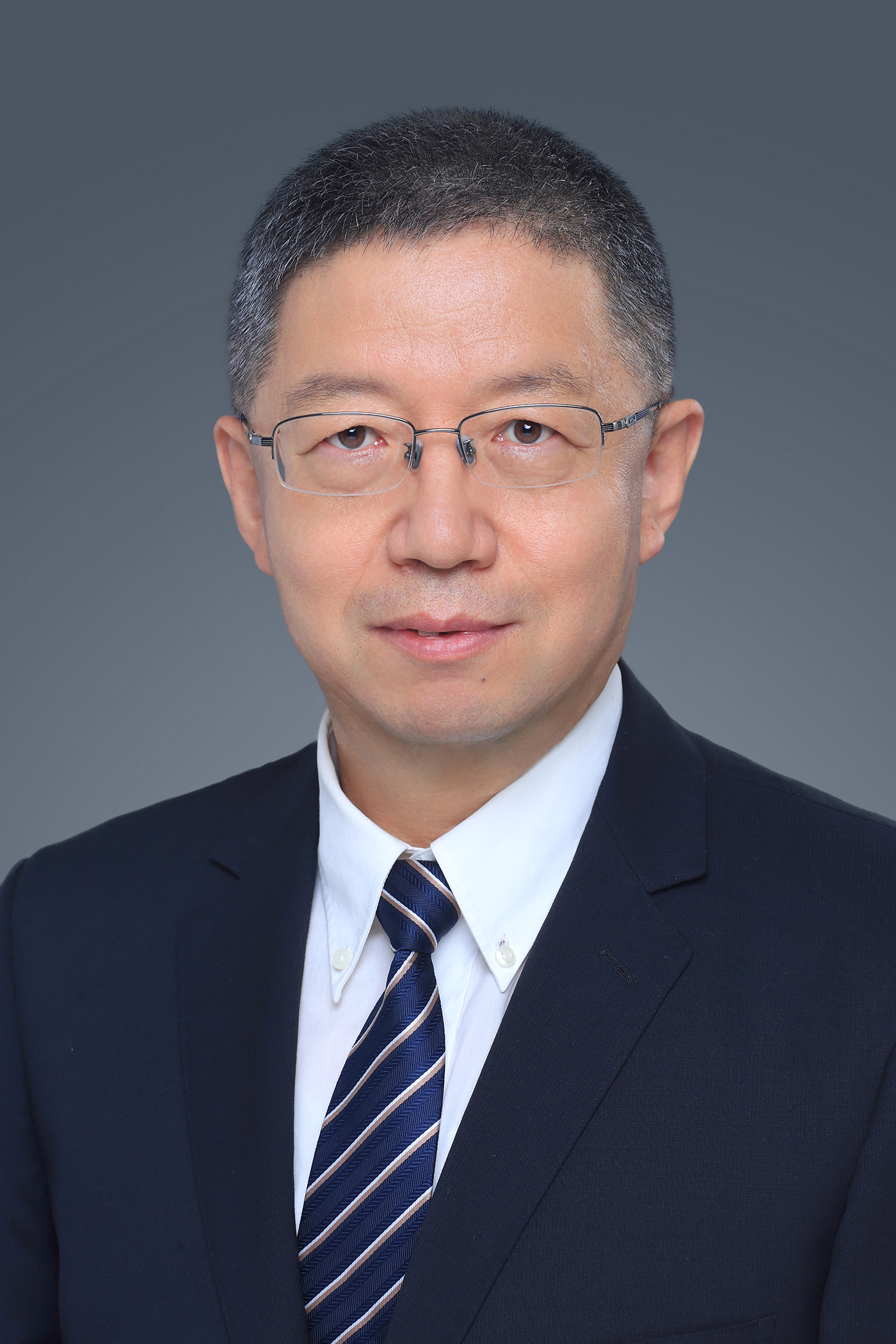}}]{Weixiao Meng}
	(SM'10) received the B.Eng., M. Eng., and Ph.D. degrees from Harbin Institute of Technology (HIT), Harbin, China, in 1990, 1995, and 2000, respectively. He worked in NTT DoCoMo as a visiting researcher from 1998 to 1999, and in University of California at Riverside as senior visiting scholar in 2017. He is now a full professor of the School of Electronics and Information Engineering of HIT. 
	His research interests include broadband wireless communications, space-air-ground integrated networks and communication and sensing integration. 
	
	He is the Chair of IEEE Communications Society Harbin Chapter, a Fellow of the China Institute of Electronics, a senior member of the IEEE ComSoc and the China Institute of Communication. He has been  an area editor for \textit{PHYCOM} journal from 2014 to 2016, an editorial board for  {\scshape IEEE Communications Surveys and Tutorials} from 2014 to 2017 and {\scshape IEEE Wireless Communications} from 2015 to 2020. He acted as Awards co-Chair of IEEE ICC 2015 and Wireless Networking Symposia co-Chair of IEEE Globecom 2015, AHSN Symposia co-Chair of IEEE Globecom 2018, leading Workshop co-Chair of IEEE ICC 2019 and IEEE ICNC 2020, AHSN Symposia co-Chair of IEEE ICC 2020. In 2005 he was honored provincial excellent returnee and selected into New Century Excellent Talents (NCET) plan by Ministry of Education (MOE), China in 2008, and the Distinguished Academic Leader of Harbin in 2015. He was the recipient of the Best Paper Award in 2021 {\scshape IEEE System Journal} and ICC 2023. From 2020 to 2022, he was selected into the Top 2\% of the World's Scientists by Stanford University.
\end{IEEEbiography}

\vspace{-30pt}
\begin{IEEEbiography}[{\includegraphics[width=1in,height=1.25in,clip,keepaspectratio]{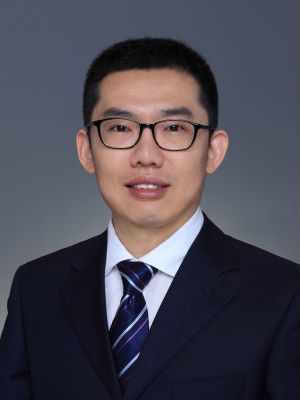}}]{Shuai Han}
	(S'11-M'12-SM'17) received the B.E., M.E. and Ph.D. degrees in Information and Communication Engineering from the Harbin Institute of Technology (HIT) in 2004, 2007 and 2011, respectively. And he completed his post-doctoral work in 2012 in Electrical and Computer Engineering at Memorial University of Newfoundland in Canada. He is currently a Full Professor with the Department of Electronics and Communication Engineering, HIT. His research interests include wireless communications, satellite IoT and integrated satellite-terrestrial communication networks. 

Dr. Han has more than twenty grants on wireless networks and positioning. He is an associate editor of IEEE China Communications, IEEE ACCESS, Journal of Communications and Information Networks (JCIN), Journal of Telemetry, Tracking and Command, Journal of Signal Processing. And has served as guest editor for many IEEE magazines and journals. He has served as a co-chair for technical symposia of international conference , IEEE GC 2023, ICC 2023, IEEE GC 2021, IEEE GC 2019, IEEE ICC 2018, IEEE VTC FALL 2016. He has also served as the TPC Chair for some international conferences, including the AICON2019 and MLICOM2018. He is a member of 2020-2021 R10 Awards \& Recognition Committee. Also, he is a senior member of IEEE Communication Society, Chair of IEEE BTS Chapter, Vice Chair of IEEE Harbin ComSoc Chapter, Vice Chair of IEEE Harbin VTS Chapter, and Vice Chair of IEEE IoT-AHSN TC.

\end{IEEEbiography}

\vspace{-30pt}
\begin{IEEEbiography}[{\includegraphics[width=1in,height=1.25in,clip,keepaspectratio]{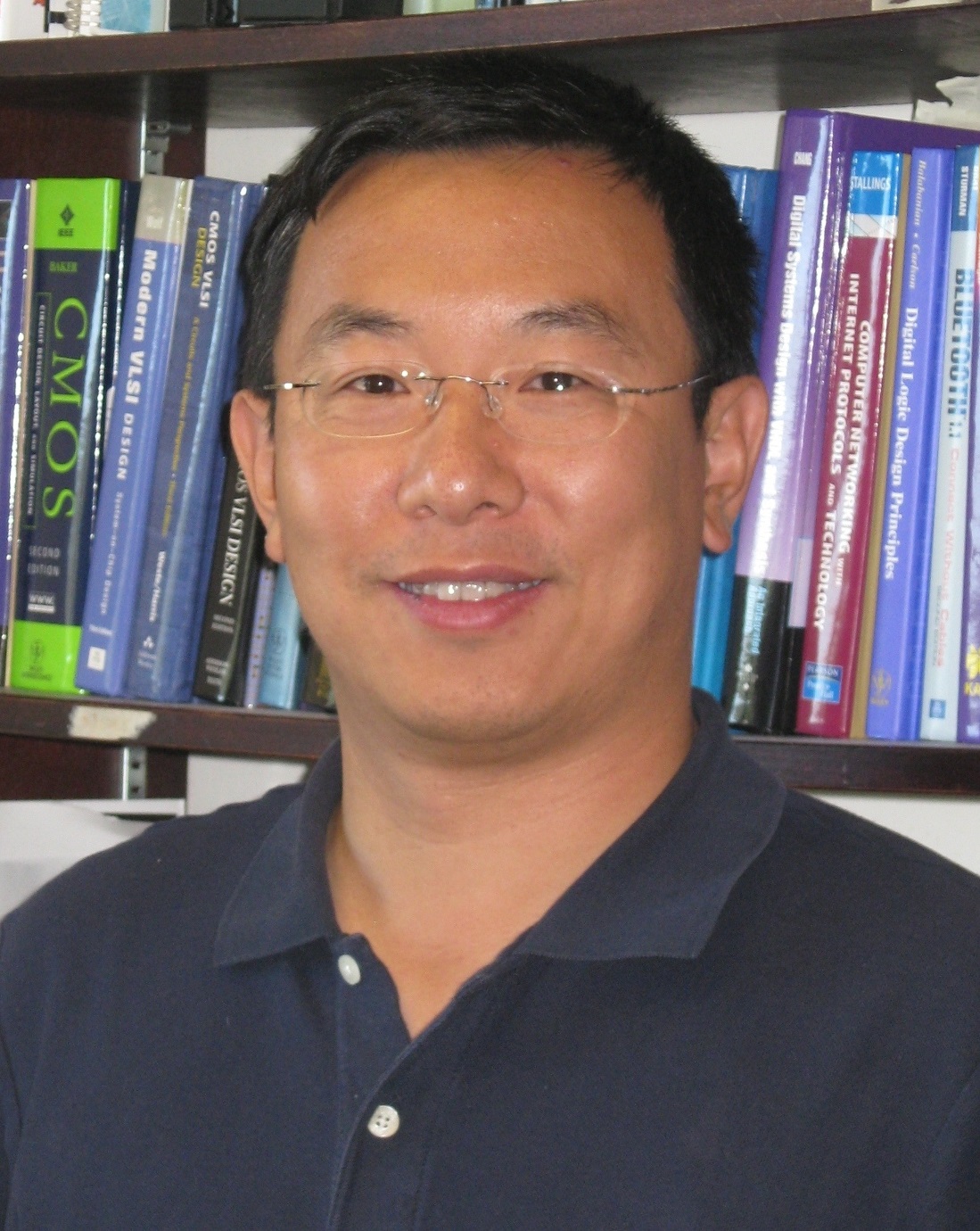}}]{Cheng Li}
	(M'04-SM'07) received his B.Eng. and M.Eng. degrees from Harbin Institute of Technology, Harbin, China, in 1992 and 1995, respectively, and his Ph.D. degree in Electrical and Computer Engineering from Memorial University, St. John's, Canada, in 2004. 

He is currently a Full Professor and Director of the School of Engineering Science of Simon Fraser University, Vancouver, Canada. He is also affiliated with the Department of Electrical and Computer Engineering, Memorial University, St. John's, NL, Canada. His research interests include wireless communications and networking, communications signal processing, underwater communications and networks, and mobile ad hoc and wireless sensor networks. He is an IEEE Communications Society Distinguished Lecturer for the 2021-23 term. He is an associate editor of the IEEE Transactions on Communications, IEEE Internet-of-Things Journal, and the IEEE Network Magazine. He has served as the General Co-Chair of the ICNC'23, Q2SWinet'20, WINCOM'19, and AICON'19, and the TPC Co-Chair for the ICNC'20, ADHOCNETS'19, ACOSIS'19, WiCON'17, MSWiM'14, WiMob'11 and QBSC'10. He also served many times as a Technical Program Co-Chair for various technical symposia/tracks of international conferences, including the IEEE GLOBECOM, ICC, WCNC, and VTC. He is the recipient of the Best Paper Award in IEEE ICC'23, Globecom'17 and ICC'10, and the Technical Achievement Award of the IEEE Communications Society Communications Software Technical Committee in 2018. Dr. Li is a registered Professional Engineer (P. Eng.) in Canada and is a Senior Member of the IEEE and a member of the IEEE Communication Society, Computer Society, Vehicular Technology Society, and Ocean Engineering Society.
\end{IEEEbiography}

\vspace{-30pt}
\begin{IEEEbiography}[{\includegraphics[width=1in,height=1.25in,keepaspectratio]{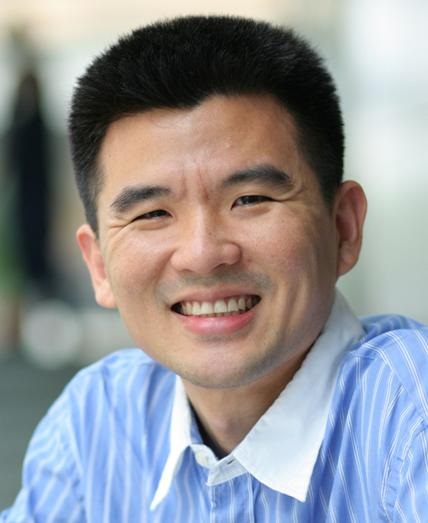}}]
	{Tony Q. S. Quek}(S'98-M'08-SM'12-F'18) received the B.E.\ and M.E.\ degrees in electrical and electronics engineering from the Tokyo Institute of Technology in 1998 and 2000, respectively, and the Ph.D.\ degree in electrical engineering and computer science from the Massachusetts Institute of Technology in 2008. Currently, he is the Cheng Tsang Man Chair Professor with Singapore University of Technology and Design (SUTD) and ST Engineering Distinguished Professor. He also serves as the Director of the Future Communications R\&D Programme, the Head of ISTD Pillar, and the Deputy Director of the SUTD-ZJU IDEA. His current research topics include wireless communications and networking, network intelligence, non-terrestrial networks, open radio access network, and 6G.

Dr.\ Quek has been actively involved in organizing and chairing sessions, and has served as a member of the Technical Program Committee as well as symposium chairs in a number of international conferences. He is currently serving as an Area Editor for the {\scshape IEEE Transactions on Wireless Communications}. 

Dr.\ Quek was honored with the 2008 Philip Yeo Prize for Outstanding Achievement in Research, the 2012 IEEE William R. Bennett Prize, the 2015 SUTD Outstanding Education Awards -- Excellence in Research, the 2016 IEEE Signal Processing Society Young Author Best Paper Award, the 2017 CTTC Early Achievement Award, the 2017 IEEE ComSoc AP Outstanding Paper Award, the 2020 IEEE Communications Society Young Author Best Paper Award, the 2020 IEEE Stephen O. Rice Prize, the 2020 Nokia Visiting Professor, and the 2022 IEEE Signal Processing Society Best Paper Award. He is an IEEE Fellow, a WWRF Fellow, and a Fellow of the Academy of Engineering Singapore.

\end{IEEEbiography}

%
%
%
%
%
%
%
%
%
%
%
	%

\end{document}